\documentclass[aps,prfluids,preprint,longbibliography,nofootinbib]{revtex4-2}

\usepackage{graphicx}
\usepackage{epstopdf}
\usepackage{amsmath}
\usepackage{amssymb}
\usepackage{dsfont}
\usepackage{booktabs}
\usepackage[dvipsnames]{xcolor}
\usepackage{caption}
\usepackage{hyperref}
\hypersetup{
	colorlinks = true,
	urlcolor   = blue,
	citecolor  = black,
	linkcolor  = black,
}

\graphicspath{{figures/}}

\definecolor{PRFblue}{RGB}{0,0,0}
\newcommand{\prfrev}[1]{{\color{PRFblue}#1}}
\begin{document}

\title{Duty-cycle modulation of the self-sustaining process by spanwise wall oscillation}

\author{Lionel Agostini}
\email{lionel.agostini@cnrs.fr}
\affiliation{Pprime Institute, Curiosity team, CNRS, Universit\'e de Poitiers, Poitiers, France}

\begin{abstract}
\prfrev{Direct numerical simulations of turbulent channel flow at $Re_{\tau,0}\approx200$ examine the
modulation of the near-wall self-sustaining process under sinusoidal and
shape-optimised spanwise wall oscillation, with the same peak wall-velocity
magnitude and oscillation period. At a period close to the reported sinusoidal optimum for the
prescribed peak velocity, the shape-optimised waveform provides a relative
increase of approximately $6\%$ in gross drag reduction over the sinusoidal
signal, at a higher actuation cost; it is therefore used as a probe of the physical mechanisms
underlying drag reduction.
Phase-resolved stresses, vorticity variances and component-enstrophy budgets
quantify the concurrent action of two tilting paths for the wall-normal vorticity associated with
the near-wall streaks: towards the streamwise direction by the mean shear, and
towards the spanwise direction by the Stokes strain. The magnitude of the Stokes
strain orders the two paths: the spanwise path prevails over most of the cycle, and
the streamwise path only while the strain is small, around its change of sign.
Within $5\le y^+\le12$, under the adopted interval criterion at $q=0.5$, the shape-optimised
waveform holds the strain at a nearly constant level for about $2/3$ of the cycle,
against $1/3$ for the sinusoidal waveform, and so approximately halves the duration
of Reversal, whilst it concentrates the phases of enhanced Reynolds
shear stress and of enhanced streamwise and wall-normal vorticity variances into shorter
intervals. The mean-shear production of streamwise enstrophy occurs mainly during
Reversal: for the shape-optimised waveform, its conditional rate is higher in every interval,
whereas its time average is $23\%$ lower, as Reversal is shorter. The lift-up
production of wall-normal enstrophy is lower at every sampled phase within $5\le y^+\le12$. These observations
support a duty-cycle description of the near-wall activity, in which the duration of each interval and the production rate within it are distinguished.}
\end{abstract}


\maketitle

\section{Introduction}
\label{sec:intro}

The reduction of turbulent skin-friction drag remains a principal objective in fluid mechanics, motivated by environmental and economic imperatives\prfrev{, and, amongst active flow-control strategies, spanwise wall actuation has received sustained attention with this aim \citep{ricco_review_2021}.} The drag-reduction potential of spanwise wall oscillation was first demonstrated experimentally by \citet{Laadhari1994}, who observed substantial attenuation of turbulent fluctuations in a boundary layer subjected to local spanwise surface motion. Subsequent experimental \prfrev{\citep{Choi1998,Ricco2004,Gouder2013}} and computational \citep{Jung1992,Baron1995,quadrio_critical_2004} investigations have established that simple temporal sinusoidal wall oscillation \prfrev{can reduce drag by up to about 40\%} at relatively low Reynolds numbers. \prfrev{The preferred waveform depends on the optimisation objective. When the net energy saving is optimised over the amplitude and period, a sinusoidal or near-sinusoidal signal gives the largest saving amongst the waveforms examined \citep{cimarelli2013prediction,guerin2026PBO}, although non-sinusoidal signals can perform better at suboptimal parameters \citep{cimarelli2013prediction}. When the gross drag reduction alone is optimised, square-wave or quasi-plateau waveforms outperform the sinusoid at a fixed amplitude \citep{cimarelli2013prediction,guerin2026PBO}, and square-wave spatial actuation has since been realised experimentally \citep{knoop2024experimental,Knoop2025}. For sinusoidal oscillation, the drag reduction collapses on the acceleration scale $W^+/T^+$ for $T^+\gtrsim100$ \citep{Ding2024}, a parameter that does not distinguish waveforms of equal peak velocity and period; non-sinusoidal waveforms that increase the cycle-integrated acceleration were proposed on this basis, without being tested.}

\prfrev{The physical basis of this drag reduction has been investigated extensively, and it is well established that the control relies upon} the generation of a transverse Stokes layer, which penetrates the near-wall region and interacts with the quasi-streamwise \prfrev{vortical structures} and velocity streaks responsible for turbulence production \citep{ricco_review_2021}. \prfrev{These near-wall structures participate in a self-sustaining process (SSP), identified by \citet{hamilton1995regeneration} and formalised theoretically by \citet{waleffe1997self}; this process resides in the buffer layer, above the viscous sublayer, and survives without input from the outer flow \citep{jimenez1999autonomous}. In vorticity terms, \citet{jimenez1999autonomous} identify the streaks with the streamwise-averaged wall-normal vorticity, whose source is the advection of spanwise vorticity, dominated by the mean shear, by the wall-normal velocity. Damping this source, or \prfrev{the linear term} through which the wall-normal velocity generates wall-normal vorticity in the presence of the mean shear, causes the near-wall turbulence to decay \citep{jimenez1999autonomous,kim2000linear}. In plane Couette flow at $Re_\tau\approx33$, spanwise wall oscillation was found to stabilise the upper-branch exact coherent states of this process and to reduce their period-averaged lift-up production \citep{bengana2022exact}.}

\prfrev{Near-wall vorticity budgets attribute the production of streamwise vorticity mainly to the tilting of wall-normal vorticity by the mean shear \citep{brooke1993origin,umair2023vorticity}. In the instantaneous streamwise-vorticity equation, that tilting appears as $\omega_y\,\partial_yu$ (with $\partial_y\equiv\partial/\partial y$, and likewise for the other coordinates and for time), accompanied by the tilting of the spanwise vorticity by the spanwise gradient of the streamwise velocity, $\omega_z\,\partial_zu$. As $\omega_y=\partial_zu-\partial_xw$ and $\omega_z=\partial_xv-\partial_yu$, the product $(\partial_zu)(\partial_yu)$ appears in the two terms with opposite signs and therefore cancels \citep{brooke1993origin}. Of the two contributions that remain, $-(\partial_xw)(\partial_yu)$ and $(\partial_xv)(\partial_zu)$, the former is the dominant net source near the wall and requires streamwise variation of the spanwise velocity \citep{brooke1993origin,umair2023vorticity}. This source produces thin sheets of streamwise vorticity rather than \prfrev{tubes}, and these sheets collapse into quasi-streamwise \prfrev{vortical structures} when stretched by the streamwise gradient $\partial_xu$ that develops as the streaks are displaced sinuously in the spanwise direction \citep{Schoppa2002}.}

\prfrev{Previous investigations of sinusoidal wall actuation have related
phase-dependent attenuation and recovery of near-wall turbulence to the
magnitude and temporal variation of the Stokes strain
\citep{touber_near-wall_2012,agostini2014spanwise,Yakeno2014}, an interpretation
since extended to spatial square-wave actuation
\citep{knoop2024experimental,Knoop2025}.
\citet{agostini2014spanwise} further inferred that the streaks do not recover
fully when the part of the period during which the strain allows their
regeneration is shorter than the time scale of streak amplification,
$t^+\approx80$, estimated from linear optimal-perturbation analysis
\citep{blesbois2013pattern}. In their simulations at
$Re_\tau=1025$ with a peak wall velocity $W^+=12$, this regeneration interval
was about $t^+=40$ at the suboptimal period $T^+=200$, selected
deliberately to accentuate the variations around the reduced-drag state. The
streaks then recovered partially, and the drag decreased and increased in
succession, whereas at $T^+=100$, close to the drag-reduction optimum for those
conditions, the interval fell to about $t^+=20$ and the streaks remained more
uniformly attenuated throughout the cycle
\citep{agostini2014spanwise,agostini_turbulence_2015}. The period, however,
also dictates the penetration of the Stokes layer and the strain near the wall
\citep{agostini2014spanwise,ricco_review_2021,quadrio2026optimal}; at a fixed
period and peak wall velocity, the waveform determines instead the distribution of the strain over the cycle, as
quantified in \S\S\ref{sec:temporal_modulation} and \ref{sec:duty_cycle}.}

\prfrev{The component-enstrophy analysis of
\citet{agostini_turbulence_2015} identified the dominant stochastic-enstrophy
production terms under sinusoidal oscillation. Near the wall, the tilting of
the wall-normal vorticity towards the spanwise direction by the Stokes strain,
together with stretching, amplifies the spanwise enstrophy, whilst the
wall-normal enstrophy, closely associated with the streaks, diminishes mainly
through a reduction of its mean-shear production. That study related the phase relationships between these terms and the
turbulent-stress response, and associated the phases of drag increase with weak
spanwise tilting of the wall-normal vorticity near the wall; this association
rested on correlations and did not quantify the relative weight of the two tilting
processes, and the streamwise-enstrophy budget was not examined. \citet{umair2023vorticity} obtained all three component budgets
under streamwise-travelling waves and homogeneous oscillation, and found the
same Stokes-strain tilting to be a significant source of spanwise enstrophy.
The relevance of vorticity to the stress response is
expressed by the kinematic identity}
\begin{equation}
 -\frac{\partial \widetilde{u^{\prime\prime}v^{\prime\prime}}^+}{\prfrev{\partial y^+}}
 =\widetilde{v^{\prime\prime}\omega_z^{\prime\prime}}^+
 -\widetilde{w^{\prime\prime}\omega_y^{\prime\prime}}^+
 \label{eq:shear_stress_vorticity}
\end{equation}
\prfrev{which relates the wall-normal gradient of the phase-coherent Reynolds shear stress to
velocity--vorticity correlations \citep{tennekes1972first}; the tilde denotes the phase-coherent part defined in \S\ref{sec:phase_analysis}, and the identity holds equally for the phase-conditional averages $\langle\cdot\rangle$ from which it is obtained by subtracting the time average. This identity,
which follows from the definition of vorticity and from incompressibility
after averaging over the statistically homogeneous streamwise and spanwise
directions, is satisfied at every height and phase. The stress and vorticity
descriptions are thereby connected, and their evolution is examined here
through the corresponding transport equations.}

\prfrev{The mean shear and the Stokes strain thus both tilt the wall-normal
vorticity, the former towards the streamwise direction
\citep{brooke1993origin,umair2023vorticity} and the latter towards the
spanwise direction \citep{Ricco2012,agostini_turbulence_2015,umair2023vorticity}. How
the relative importance of these two processes evolves during the actuation
cycle, and whether the
waveform alters the balance between the resulting production terms and the
time over which each acts, have not been examined, nor has the relation of such
a waveform-induced change to the drag reduction.}

\prfrev{The present study addresses these questions by comparing a sinusoidal
wall-velocity signal with a shape-optimised, quasi-square waveform, obtained
by policy-based optimisation (PBO) for gross drag reduction. Non-sinusoidal
waveforms entail a higher actuation cost, as noted above, and the
shape-optimised waveform is therefore employed here mainly as a probe of the physical mechanisms underlying drag
reduction. At $Re_{\tau,0}\approx200$ and a peak wall velocity $W^+=15$, the
optimisation identified $T^+\approx111$ for the shape-optimised waveform, within the range of optimal
periods reported for sinusoidal actuation
\citep{quadrio_critical_2004,guerin2025policy}. The two signals are therefore compared at this common
period and peak velocity; the optimisation and waveform definition are
detailed in \S\ref{sec:methodology}. The comparison is designed to separate
the intensity of each production term, quantified through phase-resolved
component-enstrophy budgets, from the duration over which it acts, and the
relation of these terms to the measured stress and vorticity levels is
examined over the actuation cycle.}

\prfrev{This Reynolds number allows the interaction between spanwise wall
actuation and the near-wall small-scale cycle to be examined directly, as in
earlier detailed accounts of temporal wall oscillation
\citep{Jung1992,quadrio_critical_2004,touber_near-wall_2012} and of
streamwise-travelling waves of spanwise wall velocity
\citep{quadrio_ricco_viotti_2009}. It must be acknowledged that these
conditions do not reproduce the full hierarchy of outer motions encountered at
higher Reynolds numbers, which modulate the near-wall cycle under oscillation \citep{agostini_statistical_2021}, and the present conclusions therefore concern
small-scale control in this channel-flow regime; the implications for
multiscale control are considered in \S\ref{sec:reynolds_scope}.}

\prfrev{The paper is organised as follows. Section~\ref{sec:methodology} details the direct numerical simulation (DNS)
methodology, the optimisation procedure employed to identify the quasi-square
waveform, the actuation parameters and the phase-resolved analysis framework.
Section~\ref{sec:results} presents the drag-reduction performance, the
Reynolds stresses and their production, the time-averaged vorticity profiles,
the temporal variation of the Stokes strain, and the phase-dependent turbulent
response together with the actuation intervals that organise it. Section~\ref{sec:enstrophy}
presents the phase-resolved budgets of the three stochastic-enstrophy
components and their production by the mean shear and the Stokes strain.
Section~\ref{sec:discussion} discusses the waveform comparison as a
diagnostic, the connection between the vorticity budgets and momentum
transport, the Stokes-layer response in relation to earlier studies, and the
Reynolds-number scope of the conclusions.
Section~\ref{sec:conclusions} summarises the principal conclusions, and
Appendices~\ref{app:enstrophy_equations} and \ref{app:verification} give the
stochastic-enstrophy budget equations and the quantitative checks,
respectively.}

\section{Methodology}
\label{sec:methodology}

\subsection{Numerical simulation configuration}

	Direct numerical simulations \prfrev{(DNS)} are performed using the open-source incompressible Navier--Stokes solver Xcompact3d \citep{laizet2009high,laizet2011incompact3d,bartholomew_xcompact3d_2020}. The framework employs sixth-order compact finite-difference schemes for spatial discretisation and a third-order explicit Runge--Kutta scheme for time integration. The condition of zero velocity divergence is enforced through a fractional step method, wherein a Poisson equation for the pressure is solved using three-dimensional fast Fourier transforms \citep{laizet2009high}. The baseline \prfrev{unactuated} flow is a turbulent channel flow at a friction Reynolds number $Re_{\tau,0} = u_{\tau,0} h/\nu \approx 200$ \prfrev{(exactly 201.6, with $h=1$, $\nu=1/4800$ and the nominal reference $u_{\tau,0}=0.042$ in the code units of the solver; this reference is consistent with the measured $C_{f,0}=7.94\times10^{-3}$)}, where $h$ denotes the channel half-height and $\nu$ the kinematic viscosity. \prfrev{Wall units are based on this friction velocity for all cases, including the controlled ones; peak locations are thus compared between cases at a common scaling.}
	 The simulations employ a constant-flow-rate (CFR) strategy, wherein the bulk velocity $U_b$ is maintained constant across all cases, a conceptual choice consistent with previous investigations such as \citet{cimarelli2013prediction}. In this framework, the friction Reynolds number $Re_{\tau}$ of the controlled cases decreases as drag is reduced, reflecting the \prfrev{lower} wall-shear stress required to maintain the fixed mass flow rate. Drag reduction (DR) is quantified as the relative reduction in the mean streamwise pressure gradient required to drive the flow:
	\begin{equation}
		\label{eq:dr_percent}
		DR\% = \frac{(\partial P/\partial x)_0 - (\partial P/\partial x)_{\text{control}}}{(\partial P/\partial x)_0} \times 100
	\end{equation}
	where the subscript `0' \prfrev{refers to} the \prfrev{unactuated} reference case. \prfrev{The uncertainty of the drag reduction includes the uncertainty of the unactuated reference (Appendix~\ref{app:verification}).}

	The computational domain spans $(L_x, L_y, L_z) = (24h, 2h, 6h)$ in the streamwise, wall-normal, and spanwise directions respectively, discretised using $(N_x, N_y, N_z) = (400, 221, 200)$ grid points. \prfrev{In nominal wall units, the resolution is $\Delta x^+\approx12.1$ and $\Delta z^+\approx6.0$ in the periodic directions, and the wall-normal grid is stretched from $\Delta y^+\approx0.48$ at the wall to $\Delta y^+\approx7.0$ at the centreline; the time step is $\Delta t^+\approx0.017$.}

\prfrev{The present mesh is the mesh designated S2 in \citet{guerin_preferential_2024} and in Chapter~2 of \citet{Guerin2025}, which give the complete details of the numerical implementation and its validation against established benchmarks for the unactuated channel at $Re_\tau\approx180$ (Section~3 of the former).} \prfrev{For the actuated flow, the drag reduction obtained on the present mesh deviates by 2.1\,\% from that of \citet{quadrio_critical_2004} for $T^+=200$ and $W^+=12$ (Section~4.1 of \citet{guerin_preferential_2024}). At the operating point of the sinusoidal case, the wall-motion power, 119\,\% of the pumping power of the unactuated flow, agrees with the value obtained from cases 15, 16, 23 and 24 of \citet{quadrio_critical_2004} ($T^+=100$ and 125, $W^+=12$ and 18) by scaling with $W^{+2}$, and the gross drag reduction, $38.4\,\%$, lies within the range of these cases, $32.5$--$39.3\,\%$. No mesh refinement was made for the vorticity statistics of the actuated flows.} \prfrev{Below $y^+=30$, the grid spacings of the controlled flows do not exceed $6\eta$ in the streamwise, $3\eta$ in the spanwise and $0.4\eta$ in the wall-normal direction, where $\eta$ is the Kolmogorov length estimated from the pseudo-dissipation. For the unactuated flow, the r.m.s. vorticity fluctuations agree with the DNS of \citet{moser1999} at $Re_\tau=180$ to within 2\,\% for the wall-normal and spanwise components and 6\,\% for the streamwise component over $5\le y^+\le30$, and the vorticity budgets close with the residuals given in Appendix~\ref{app:verification}.}

\subsection{Generation of the shape-optimised waveform via Policy-Based Optimisation}

\begin{figure}
	\centering
	\includegraphics[width=0.6\columnwidth]{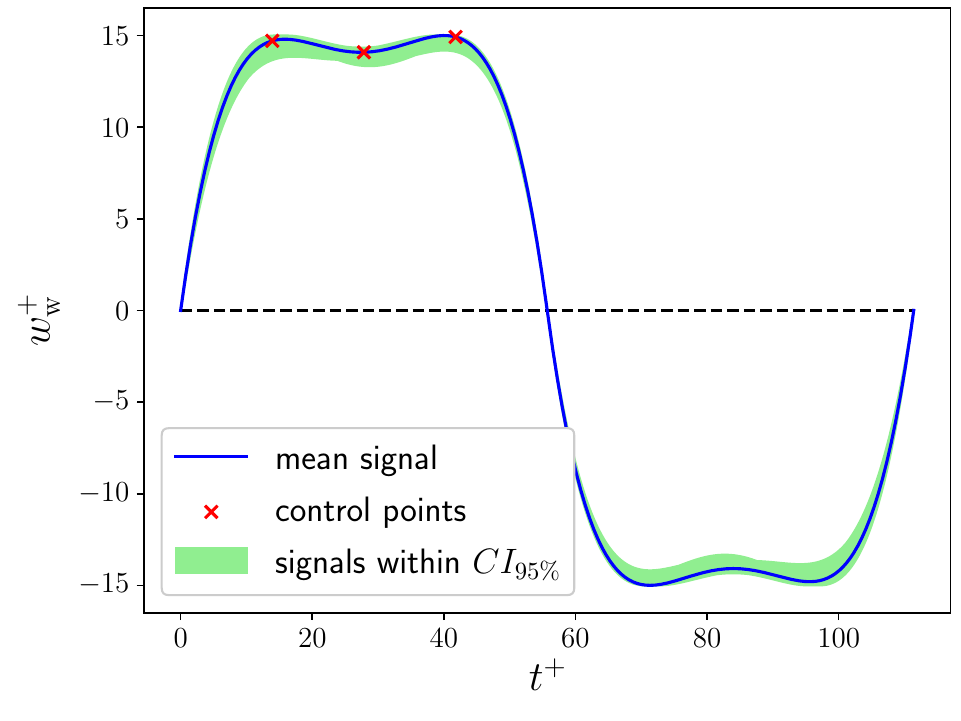}
	\caption{\prfrev{Shape-optimised spanwise wall velocity at $W^+=15$ and
		$T^+ \approx 111$, identified by PBO
		\citep{guerin2026PBO}. The shaded region represents the 95\% uncertainty band
		of the optimisation model. Crosses: intermediate control points of the parameterisation.}}
	\label{fig:waveform_shape}
\end{figure}

\prfrev{The shape-optimised waveform employed in this study
(Figure~\ref{fig:waveform_shape}) is taken from \citet{guerin2026PBO} and is
the converged outcome of a systematic policy-based optimisation (PBO)
procedure rather than an ad hoc choice. That study extended to
non-sinusoidal waveforms the PBO methodology established by
\citet{guerin2025policy} for sinusoidal actuation: candidate signals are
evaluated by implicit large-eddy simulation (LES), the parameter distribution
is updated by policy-gradient ascent, and the selected configuration is
validated by DNS. The optimisation itself is not the subject of the present
paper: its parameter space, convergence history and the candidate waveforms
evaluated are documented in \citet{guerin2026PBO} and, in detail, in Chapters 4
and 5 of \citet{Guerin2025}.}

	Two complementary optimisation problems were addressed in \citet{guerin2026PBO},
	distinguished by the choice of objective function.  In the first, the objective
	is the gross drag-reduction rate $P_{\text{sav}}$, with the peak velocity
	amplitude fixed at $W^+ = 15$ whilst the oscillation period $T^+$ and the
	waveform shape parameters $(a_1, a_2)$ were left free; the optimisation therefore
	sought:
\begin{equation}
	\label{eq:opt_objective}
	\prfrev{(T^+,a_1,a_2)_{\mathrm{opt}}} = \operatorname*{arg\,max}_{\prfrev{(T^+,\,a_1,\,a_2)}}
	\left[ \frac{\tau_{0} - \tau(w_w^+)}{\tau_{0}} \right]\prfrev{,}
\end{equation}
where $\tau(w_w^+)$ \prfrev{is} the wall-shear stress under control and
$\tau_0$ the \prfrev{unactuated} reference\prfrev{; the bound $|w_w^+|\le W^+$
and the periodicity of the waveform are satisfied by construction of the
parameterisation described below}.  In the second, the objective is the net
	energetic gain $P_{\text{net}} = P_{\text{sav}} - P_{\text{req}}$, which includes
	the power expenditure associated with driving the wall motion, with both
	amplitude and period free.  These two problems produce substantially different
	outcomes: optimisation for $P_{\text{sav}}$ converges to a quasi-square-wave
	(plateau-and-impulse) topology, whereas optimisation for $P_{\text{net}}$
	converges to a near-sinusoidal waveform, \prfrev{both outcomes agreeing with
	the discrete sampling of \citet{cimarelli2013prediction}. The present study
	examines the first objective, gross drag reduction.}

	The waveform was parameterised by \prfrev{Lagrange interpolation over the
	first half-period}, with five control points. To ensure
	continuity and periodicity, the first and last control points were fixed at
	zero, whilst \prfrev{the three intermediate points, shown as crosses in
	Figure~\ref{fig:waveform_shape},} were governed by two free parameters
	$(a_1, a_2)$ as described in \citet{guerin2026PBO}.
\prfrev{With the phase defined as $t^*=t^+/T^+$, the five control points are located at
$t^*=0$, $1/8$, $1/4$, $3/8$ and $1/2$, where the normalised amplitudes are
$0$, $a_1$, $a_2(1-a_1)$, $(1-a_1)(1-a_2)$ and $0$. The fourth-order
Lagrange polynomial through these points is scaled to a maximum of $W^+=15$, and the second half-period is the time-mirrored and sign-reversed
first half, $w_w(t^+)=-w_w(T^+-t^+)$. The search covered $T^+\in[10,300]$ and
$(a_1,a_2)\in[0,1]^2$, and converged within 480 LES evaluations to $T^+\approx111$, $a_1=0.336481$ and $a_2=0.485570$, an optimum within this parameterisation; see Supplemental Material at [URL will be inserted by publisher] for the executed nodal values, the polynomial coefficients and the waveform over one period.}

\prfrev{The optimum of this search, $T^+\approx111$, is within the range of periods
at which sinusoidal actuation is most effective. For a given amplitude at
$Re_\tau\approx200$, \citet{quadrio_critical_2004} obtained the maximum drag
reduction at $T^+=100$--125, and the PBO of \citet{guerin2025policy}
identified an optimal period in the range $T^+=110$--125. This optimum
is commonly interpreted as reflecting two opposing constraints. At substantially shorter periods, the
Stokes layer, whose thickness scales as $(\nu T)^{1/2}$, remains confined to
the viscous sublayer, below the streaks at the bottom of the buffer layer. At
substantially longer periods, the streaks have time to re-establish within
each cycle and to reorient with the slowly varying Stokes strain
\citep{touber_near-wall_2012,blesbois2013pattern,agostini2014spanwise}.
\citet{Yakeno2014} attributed the optimal period to a trade-off between the
suppression of ejections, which dominates up to the optimum, and the
enhancement of sweeps by the spanwise tilting of the near-wall structures,
which prevails at longer periods.}

\subsection{Wall actuation parameters}
\label{sec:actuation}
	Two distinct spanwise wall oscillation strategies are examined. The sinusoidal
	baseline employs harmonic temporal actuation \prfrev{$w_w^+=W^+\sin(2\pi t^*)$},
	whilst the shape-optimised configuration implements the quasi-square-wave profile
	described in the preceding subsection. Both operate at $T^+ \approx 111$ and
	$W^+ = 15$\prfrev{; the two cases therefore differ only in waveform shape}. \prfrev{The two walls move in phase with identical velocity; the mean spanwise velocity is consequently symmetric, and the Stokes strain antisymmetric, about the centreline.}

\prfrev{The wall acceleration determines how the actuation is imparted to the
near-wall flow. As no spanwise pressure gradient is imposed and the wall-normal gradient
of the Reynolds stress $\langle v''w''\rangle$ vanishes at the wall, the
phase-averaged spanwise momentum equation reduces there to
$\partial_{t^+}w_w^+=\partial_{y^+}^2\langle w\rangle^+|_{y=0}
=\partial_{y^+}G^+|_{y=0}$, where $G=\partial_y\langle w\rangle$
is the Stokes strain; the acceleration therefore sets the flux of Stokes strain
from the wall into the flow.
Both waveforms share the acceleration scale $W^+/T^+\approx0.135$, whereas
their peak accelerations differ. For the sinusoidal waveform,}
\begin{equation}
\prfrev{\max\left|\frac{\partial w_w^+}{\partial t^+}\right|
 =\frac{2\pi W^+}{T^+}\approx0.85,}
\end{equation}
\prfrev{whereas the shape-optimised waveform reaches approximately 2.7 at the wall-velocity reversals, about 3.1 times the sinusoidal value. The
time-averaged magnitudes of the two accelerations, in contrast, differ by approximately 5\,\%
(0.54 and 0.56, respectively), as both waveforms reverse the wall
velocity between $-W^+$ and $W^+$ in each half-period, the excess of the shape-optimised waveform being the additional travel of its non-monotone profile; their r.m.s. values differ by 53\,\%. For the
shape-optimised waveform, the wall acceleration, and with it the flux of
Stokes strain into the flow, is therefore largely confined to short intervals
around the reversals. Its magnitude remains below $1/10$ of its peak over
58\,\% of the cycle, whereas the sinusoidal acceleration, which varies as
$\cos(2\pi t^*)$, remains below $1/10$ of its own peak over only 6\,\%. For a periodic waveform, the mean spanwise motion of the turbulent flow is approximated by the laminar
Stokes solution, built by the harmonic superposition of the waveform \citep{cimarelli2013prediction};
the waveform therefore prescribes the distribution of the Stokes strain over the cycle. The Stokes strain is thus expected to reverse
more sharply and to vary slowly between the reversals; the resulting strain is
characterised in \S\ref{sec:temporal_modulation} and
Table~\ref{tab:forcing_verification}, and the durations of these intervals in
\S\ref{sec:duty_cycle}.}

\prfrev{The intervals around the reversals of the Stokes strain and those
between them are termed Reversal and Displacement, respectively. During
Reversal, the strain changes rapidly and passes through zero; during
Displacement, it varies slowly and remains close to its maximum magnitude. As
the actuation is transmitted from the wall by viscous diffusion through the
Stokes layer, the strain at a given wall-normal location lags the wall motion;
the two intervals are therefore defined locally, from the rate of change of the
Stokes strain at each $y^+$, rather than from the wall motion
(\S\ref{sec:duty_cycle}).}

\subsection{Phase-resolved analysis framework}
\label{sec:phase_analysis}

	The periodic nature of the wall actuation introduces a deterministic temporal
	structure into the turbulent flow; a decomposition framework capable of separating
	this phase-locked response from the underlying stochastic turbulence is therefore
	required.  Standard Reynolds decomposition, which separates the instantaneous
	velocity $u_i$ into a time average $\overline{u_i}$ and a fluctuation $u_i'$, is
	insufficient for this purpose, as it cannot distinguish between fluctuations
	correlated with the actuation phase and fluctuations of genuinely stochastic origin.
	A triple decomposition is therefore adopted, wherein the instantaneous \prfrev{velocity} is
	expressed as
\begin{equation}
	u_i = \overline{u_i} + \widetilde{u_i}(t^*) + u_i^{\prime\prime}(t^*),
\end{equation}
	comprising three components: the time average $\overline{u_i}$; the phase-coherent
	fluctuation $\widetilde{u_i}(t^*)$, which captures the deterministic oscillation
	induced by the wall actuation; and the stochastic fluctuation
	$u_i^{\prime\prime}(t^*)$, \prfrev{whose phase average vanishes, whilst its
	statistics depend on phase}.

\begin{figure}
	\centering
	\includegraphics[width=0.8\textwidth]{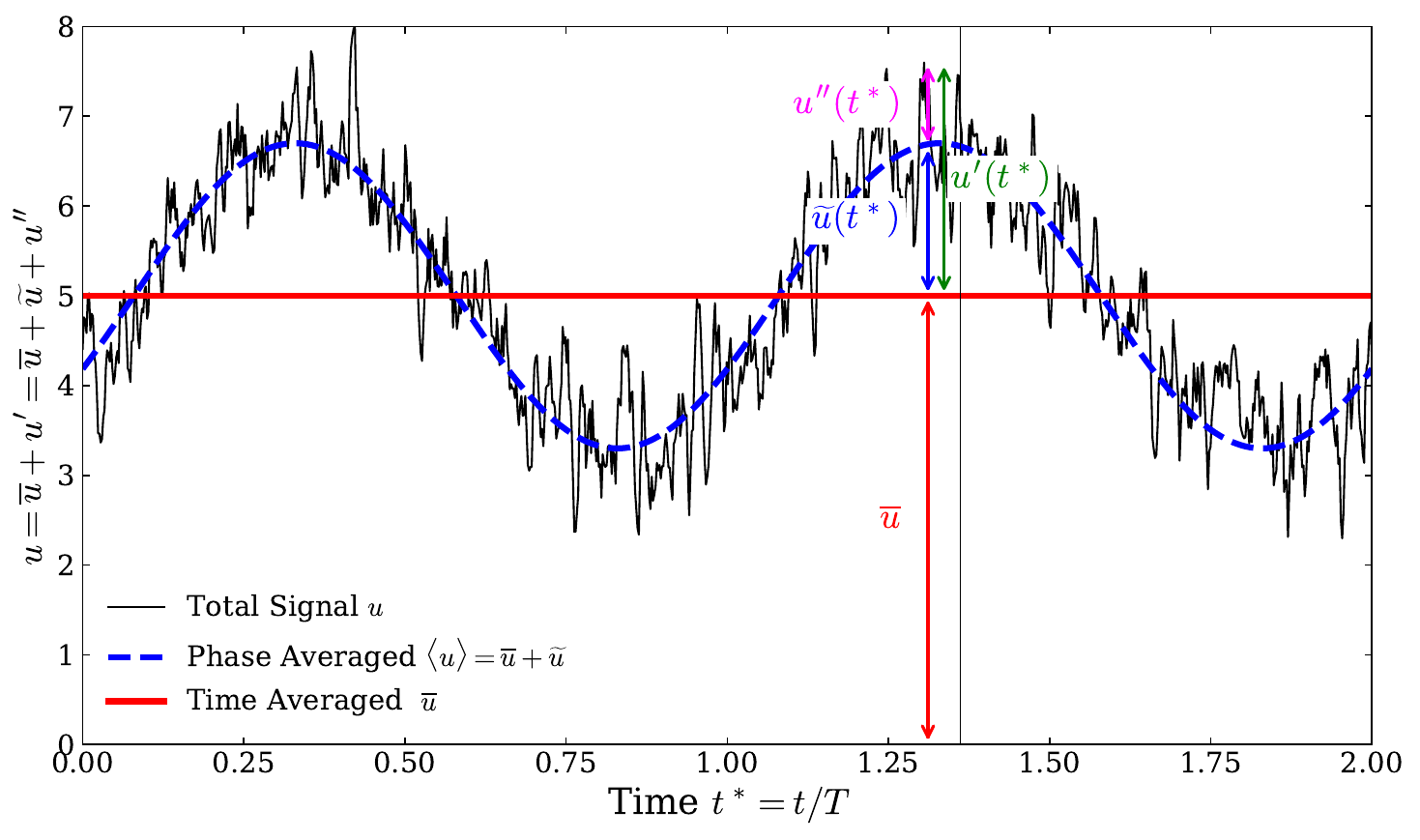}
	\caption{\prfrev{Triple decomposition for periodically actuated channel flow:
		total signal $u$ (solid black), time average $\overline{u}$ (solid red),
		phase-averaged component $\langle u \rangle$ (dashed blue). Arrows indicate the
		phase-coherent component $\widetilde{u}(t^*)$ (blue) and the stochastic
		fluctuation $u^{\prime\prime}$ (magenta), whose sum is the Reynolds fluctuation $u'$ (green).}}
	\label{fig:signal_decomposition}
\end{figure}

	The phase-coherent component is extracted by ensemble averaging at fixed
	phase $t^*\in[0,1]$,
	across $N_T$ complete cycles\prfrev{, and over the homogeneous streamwise and spanwise directions}:
\begin{equation}
	\prfrev{\langle u_i \rangle(y,t^*) = \frac{1}{N_T L_x L_z} \sum_{n=0}^{N_T-1} \int_0^{L_x}\!\!\int_0^{L_z} u_i(x,y,z,t^* T + nT)\,\mathrm{d}z\,\mathrm{d}x.}
\end{equation}
	For notational brevity this phase-conditional average is denoted $\langle\cdot\rangle$
	throughout.  The phase-coherent component then follows as
	$\widetilde{u_i}(t^*) = \langle u_i \rangle(t^*) - \overline{u_i}$, satisfying
	\prfrev{$\overline{\widetilde{u_i}}=0$} by construction, and the stochastic
	fluctuation is $u_i^{\prime\prime} = u_i - \langle u_i \rangle$.
	Figure~\ref{fig:signal_decomposition} illustrates these three components
	graphically.

	This decomposition is applied to any flow quantity $A$, including the Reynolds
	stresses and vorticity variances examined in subsequent sections, with
	$\widetilde{A}(t^*) = \langle A \rangle(t^*) - \overline{A}$.  For variance
	quantities, positive values of $\widetilde{A^{\prime\prime}A^{\prime\prime}}$
	identify phases during which turbulence intensity exceeds \prfrev{its time average};
	\prfrev{negative values identify phases during which the intensity is below the time average of the same controlled case. Such phases correspond to weaker activity within the cycle; they imply neither suppression below the unactuated flow nor an interruption of the SSP.}  This differential
	representation isolates the cyclic modulation of turbulent activity provoked by
	the wall actuation.
	The same decomposition is applied to the vorticity
	\prfrev{$\omega_i=\epsilon_{ijk}\,\partial_ju_k$}.  The streamwise component
	$\omega_x$ characterises the quasi-streamwise \prfrev{vortical structures}; the wall-normal component
	$\omega_y$\prfrev{, whose streamwise average is associated with the streaks (\S\ref{sec:intro}), appears in the velocity--vorticity correlation}
	$\widetilde{w^{\prime\prime}\omega_y^{\prime\prime}}$ appearing in
	equation~(\ref{eq:shear_stress_vorticity}); and the mean spanwise component \prfrev{equals}
	\prfrev{$\overline{\omega}_z=-\partial_y\overline{u}$} in channel flow.
\prfrev{Only its wall value is related to the skin-friction coefficient,
$C_f=2\nu(\partial_y\overline{u})|_{y=0}/U_b^2$; away from the wall,
it carries the local mean shear, whereas the Reynolds shear stress contributes
to the friction through the weighted wall-normal integral of the
Fukagata--Iwamoto--Kasagi (FIK) identity \citep{fukagata2002}.}
	The physical significance of each component for the drag-reduction process is
	developed in \prfrev{\S\S\ref{sec:results} and \ref{sec:enstrophy}}.

\prfrev{Two sets of phase-averaged statistics, obtained from the same
simulations over about 45 actuation periods, are used for the velocity statistics and the budgets. The first was
accumulated by the solver in 400 phase bins per period and provides the phase
maps of the velocity statistics. The second is an ensemble of instantaneous
fields stored at 32 equally spaced phases per period, from which the
correlations involving velocity gradients were formed. It provides the
time-averaged vorticity profiles and budgets, the phase maps of the budget
terms, the phase-resolved vorticity variances and the checks of
Appendix~\ref{app:verification}. The phase-averaged vorticity of
Figure~\ref{fig:vorticity_phase_variation} was obtained from snapshots of the
same simulations, sorted into 32 phase bins per period, without an established common averaging window with the two sets above. The comparisons based
on the 32-phase ensemble are descriptive, as its near-wall vorticity
covariances are not strictly stationary; this test and the sampling of the
three sets are detailed in Appendix~\ref{app:verification}.}

\prfrev{When a production term is the product of two quantities that both
vary over the cycle, its time average is not, in general, the product of
their time averages. For a production term $P=C\,\Gamma$, where $C$ is a
correlation and $\Gamma$ a mean gradient, such as the mean shear
$S=\partial_y\langle u\rangle$ or the Stokes strain $G$, the time average}
\begin{equation}
\prfrev{\overline P=\overline C\,\overline\Gamma+\overline{\widetilde C\,\widetilde\Gamma}}
\label{eq:phase_covariance}
\end{equation}
\prfrev{contains, in addition to the product of the time averages, the time
average $\overline{\widetilde C\,\widetilde\Gamma}$ of the product of the
phase-coherent components, i.e.\ their phase covariance. Every time-averaged
production term of the present paper is formed by multiplying $C$ and
$\Gamma$ at each phase before averaging over the cycle, and therefore
includes this covariance.}

\prfrev{The notation of the budget analysis is thus as follows. Angle brackets
denote the phase-conditional average at fixed $t^*$, an overbar the average over
the cycle, $\overline A=\int_0^1\langle A\rangle\,\mathrm{d}t^*$ (evaluated
from the 32 phase samples of the ensemble), and a tilde the phase-coherent part.
In the time-averaged budget figures and in the quantitative comparisons of
\S\ref{sec:enstrophy}, a term written with an overbar, e.g.\ $\overline{\mathrm{Z1}}$,
is the cycle average of the complete product of Eq.~(\ref{eq:phase_covariance}),
whereas the phase maps show the same term without an overbar. The average of a
quantity $A$ over a phase interval $I$ is written $\overline A^{\,I}$
(\S\ref{sec:duty_cycle}) and is not the cycle average. As $\overline w=0$, the
cycle average of the Stokes strain vanishes, $\overline G=0$: for the terms
Y1, Z1, X1 and $-\mathrm{X0}$, whose mean gradient is $G$ or $\partial_yG$, the
product of the separately averaged factors is zero, and the cycle average is the
phase covariance alone. For $-\mathrm{Y2}$ and $-\mathrm{X2}$, in which the
mean shear $S$ varies by about 2\,\% over the cycle (\S\ref{sec:enstr_wy}), the
complete-product average likewise retains the covariance with this weakly varying shear.}

\section{Results}
\label{sec:results}

\prfrev{All wall units use the fixed nominal reference $u_{\tau,0}=0.042$
defined in \S\ref{sec:methodology}, common to both controlled cases. The
sinusoidal and shape-optimised waveforms achieve gross drag reductions of
$38.4\pm0.5\,\%$ and $40.7\pm0.4\,\%$, respectively, the uncertainties being
approximate 95\,\% intervals (Table~\ref{tab:drag_verification}). Their
difference, $2.3\pm0.6$ percentage points, corresponds to a relative improvement of approximately 6\% over a
sinusoidal actuation that is already close to optimal at this amplitude. The
cycle-averaged power supplied by the wall motion,
$-\nu\,\overline{w_w\,\partial_n\langle w\rangle}$, evaluated at each wall
with $n$ directed into the fluid, is, however, more than 1.5 times as large for the
shape-optimised waveform, an increase substantially larger than the
corresponding increase in gross drag reduction. The
subsections that follow examine the changes in the flow that accompany the
additional drag reduction.}

\subsection{Reynolds stress modulation and production dynamics}
\label{sec:reynolds_stress}

Attention is first directed to the \prfrev{Reynolds stresses and their production}, as the streamwise
Reynolds stress $\overline{u^{\prime\prime}u^{\prime\prime}}^+$ constitutes the
primary signature of the \prfrev{near-wall streaks, which are closely associated
with the turbulent skin-friction drag}.
\prfrev{The present subsection examines the streamwise stress, its production term and the relation between the two over the actuation cycle}.

\begin{figure}
	\centering
	\begin{minipage}{0.48\textwidth}
		\centering
		\includegraphics[width=\textwidth]{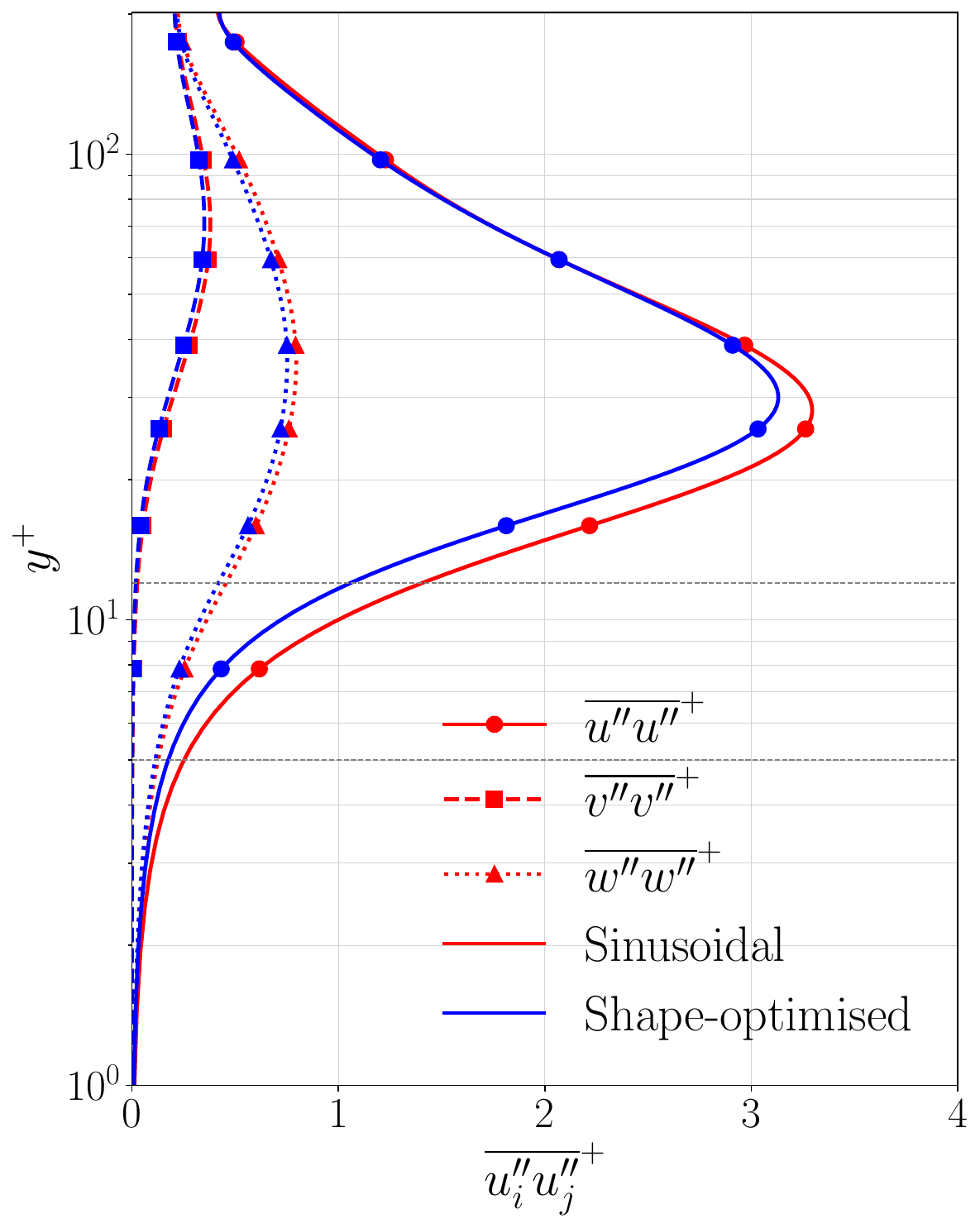}
		\vspace{0.2cm}
		\centerline{(a)}
	\end{minipage}%
	\hfill
	\begin{minipage}{0.48\textwidth}
		\centering
		\includegraphics[width=\textwidth]{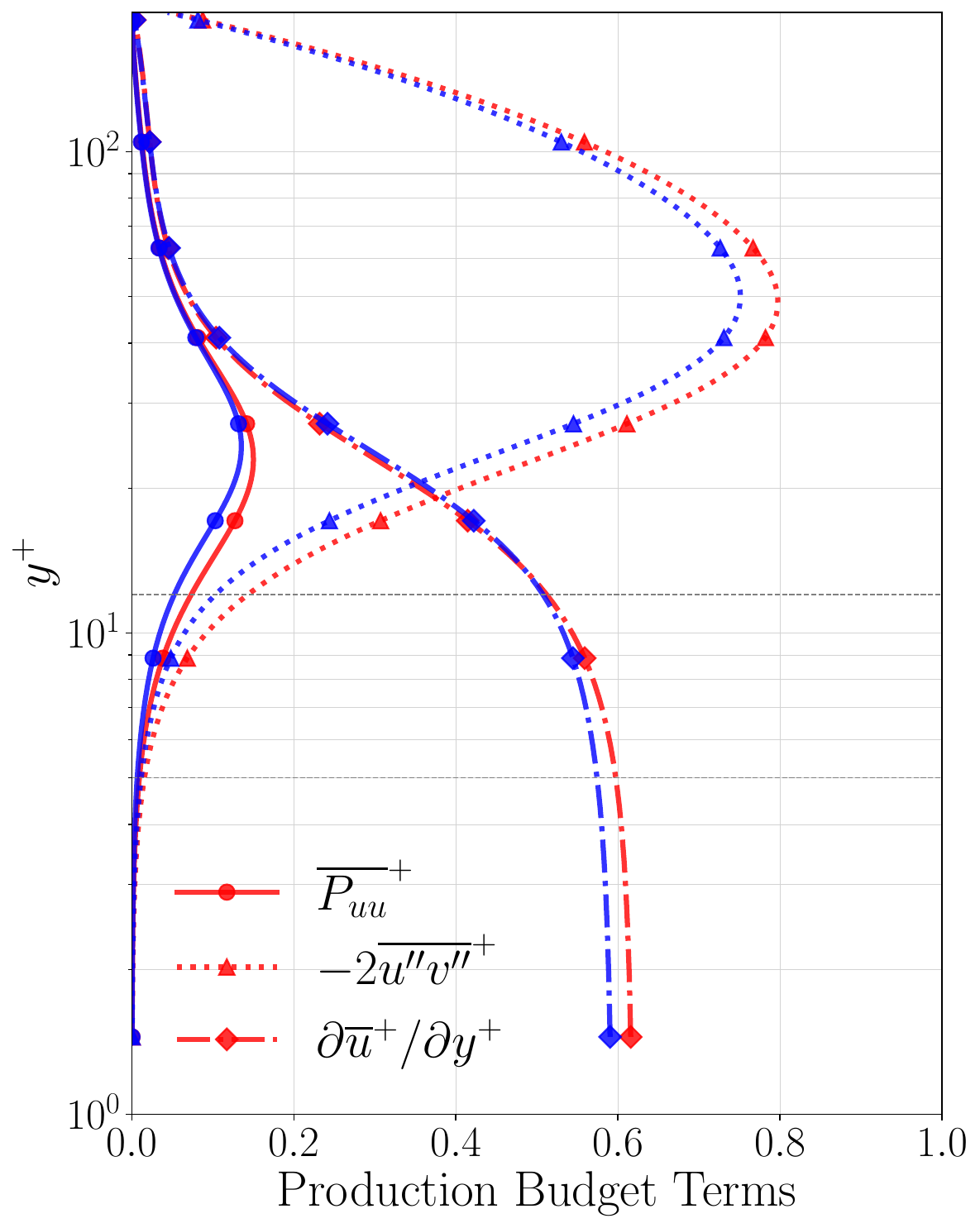}
		\vspace{0.2cm}
		\centerline{(b)}
	\end{minipage}
	\caption{
		Wall-normal profiles of (a) \prfrev{the normal stresses $\overline{u^{\prime\prime}u^{\prime\prime}}^+$, $\overline{v^{\prime\prime}v^{\prime\prime}}^+$ and $\overline{w^{\prime\prime}w^{\prime\prime}}^+$} and (b) streamwise Reynolds stress production \prfrev{$\overline{P_{uu}}^+$, with the time-averaged stress $-2\overline{u^{\prime\prime}v^{\prime\prime}}^+$ and mean-velocity gradient $\partial\overline{u}^+/\partial y^+$, whose product is not, in general, $\overline{P_{uu}}^+$,} for \prfrev{the sinusoidal and shape-optimised cases; the unactuated reference is not plotted. Reference lines at $y^+=5$ and 12 delimit the near-wall comparison band.}
	}
	\label{fig:reynolds_stress_comparison}
\end{figure}

\begin{figure}
	\centering
	\begin{minipage}{0.48\textwidth}
		\centering
		\includegraphics[width=\textwidth]{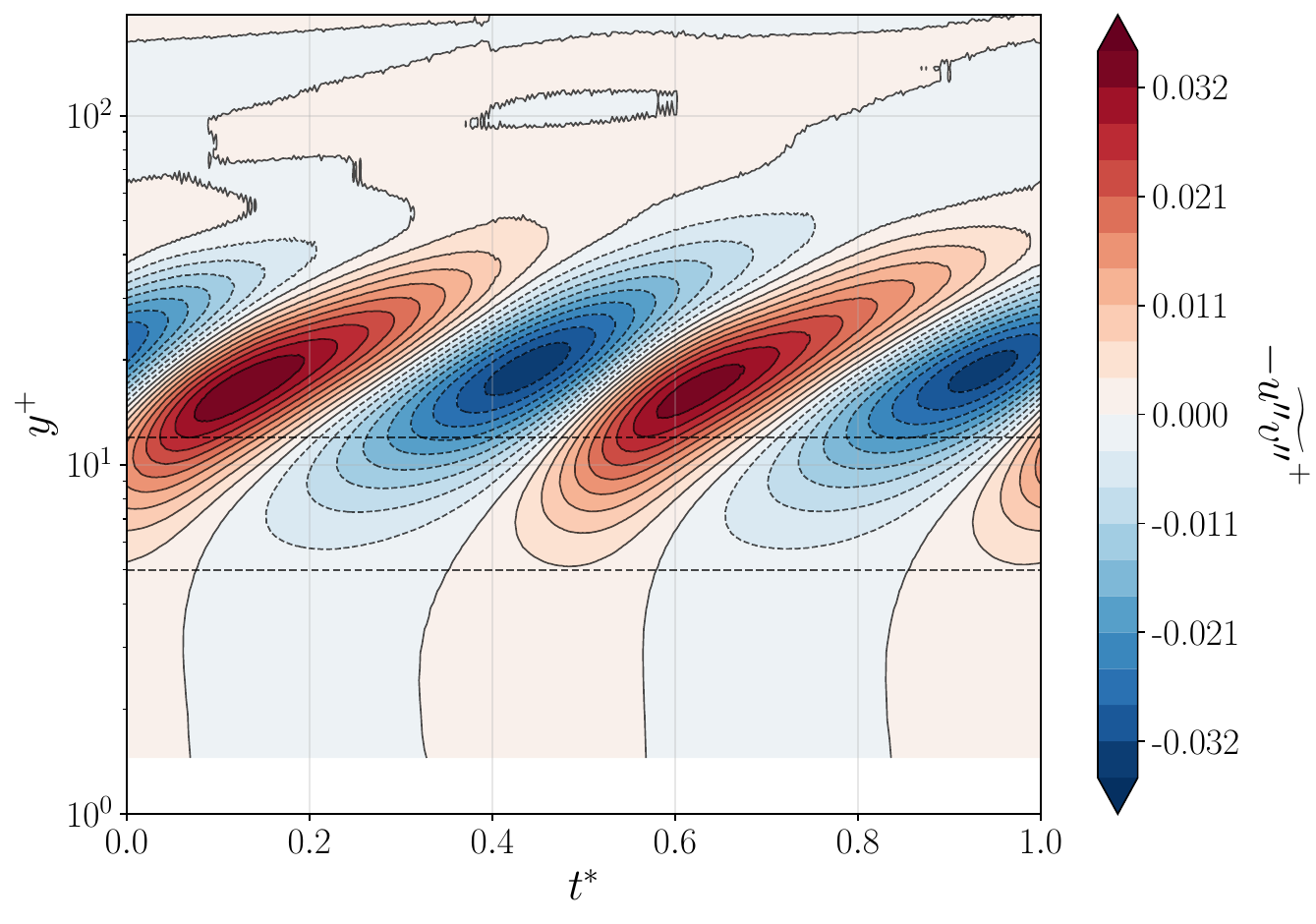}
		\centerline{(a)}
	\end{minipage}%
	\hfill
	\begin{minipage}{0.48\textwidth}
		\centering
		\includegraphics[width=\textwidth]{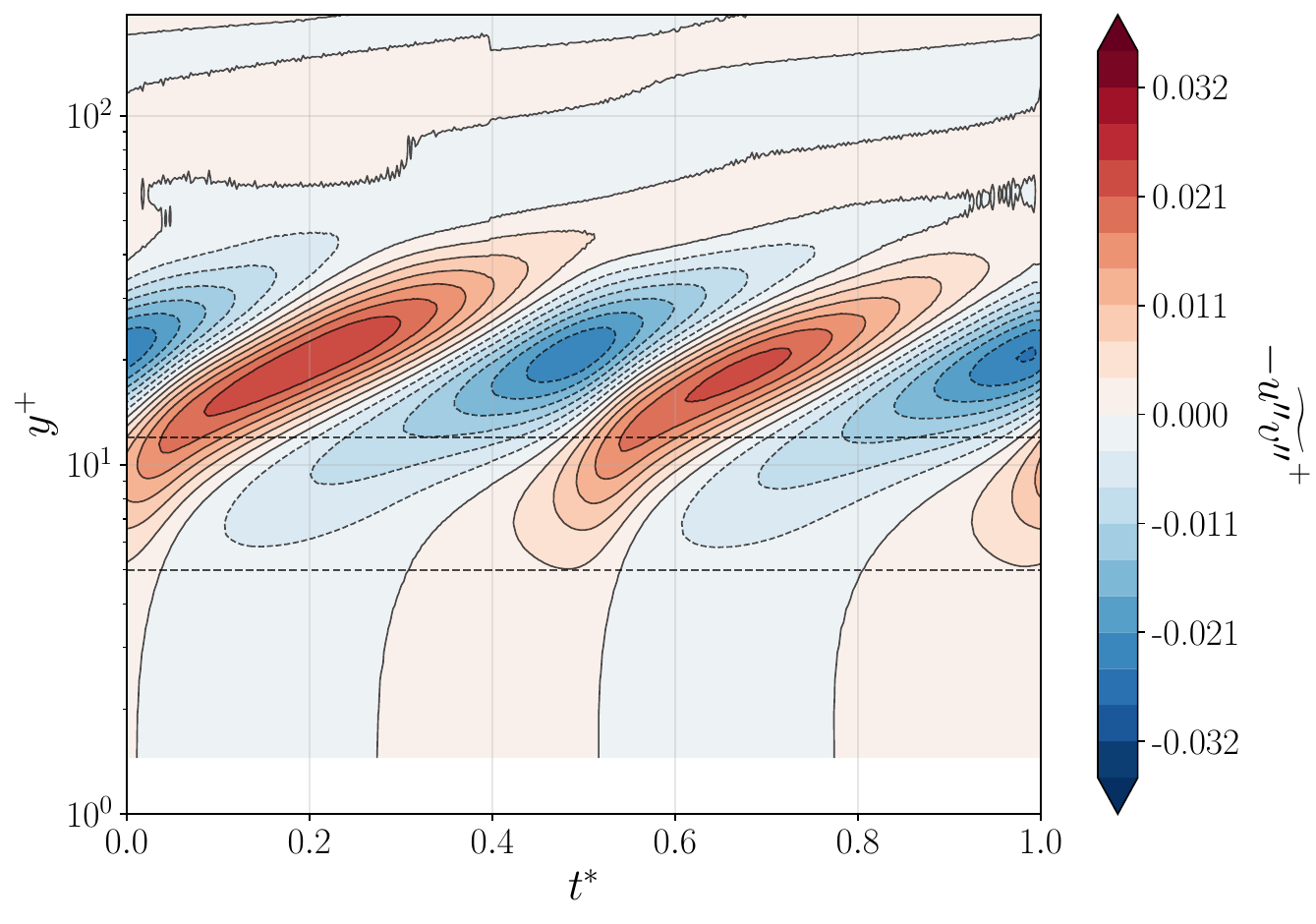}
		\centerline{(b)}
	\end{minipage}        
	
	\begin{minipage}{0.48\textwidth}
		\centering
		\includegraphics[width=\textwidth]{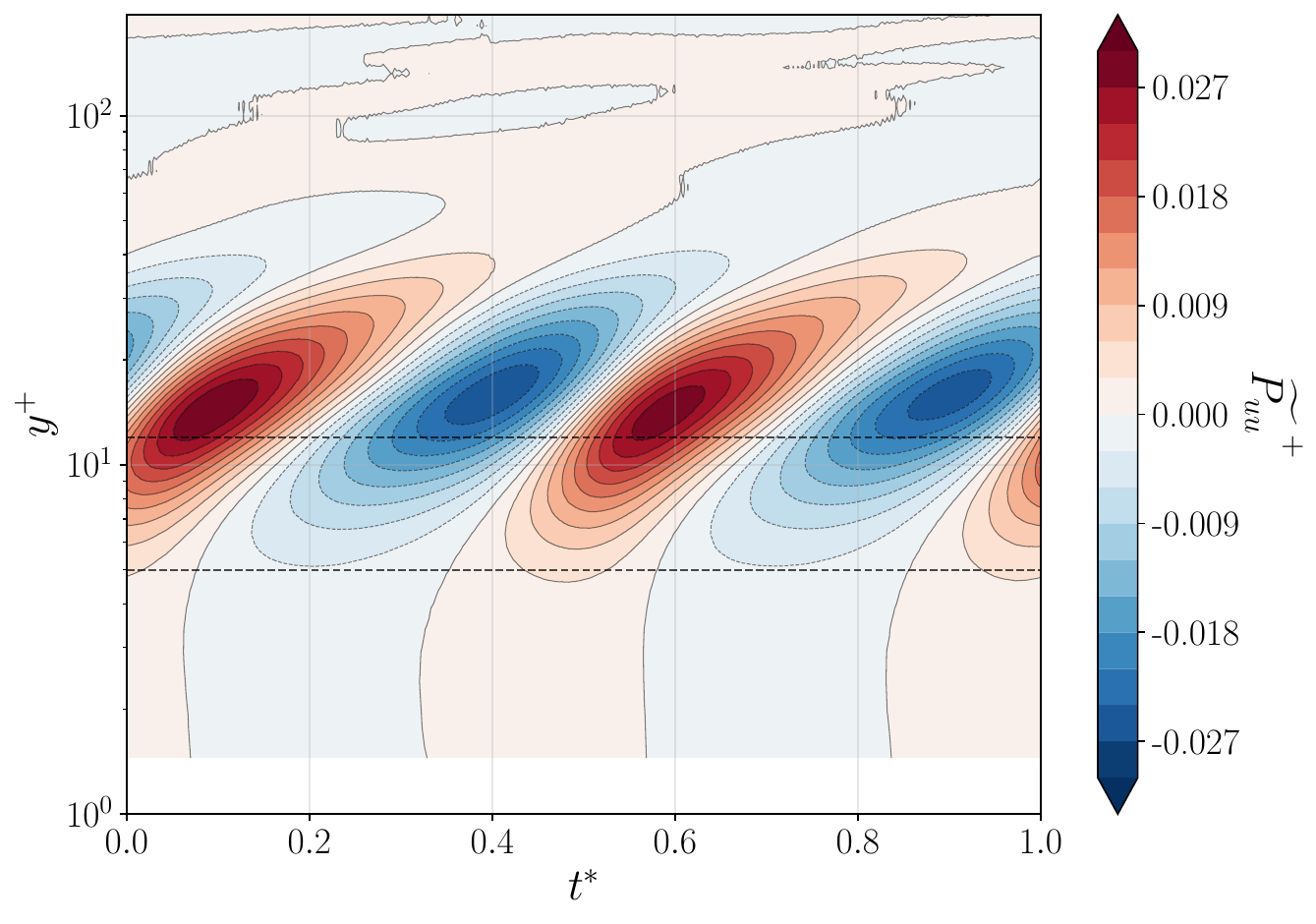}
		\centerline{(c)}
	\end{minipage}%
	\hfill
	\begin{minipage}{0.48\textwidth}
		\centering
		\includegraphics[width=\textwidth]{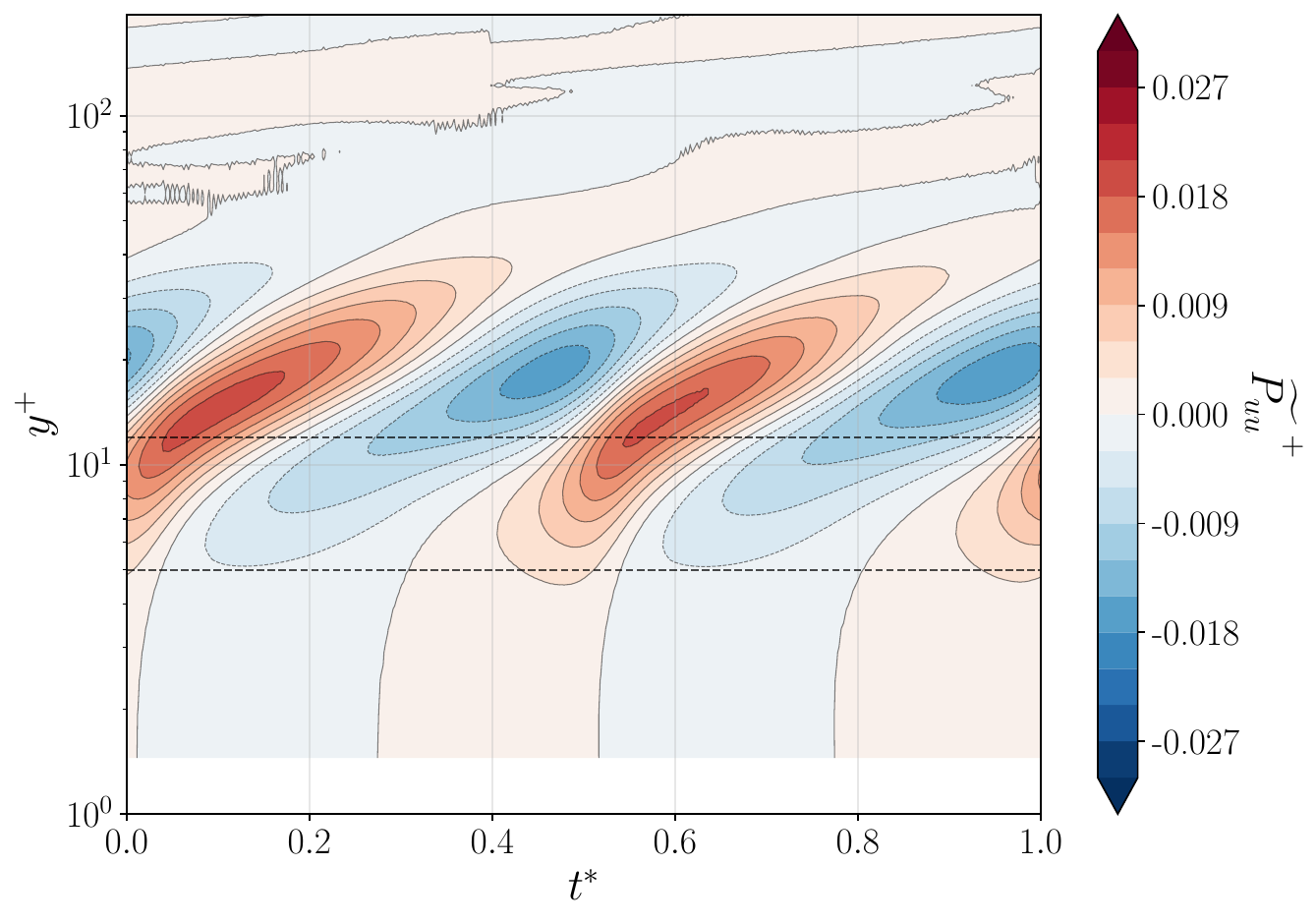}
		\centerline{(d)}
	\end{minipage}
	
	\begin{minipage}{0.48\textwidth}
		\centering
		\includegraphics[width=\textwidth]{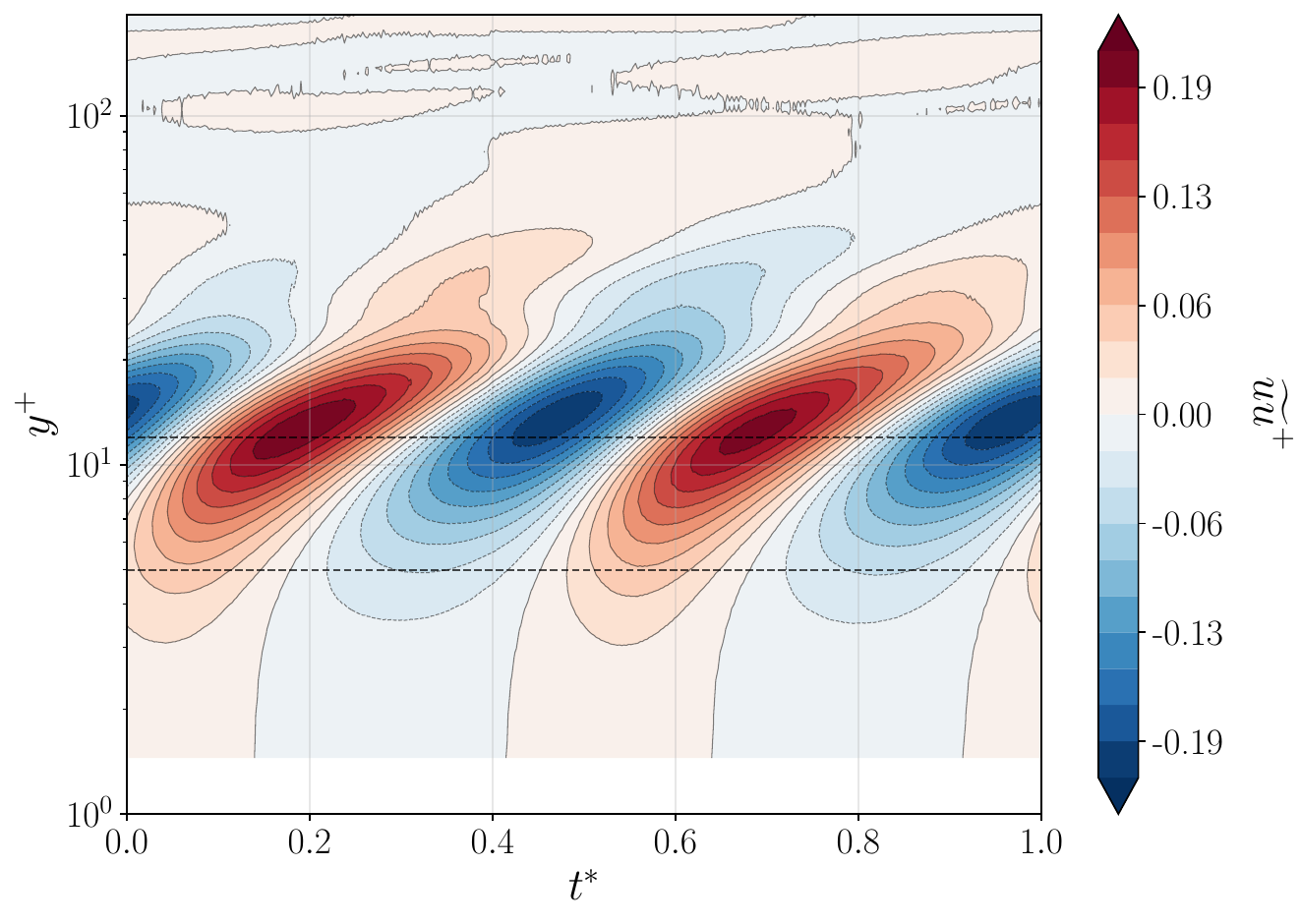}
		\centerline{(e)}
	\end{minipage}%
	\hfill
	\begin{minipage}{0.48\textwidth}
		\centering
		\includegraphics[width=\textwidth]{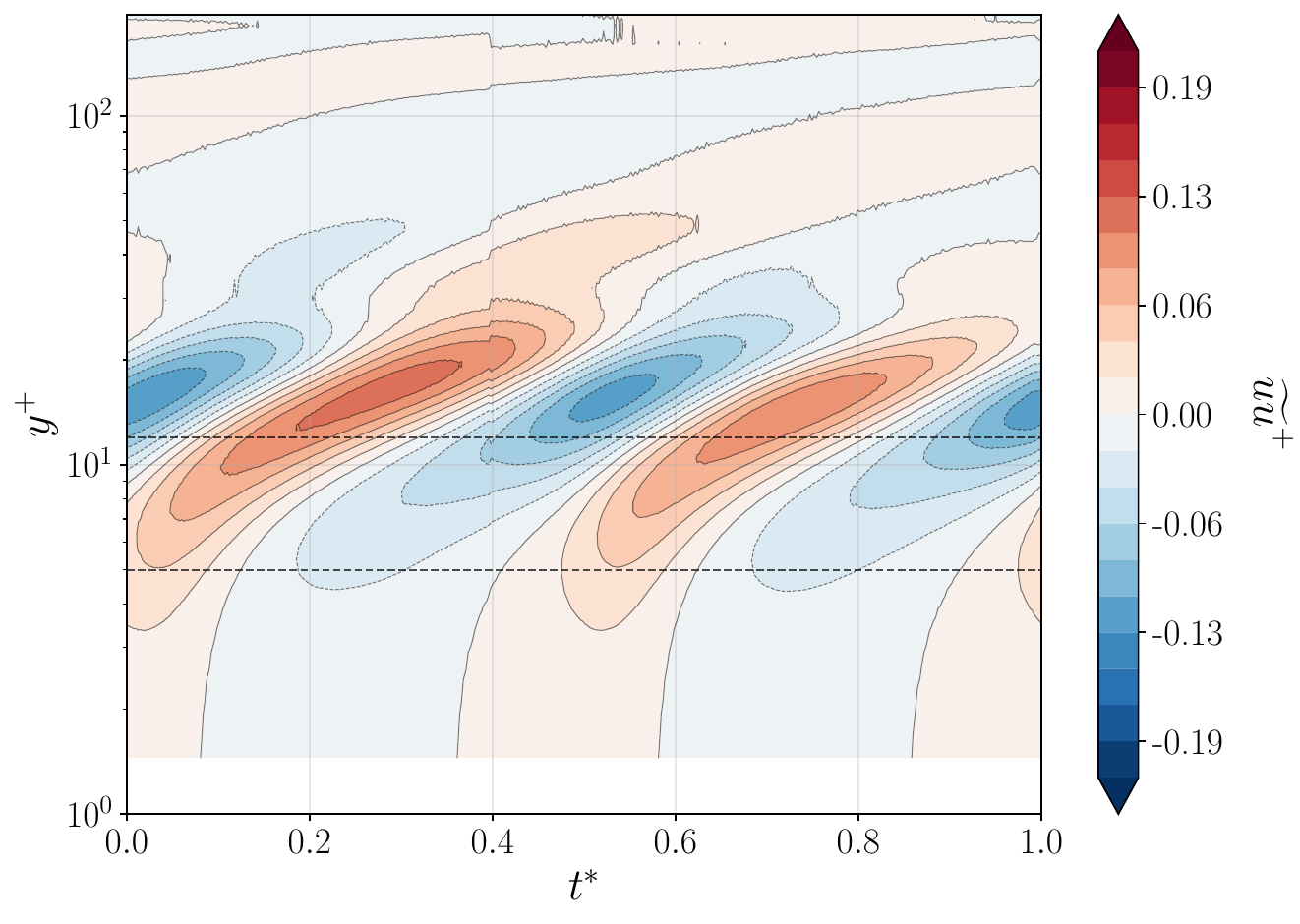}
		\centerline{(f)}
	\end{minipage}
	
	\caption{
		\prfrev{Phase-coherent Reynolds stress quantities as functions of actuation phase $t^*$ and wall-normal distance $y^+$; sinusoidal actuation on the left, shape-optimised on the right.
		Top row: Reynolds shear stress $-\widetilde{u^{\prime \prime}v^{\prime \prime}}^+$ (a, b);
		middle row: production $\widetilde{P_{uu}}^+$ (c, d);
		bottom row: streamwise Reynolds stress $\widetilde{u^{\prime \prime}u^{\prime \prime}}^+$ (e, f). Reference lines at $y^+=5$ and 12 delimit the near-wall comparison band.}
	}
	\label{fig:phase_comparison}
\end{figure}

The influence of spanwise wall actuation upon the Reynolds stress is
examined through comparison of the wall-normal distribution of the normal-stress
components (Figure~\ref{fig:reynolds_stress_comparison}(a)).  Both actuated
configurations exhibit substantial attenuation of
$\overline{u^{\prime\prime}u^{\prime\prime}}^+$ relative to the unactuated
baseline throughout the near-wall region, with peak values reduced from
$\overline{u^{\prime\prime}u^{\prime\prime}}^+_{\max} \approx \prfrev{7.0}$ in the
unactuated baseline to \prfrev{approximately 3.3 and 3.1} for sinusoidal and
shape-optimised actuation, respectively.  This reduction is most pronounced
\prfrev{in the part of the buffer layer below $y^+\approx25$, where the velocity
streaks of the unactuated flow attain their maximum intensity, and is largest
near $y^+\approx12$}.
The shape-optimised configuration achieves systematically lower
$\overline{u^{\prime\prime}u^{\prime\prime}}^+$ across the entire near-wall
region, indicating \prfrev{a stronger attenuation of the} streak-associated turbulent
fluctuations.  Similar trends are observed for the wall-normal and spanwise
stress components, \prfrev{whose unactuated peak values of approximately 0.76 and
1.28 decrease to 0.38 and 0.80 for the sinusoidal waveform and to 0.35 and 0.75
for the shape-optimised waveform. The phase-dependent comparisons of the
following sections are made within the near-wall band $5\le y^+\le12$,
delimited by reference lines in
Figures~\ref{fig:reynolds_stress_comparison} and \ref{fig:phase_comparison}
and in the subsequent phase maps, and situated below the wall-normal location
of the variance maximum of both controlled cases.}  The
question of how actuation achieves this attenuation of
$\overline{u^{\prime\prime}u^{\prime\prime}}^+$ constitutes the subject of the
analysis that follows.

\prfrev{The production term $P_{uu}^+ = -2\langle u^{\prime\prime}v^{\prime\prime}\rangle^+\,\partial_{y^+}\langle u\rangle^+$} quantifies the rate at which energy is extracted from the mean flow and \prfrev{supplied} to the streamwise-velocity fluctuations\prfrev{. In the transport equation of the streamwise stress, production is the principal source, and $\overline{u^{\prime\prime}u^{\prime\prime}}^+$ reflects its balance with the dissipation, pressure, turbulent-transport and viscous-diffusion terms (Appendix~\ref{app:verification}). The peak production is approximately 10\,\% lower for the shape-optimised waveform than for the sinusoidal waveform (Figure~\ref{fig:reynolds_stress_comparison}(b)). The ratio of the peak values of $\overline{u^{\prime \prime} u^{\prime \prime}}^+$ and $\overline{P_{uu}}^+$ defines a characteristic replenishment timescale, $\tau^+ = \overline{u^{\prime \prime} u^{\prime \prime}}^+_{\max}/\overline{P_{uu}}^+_{\max} \approx 22.5$ wall-time units, which represents the time required for production to replenish the observed stress level at statistical equilibrium. As production peaks nearer the wall than the variance, at $y^+\approx24$ and 30, respectively, $\tau^+$ is an order-of-magnitude estimate rather than a local timescale. Moreover, $\tau^+$ is a timescale, not a delay: when production varies periodically, the variance in a linear relaxation model lags it by less than $\tau^+$, as estimated below.}

The time-averaged profiles conveyed by Figure~\ref{fig:reynolds_stress_comparison} \prfrev{show lower Reynolds stresses for the shape-optimised waveform}; however, they do not reveal when during the actuation cycle this \prfrev{attenuation} occurs. To examine how the temporal structure of the actuation waveform modulates the turbulent momentum transport, the phase-coherent Reynolds shear stress $-\widetilde{u^{\prime \prime}v^{\prime \prime}}^+$, the phase-coherent production $\widetilde{P_{uu}}^+$, and the phase-coherent streamwise stress $\widetilde{u^{\prime \prime}u^{\prime \prime}}^+$ are presented in Figure~\ref{fig:phase_comparison} as functions of actuation phase $t^*$ and wall-normal distance $y^+$. The Reynolds shear stress $-\overline{u^{\prime \prime}v^{\prime \prime}}$ represents the turbulent transport of streamwise momentum in the wall-normal direction, a process accomplished through sweep and ejection events that constitute the primary pathway \prfrev{of momentum transport} between the near-wall region and the outer flow. The phase-coherent component $-\widetilde{u^{\prime \prime}v^{\prime \prime}}^+$ (Figure~\ref{fig:phase_comparison}(a,b)) isolates the portion of this transport that is correlated with the actuation phase. For sinusoidal actuation (Figure~\ref{fig:phase_comparison}(a)), the phase-coherent shear stress exhibits a smooth, quasi-harmonic variation throughout the cycle, with regions of \prfrev{large} magnitude distributed over extended temporal intervals. In contrast, \prfrev{below $y^+\approx15$,} the shape-optimised waveform
(Figure~\ref{fig:phase_comparison}(b)) compresses the \prfrev{phases} of enhanced
$-\widetilde{u^{\prime\prime}v^{\prime\prime}}^+$ into narrower \prfrev{intervals
shortly after the reversals of the wall velocity, whilst the stress remains below its time average during the longer plateaus of the wall velocity: at
$y^+\approx12$, it exceeds its time average over 41\% of the cycle, against 46\%
for the sinusoidal waveform. Such a compression in time is consistent with
shorter intervals of effective momentum transport by sweep and ejection events,
although the phase-coherent stress alone does not identify the events
themselves. The relation between these intervals and the Reversal and
Displacement intervals is examined in \S\ref{sec:duty_cycle}.}

The phase-coherent production $\widetilde{P_{uu}}^+$ (Figure~\ref{fig:phase_comparison}(c,d)) exhibits spatio-temporal patterns that closely resemble the patterns of the Reynolds shear stress, a correspondence that arises directly from \prfrev{the definition of $P_{uu}$}. As the mean velocity gradient \prfrev{$\partial_y\langle u\rangle$} varies \prfrev{by less than 5\% over the cycle below $y^+\approx15$}, the phase-coherent production is predominantly governed by the phase-coherent shear stress. \prfrev{For the shape-optimised waveform, production therefore inherits the
narrower intervals of enhanced shear stress, and its departures from the time average are also smaller than for the sinusoidal waveform.}

\prfrev{As stated in \S\ref{sec:phase_analysis}, negative values in these maps
correspond to phases below the time average of each case, not to a suppression
below the unactuated flow. Equally, positive values do not indicate a return to
the unactuated level. Where the streamwise stress is largest, i.e.\ at the
wall-normal location of its time-averaged maximum, $y^+\approx28$ and 30 for
the two waveforms, its phase average in the separate phase ensemble
(\S\ref{sec:phase_analysis}) varies over the cycle by less than 4\,\%, around 3.3 and 3.1,
respectively, and remains well below the
unactuated values of about 5.2 and 4.9 at these wall-normal locations at every
phase, the unactuated peak of 7.0 being located near $y^+\approx14$. Over $5\le y^+\le12$, the band-averaged $\langle u''u''\rangle^+$ and $\langle v''v''\rangle^+$ remain below 0.23 of the unactuated values and $\langle w''w''\rangle^+$ below 0.70, at every phase and for both waveforms. The modulation of $\langle u''u''\rangle^+$ is considerably stronger
nearer the wall, where its peak-to-peak amplitude reaches approximately 40\%
and 30\% of the respective time averages at $y^+\approx8$.}

\prfrev{The maps in Figure~\ref{fig:phase_comparison}(c--f) reveal that the
streamwise variance follows production with a lag, as is apparent along a
horizontal line at fixed $y^+$. Both quantities repeat every half-period, as the flow
responds in the same way to either direction of wall motion and the wall
velocity changes sign after half a period. The order of magnitude of
this lag may
be estimated by modelling the response of the variance as a linear relaxation
towards production, with the replenishment timescale $\tau^+$:}
\begin{equation}
\label{eq:relaxation}
\prfrev{\frac{\partial\langle u^{\prime\prime}u^{\prime\prime}\rangle^+}{\partial t^+}
 =P_{uu}^+-\frac{\langle u^{\prime\prime}u^{\prime\prime}\rangle^+}{\tau^+}.}
\end{equation}
\prfrev{If production is represented by its dominant component, at twice the
actuation frequency, i.e.\ at the angular frequency $\omega^+=4\pi/T^+$, this
component may be written
$\hat P\,\mathrm{e}^{\mathrm{i}\omega^+t^+}$. As Eq.~(\ref{eq:relaxation}) is
linear, the variance then oscillates as $\hat A\,\mathrm{e}^{\mathrm{i}\omega^+t^+}$,
with $\hat A=\tau^+\hat P/(1+\mathrm{i}\,\omega^+\tau^+)$, whose phase lags
the phase of production by $\arctan(\omega^+\tau^+)$. Division of this phase lag by
the angular frequency gives the delay}
\begin{equation}
\prfrev{\Delta t^+=\frac{1}{\omega^+}\arctan\left(\omega^+\tau^+\right)
 =\frac{T^+}{4\pi}\arctan\left(\frac{4\pi\tau^+}{T^+}\right)\approx10.6,}
\end{equation}
\prfrev{which, in this model, cannot exceed $\pi/(2\omega^+)=T^+/8\approx14$, whatever the
value of $\tau^+$, as the phase lag remains below $\pi/2$. The phase offsets of this component, measured at the height of its time-mean maximum ($y^+=28$ and 30), are 12.6
and 13.9 wall-time units for the sinusoidal and shape-optimised waveforms,
respectively (Appendix~\ref{app:verification}), slightly above this estimate
and close to its upper bound within the sampling uncertainty. The streamwise variance therefore does not
follow production instantaneously; the relaxation estimate gives the order of magnitude of the delay but lies below it. The difference is examined with the budget of
$\langle u^{\prime\prime}u^{\prime\prime}\rangle$ at the wall-normal location of the maximum of its time average (Appendix~\ref{app:verification}). In
Eq.~(\ref{eq:relaxation}), the losses are proportional to
$\langle u^{\prime\prime}u^{\prime\prime}\rangle$ and reach their maximum with
it, whereas dissipation, the largest loss on time average, follows production
with a delay of 3 to 4 wall-time units. The subtraction of dissipation
therefore reduces the amplitude of the modulated production by
approximately a quarter, with little change in its phase, and the integration
of production minus dissipation gives lags of 12.7 and 12.4 wall-time units,
respectively, within 1.5 wall-time units of the corresponding measured
offsets. Of the remaining terms, the pressure and viscous-diffusion
terms increase the lag and turbulent transport reduces it; with all terms, the
integration gives 12.4 and 13.4 wall-time units. The phase relation and the
budget thus support the interpretation that production drives the modulation of
the variance, although they do not constitute a demonstration of causality.}

\prfrev{Through production, the phase modulation of the streak intensity is
thus associated with the phase modulation of the Reynolds shear stress, whose time average contributes to the friction through the weighted wall-normal integral of the FIK identity
\citep{fukagata2002}. The mechanism driving the pronounced phase variation
of this shear stress, however, remains to be established.
\citet{agostini_turbulence_2015} highlighted the central role of vorticity
tilting and stretching in drag reduction for sinusoidal actuation. Following this
approach, the subsequent sections examine the interaction of the Stokes layer
with near-wall turbulence in terms of vorticity dynamics, in order to
characterise the phase-dependent production terms of the component
enstrophies, as formalised in \S\ref{sec:enstrophy}.}

\subsection{\prfrev{Time-averaged vorticity profiles}}
\label{sec:mean_vorticity}
\begin{figure}
	\centering
	\begin{minipage}{0.48\textwidth}
		\centering
		\includegraphics[width=\textwidth]{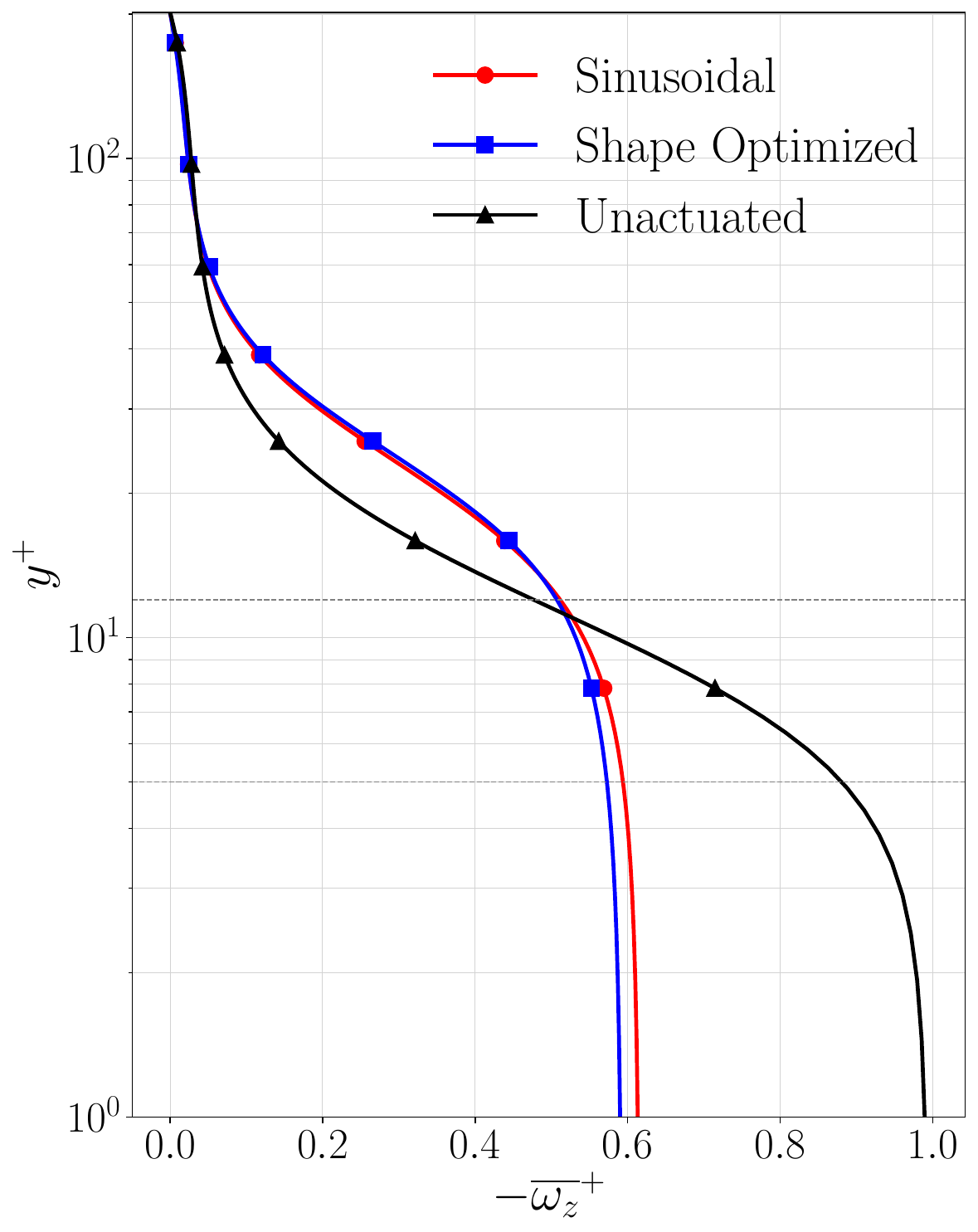}
		\vspace{0.2cm}
		\centerline{(a)}
	\end{minipage}%
	\hfill
	\begin{minipage}{0.48\textwidth}
		\centering
		\includegraphics[width=\textwidth]{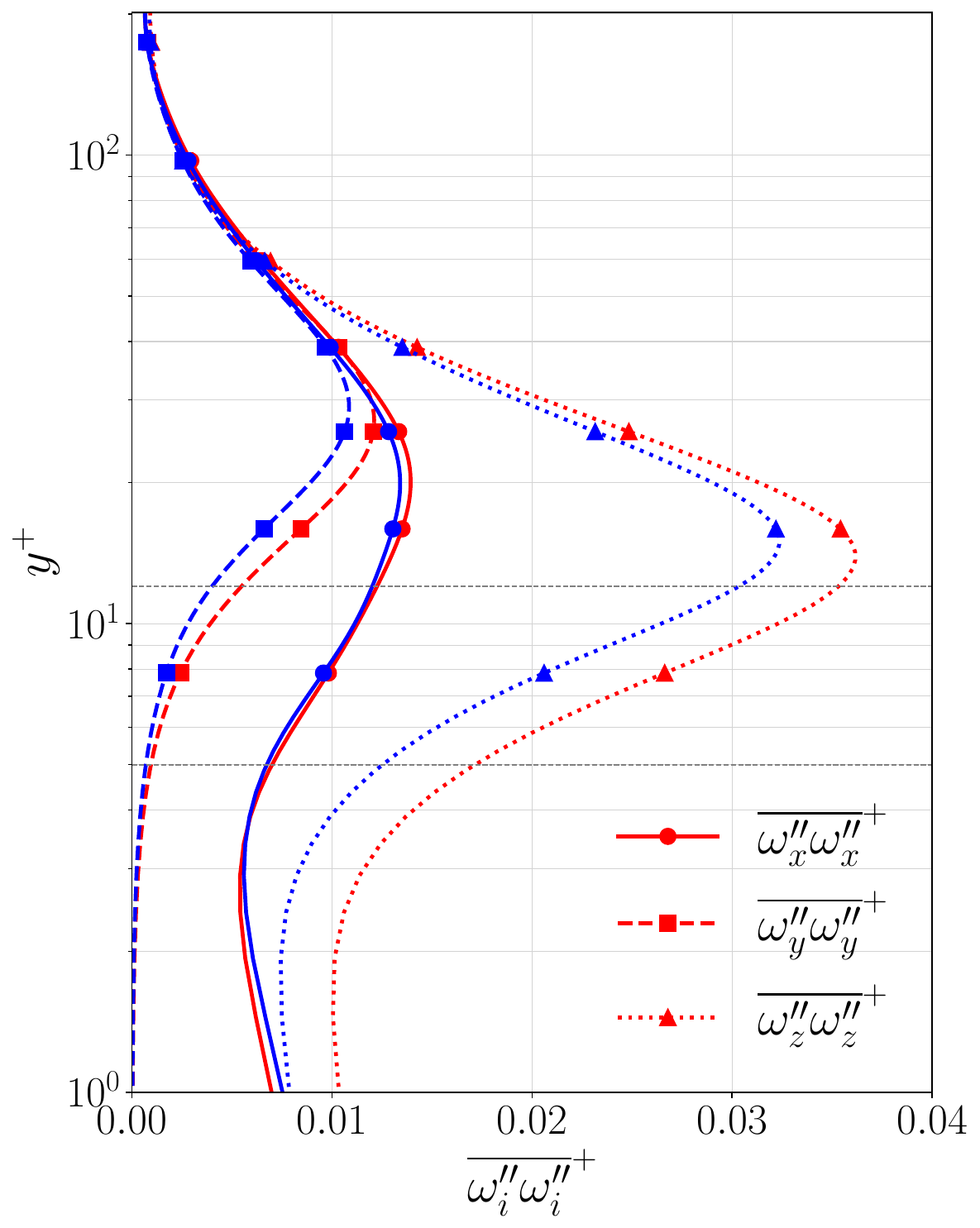}
		\vspace{0.2cm}
		\centerline{(b)}
	\end{minipage}
	\caption{
		Wall-normal profiles of (a) mean spanwise vorticity \prfrev{\prfrev{$-\overline{\omega_z}^+=\partial_{y^+}\overline{u}^+$}}
		and (b) vorticity variance components $\overline{\omega_i^{\prime\prime}\omega_i^{\prime\prime}}^+$
		for \prfrev{the sinusoidal and shape-optimised cases, both from the separate phase ensemble (\S\ref{sec:phase_analysis}). Panel~(a) also includes the unactuated profile, from the time statistics of the unactuated simulation. Reference lines at $y^+=5$ and 12 delimit the near-wall comparison band.}
	}
	\label{fig:vorticity_comparison}
\end{figure}

\prfrev{The vorticity field is first characterised by its time average. The
wall-normal distributions of the mean spanwise vorticity and of the variance
of each vorticity component are conveyed by
Figure~\ref{fig:vorticity_comparison}. In channel flow,
\prfrev{$-\overline{\omega_z}^+=\partial_{y^+}\overline{u}^+$} is the mean shear,
and its wall value is the time-averaged wall shear stress, which determines
the drag under constant flow rate; this wall value is lower for the
shape-optimised waveform (Figure~\ref{fig:vorticity_comparison}(a)), in
agreement with its larger drag reduction. The actuated profiles intersect the
unactuated profile at $y^+\approx11$, above which they exceed it. The
intersection is consistent with the constant flow rate: as the bulk velocity
is fixed, a reduction of the mean velocity gradient near the wall lowers the
mean velocity in that region and must be accompanied by a larger gradient
further from the wall.}

\prfrev{Over the near-wall band $5\le y^+\le12$, the time-averaged vorticity variances
(Figure~\ref{fig:vorticity_comparison}(b)) are lower for the shape-optimised
waveform in the wall-normal and spanwise components
($\overline{\omega_y^{\prime\prime}\omega_y^{\prime\prime}}^+$ and
$\overline{\omega_z^{\prime\prime}\omega_z^{\prime\prime}}^+$), by about 28\% and
21\%, respectively, whereas the streamwise variance
($\overline{\omega_x^{\prime\prime}\omega_x^{\prime\prime}}^+$) differs by only
about 2\%. The reduction of the wall-normal variance, the
component with which the streaks are associated (\S\ref{sec:intro}) and by
which \citet{du2002drag} \prfrev{quantified} the weakening of the streaks under spanwise
actuation, is consistent with the lower streamwise stress of the shape-optimised waveform
(\S\ref{sec:reynolds_stress}). The variances alone, however, do not identify
the mechanism of these changes; the budgets of \S\ref{sec:enstrophy} examine
the corresponding production terms.}

\subsection{Temporal modulation of the Stokes strain}
\label{sec:temporal_modulation}

\begin{figure}
	\centering
	\begin{minipage}{0.48\textwidth}
		\centering
		\includegraphics[width=\textwidth]{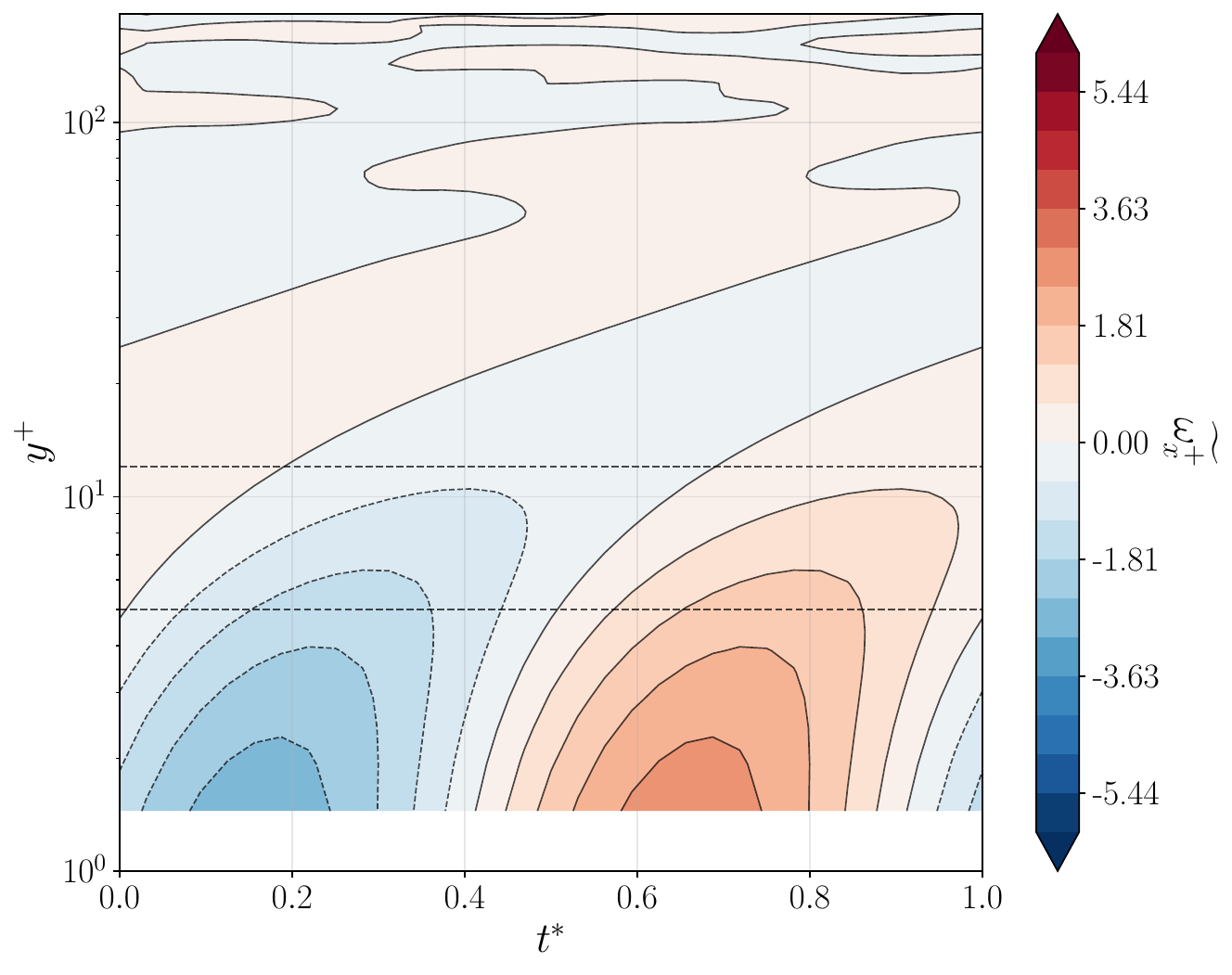}
		\vspace{0.2cm}
		\centerline{(a)}
	\end{minipage}
	\hfill
	\begin{minipage}{0.48\textwidth}
		\centering
		\includegraphics[width=\textwidth]{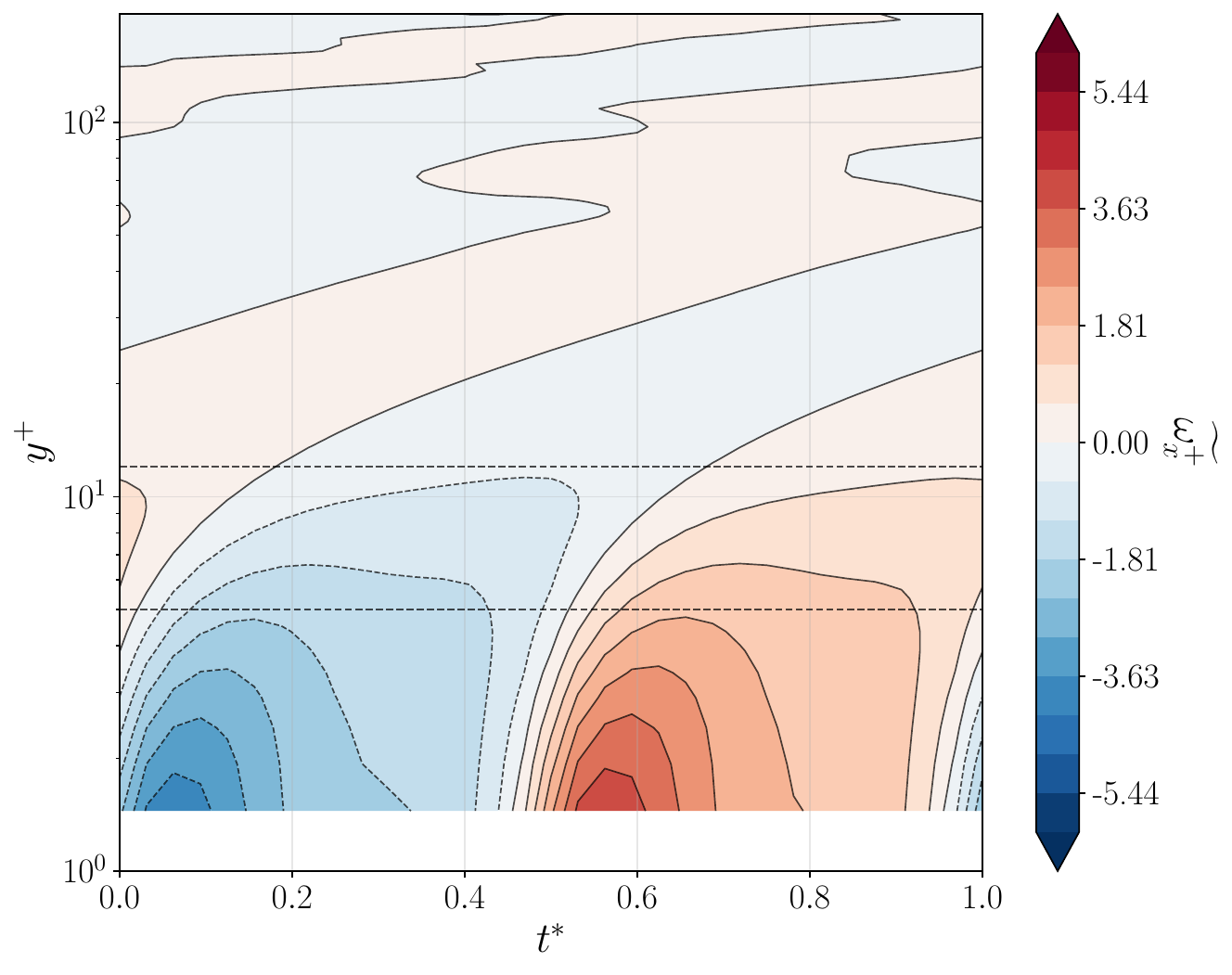}
		\vspace{0.2cm}
		\centerline{(b)}
	\end{minipage}
	
	\begin{minipage}{0.48\textwidth}
		\centering
		\includegraphics[width=\textwidth]{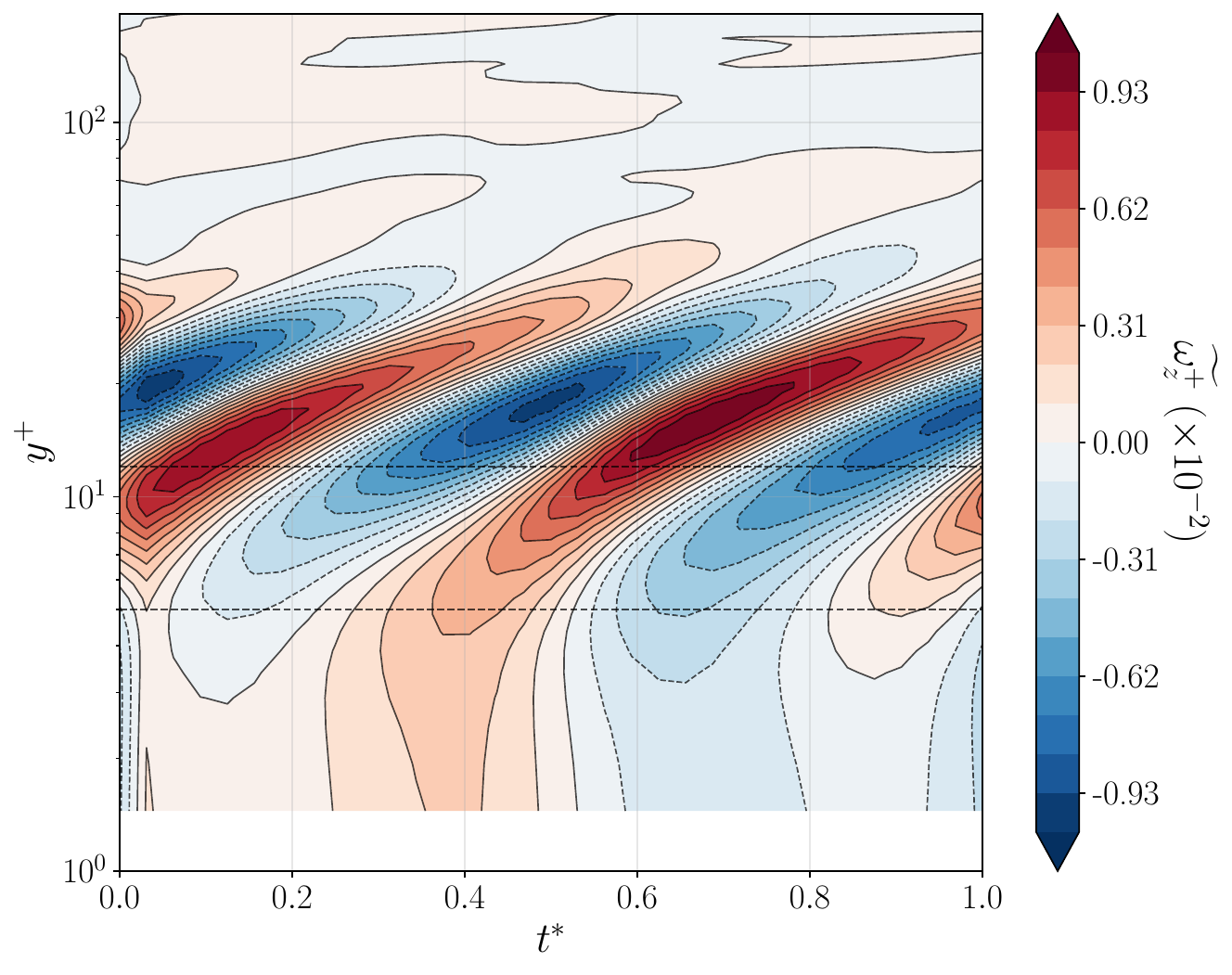}
		\vspace{0.2cm}
		\centerline{(c)}
	\end{minipage}
	\hfill
	\begin{minipage}{0.48\textwidth}
		\centering
		\includegraphics[width=\textwidth]{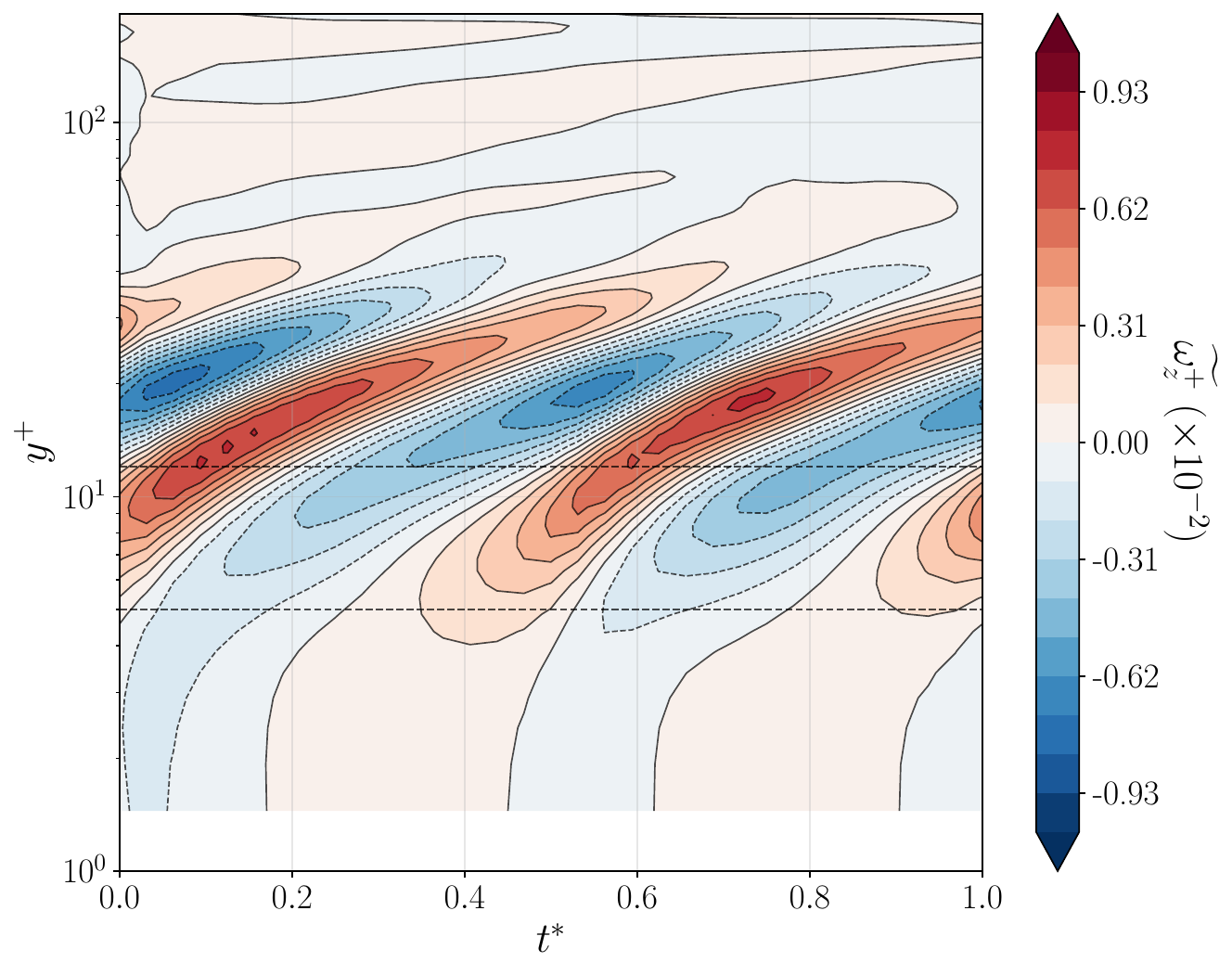}
		\vspace{0.2cm}
		\centerline{(d)}
	\end{minipage}
	\caption{\prfrev{Phase-coherent vorticity components as functions of actuation phase $t^*$ and wall-normal distance $y^+$; sinusoidal actuation on the left, shape-optimised on the right.
		Top row: streamwise component $\widetilde{\omega_x}^+$ (a, b);
		bottom row: spanwise component $\widetilde{\omega_z}^+$ (c, d).
		Reference lines at $y^+ = 5$ and $y^+ = 12$ delimit the near-wall comparison band.}}
	\label{fig:vorticity_phase_variation}
\end{figure}

The phase-coherent streamwise vorticity $\widetilde{\omega_x}^+$ \prfrev{coincides
with the Stokes strain: as the phase average is homogeneous in the spanwise
direction and the time average of $\langle w\rangle$ vanishes,
$\widetilde{\omega_x}=\partial_y\widetilde{w}=\partial_y\langle w\rangle=G$}. \prfrev{The spanwise component
$\widetilde{\omega_z}^+$ represents the phase-coherent modulation of the local
mean shear.}
Figure~\ref{fig:vorticity_phase_variation} presents these two components as
functions of actuation phase $t^*$ and wall-normal distance $y^+$ for
sinusoidal (left) and shape-optimised (right) configurations.  For sinusoidal
actuation (Figure~\ref{fig:vorticity_phase_variation}(a)), the phase variation
at the wall is characteristically harmonic, directly reflecting the imposed
oscillation, with a systematic phase lag towards later times observed at
increasing wall-normal distances as a consequence of viscous diffusion.  The
shape-optimised configuration
(Figure~\ref{fig:vorticity_phase_variation}(b)) exhibits fundamentally
different temporal behaviour: pronounced variations in $\widetilde{\omega_x}^+$
occur at each directional reversal of wall velocity, followed by rapid decay
and extended plateau regions wherein the Stokes strain remains approximately
constant throughout a substantial portion of the actuation cycle.  The
penetration depth of the Stokes layer remains comparable in both cases, with
modulation of appreciable amplitude confined to the near-wall region
($y^+ < 30$)\prfrev{: the higher harmonics of the shape-optimised waveform decay
within $y^+\approx10$, and the envelope of the spanwise velocity falls to 10\%
of $W^+$ near $y^+\approx14.5$ for both waveforms. The two waveforms
therefore differ in the temporal distribution of the strain within this
layer. Within $5\le y^+\le12$, the peak absolute values of the wall-normal
curvature of $\langle w\rangle$, $|\partial_yG|$, and of the rate of change
of the Stokes strain, $|\partial_tG|$, are
1.7 and 2.4
times larger for the shape-optimised waveform
(Table~\ref{tab:forcing_verification}). The shape-optimised waveform thus
changes at least two features at once, the time over which $G$ is maintained near its maximum and the sharpness of
its reversals; the comparison of the two waveforms therefore cannot attribute a difference in the flow response to either feature separately. The phase-conditional analysis of
\S\S\ref{sec:duty_cycle} and \ref{sec:enstrophy} distinguishes, within each
cycle, the intervals in which the strain is maintained from those in which it
reverses; the durations of these intervals, and the strain level sustained
during them, are quantified in \S\ref{sec:duty_cycle}. The conditional analysis does not,
however, separate the effect of the duration of these intervals from the effect of
the sharpness of the reversals, which can also modify the response within
each interval.}

\prfrev{The phase-coherent spanwise vorticity
(Figure~\ref{fig:vorticity_phase_variation}(c,d)) reaches its maxima shortly
before the zero crossings of the Stokes strain $G$ and its minima shortly after
them; the phases of these extrema shift with distance from the wall. Its modulation is in phase opposition between the wall and the buffer
layer ($y^+\approx12$); as for the time-averaged profiles
(\S\ref{sec:mean_vorticity}), the constant flow rate requires a change of sign. The
modulation also reflects the two features that distinguish the waveforms, the
sharpness of the reversals and the time over which $G$ is maintained: the
sinusoidal waveform produces a smooth, quasi-harmonic modulation throughout the
cycle, whereas the shape-optimised waveform confines the strongest excursions
to brief events at its sharper reversals, separated by longer intervals of weak
modulation, over which $G$ is maintained.}

\prfrev{The drag varies only weakly over the cycle: the phase-averaged wall
shear stress varies peak to peak by 0.32\% and 0.27\% of its time average for the
sinusoidal and shape-optimised waveforms, respectively (separate phase
ensemble, Appendix~\ref{app:verification}; the response of $\tau_w$ to the stress gradient is discussed in \S\ref{sec:causal_chain}). These variations are small
compared with the difference between the two waveforms, whose time-averaged
wall shear stresses differ by 3.8\% of the sinusoidal value, whereas the relative improvement of
approximately 6\% quoted in \S\ref{sec:results} refers to the drag reduction itself. This weak modulation is consistent
with actuation near the optimal period: \citet{agostini2014spanwise} found
distinct oscillations of the drag at the suboptimal period $T^+=200$, whereas
at $T^+=100$, close to the optimum, the streaks remained more uniformly
attenuated over the cycle (\S\ref{sec:intro}). In contrast, the Reynolds stresses
(\S\ref{sec:reynolds_stress}) and the vorticity variances at $y^+<30$ are
strongly modulated over the cycle; the following subsections examine the
modulation of the variances.}

\subsection{Phase-dependent turbulent response}
\label{sec:phase_turbulent}
\begin{figure}
	\centering
	\begin{minipage}{0.48\textwidth}
		\centering
		\includegraphics[width=\textwidth]{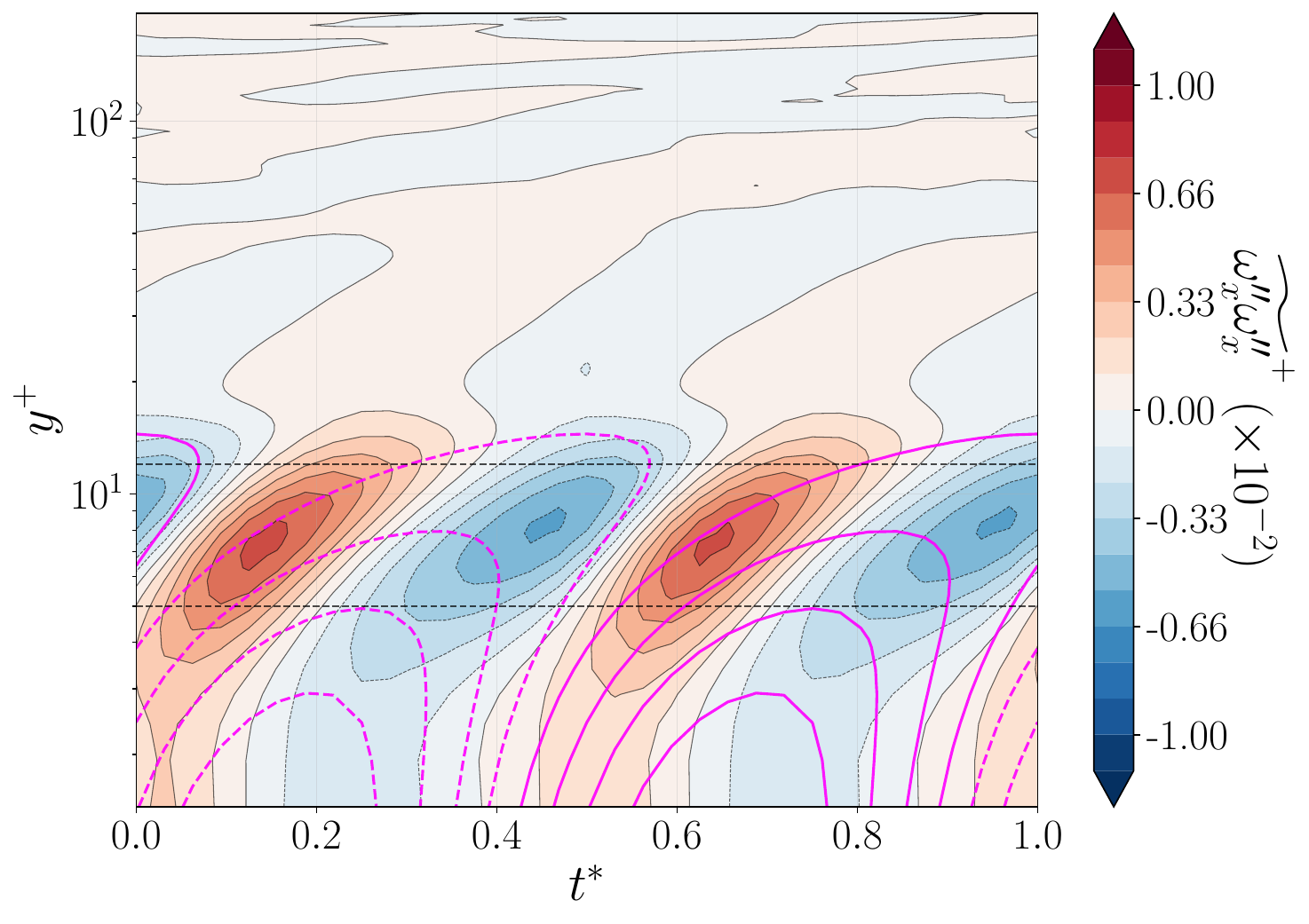}
		\centerline{(a)}
	\end{minipage}
	\hfill
	\begin{minipage}{0.48\textwidth}
		\centering
		\includegraphics[width=\textwidth]{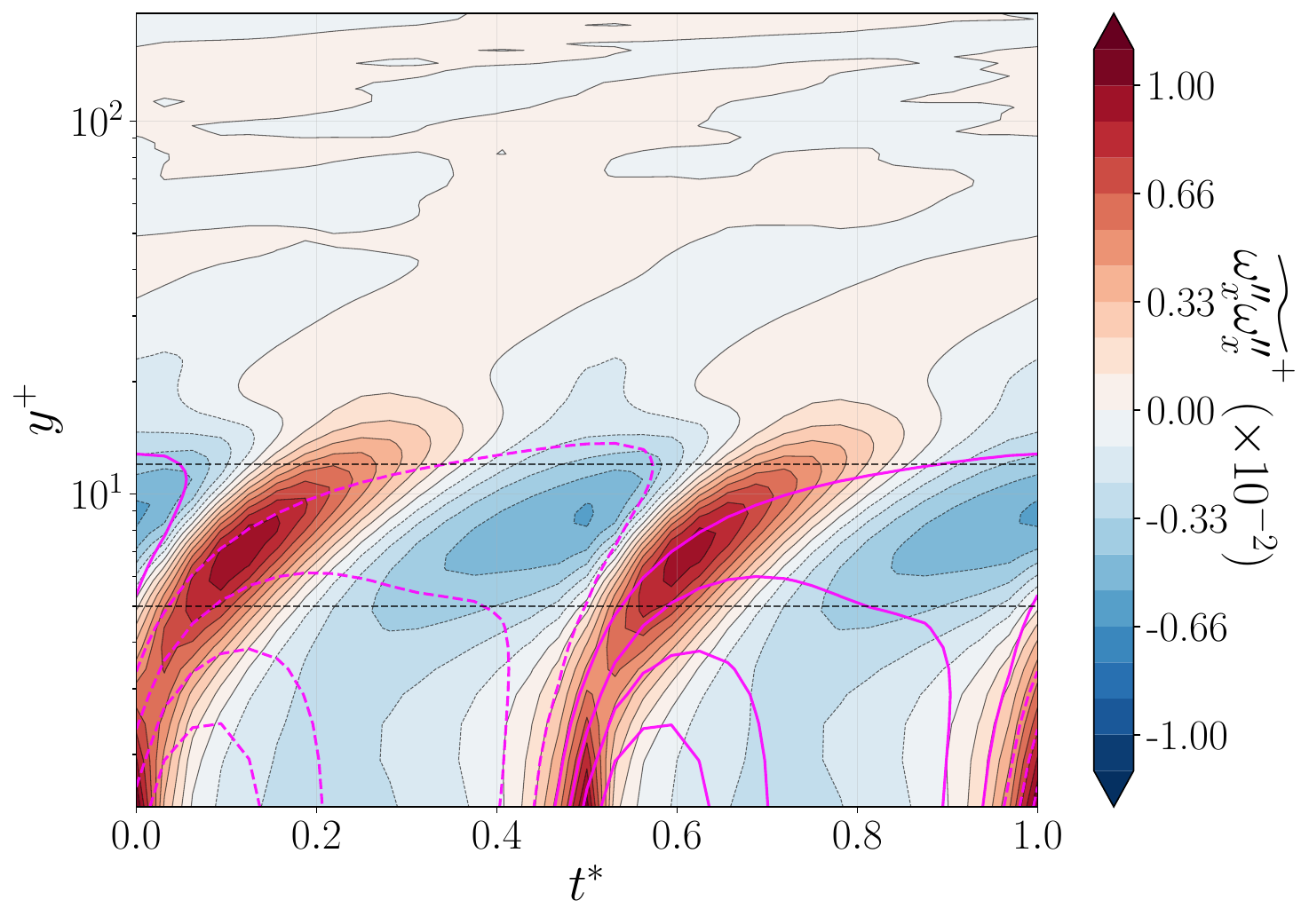}
		\centerline{(b)}
	\end{minipage}
	
	\begin{minipage}{0.48\textwidth}
		\centering
		\includegraphics[width=\textwidth]{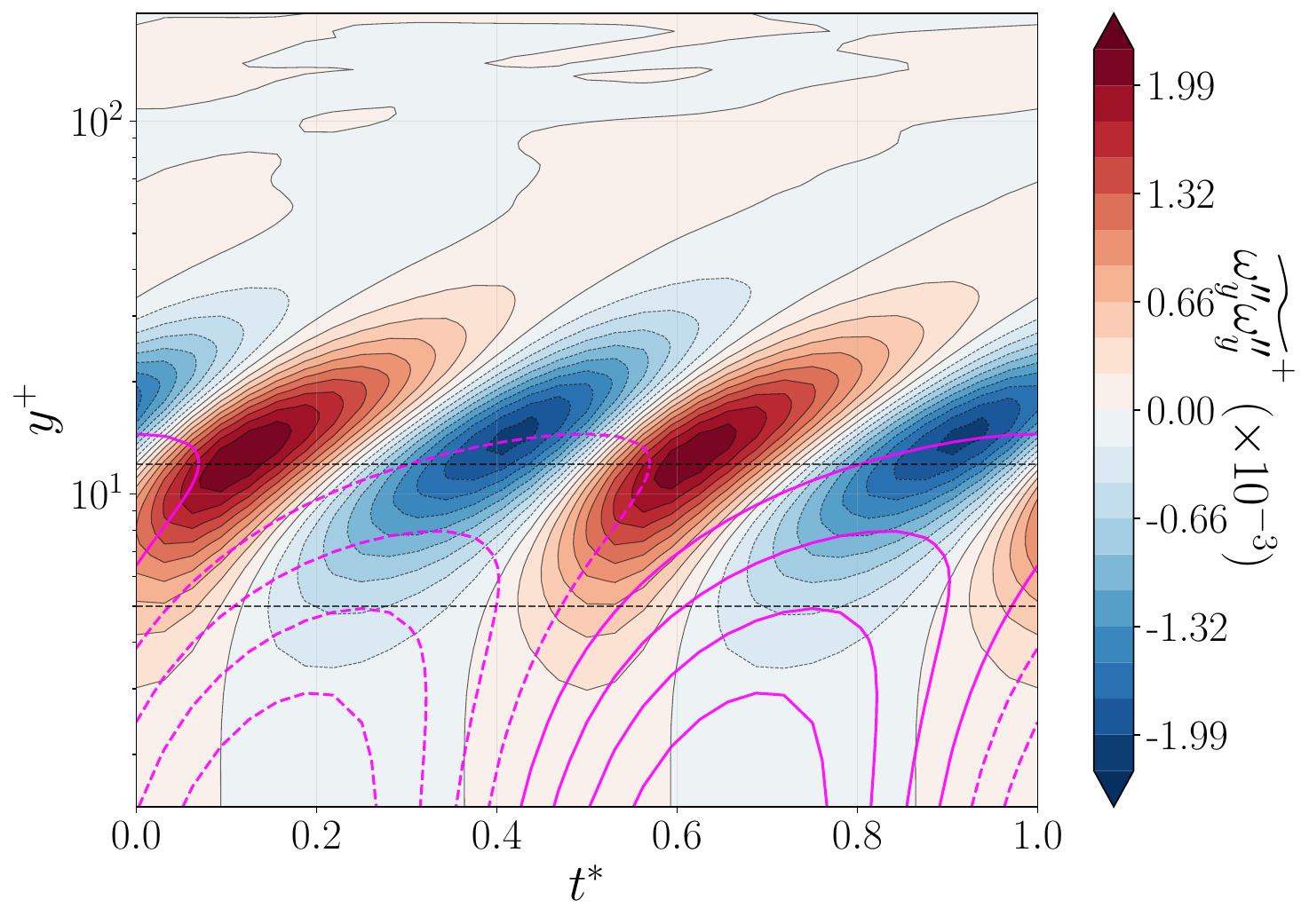}
		\centerline{(c)}
	\end{minipage}
	\hfill
	\begin{minipage}{0.48\textwidth}
		\centering
		\includegraphics[width=\textwidth]{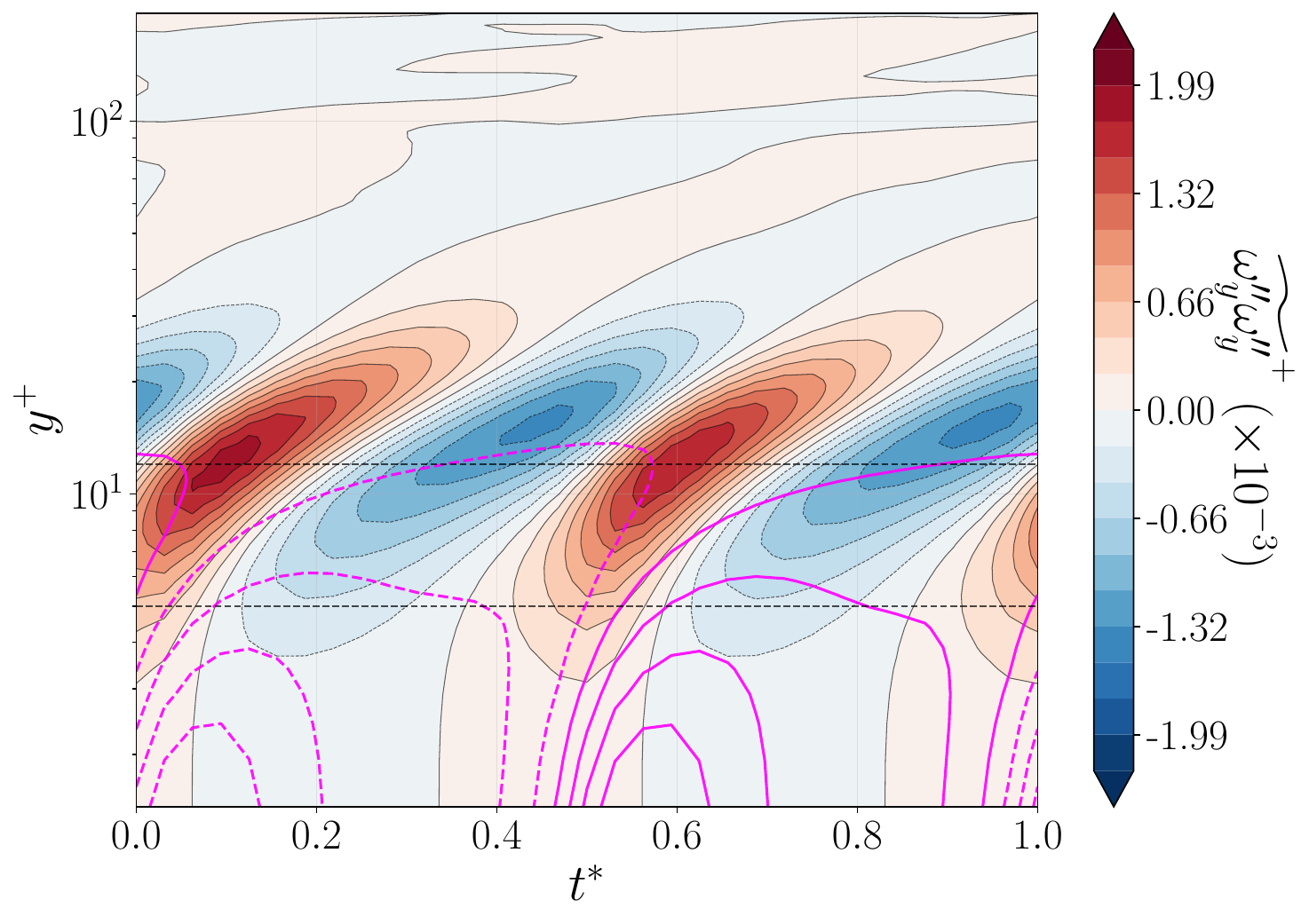}
		\centerline{(d)}
	\end{minipage}
	
	\begin{minipage}{0.48\textwidth}
		\centering
		\includegraphics[width=\textwidth]{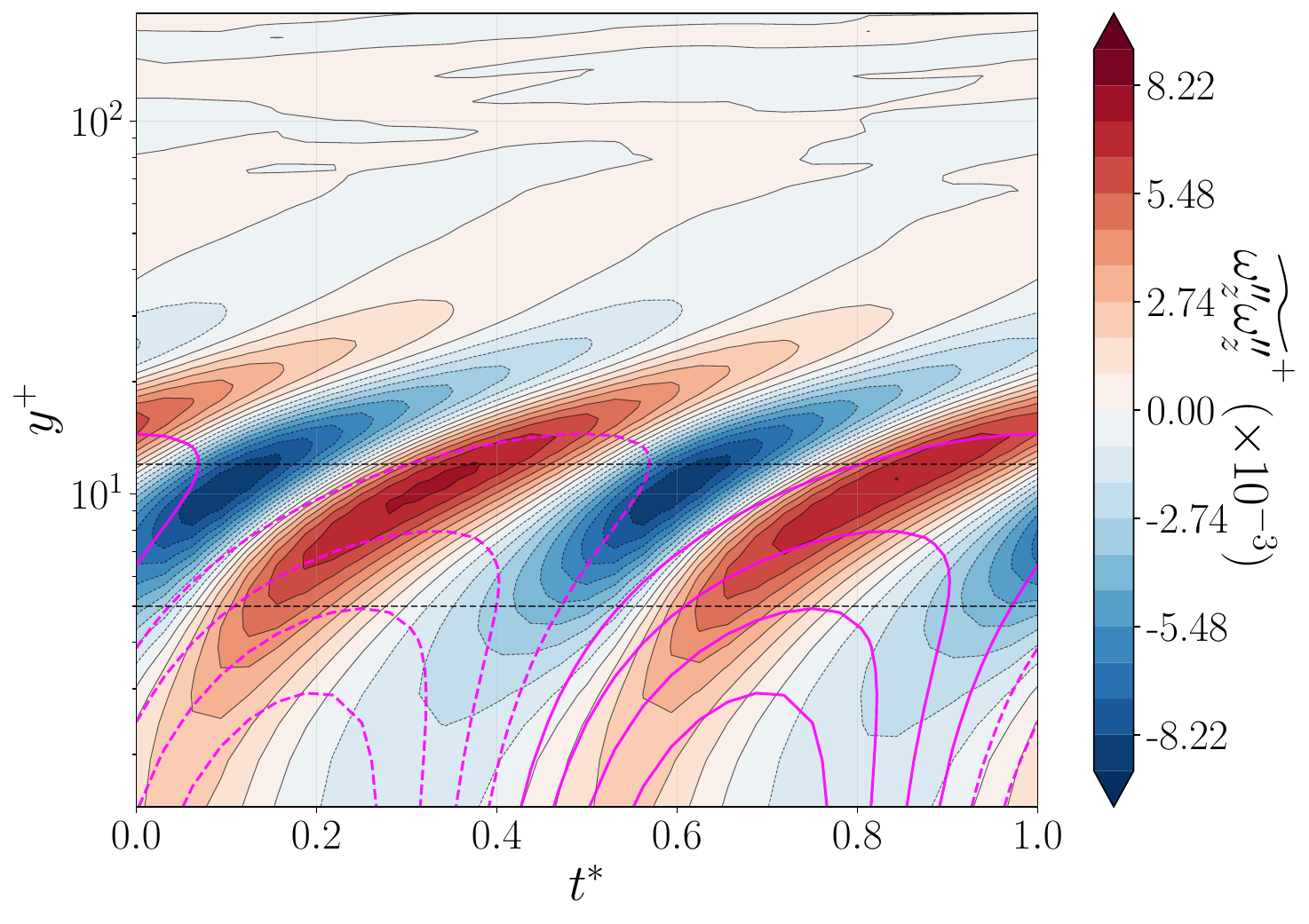}
		\centerline{(e)}
	\end{minipage}
	\hfill
	\begin{minipage}{0.48\textwidth}
		\centering
		\includegraphics[width=\textwidth]{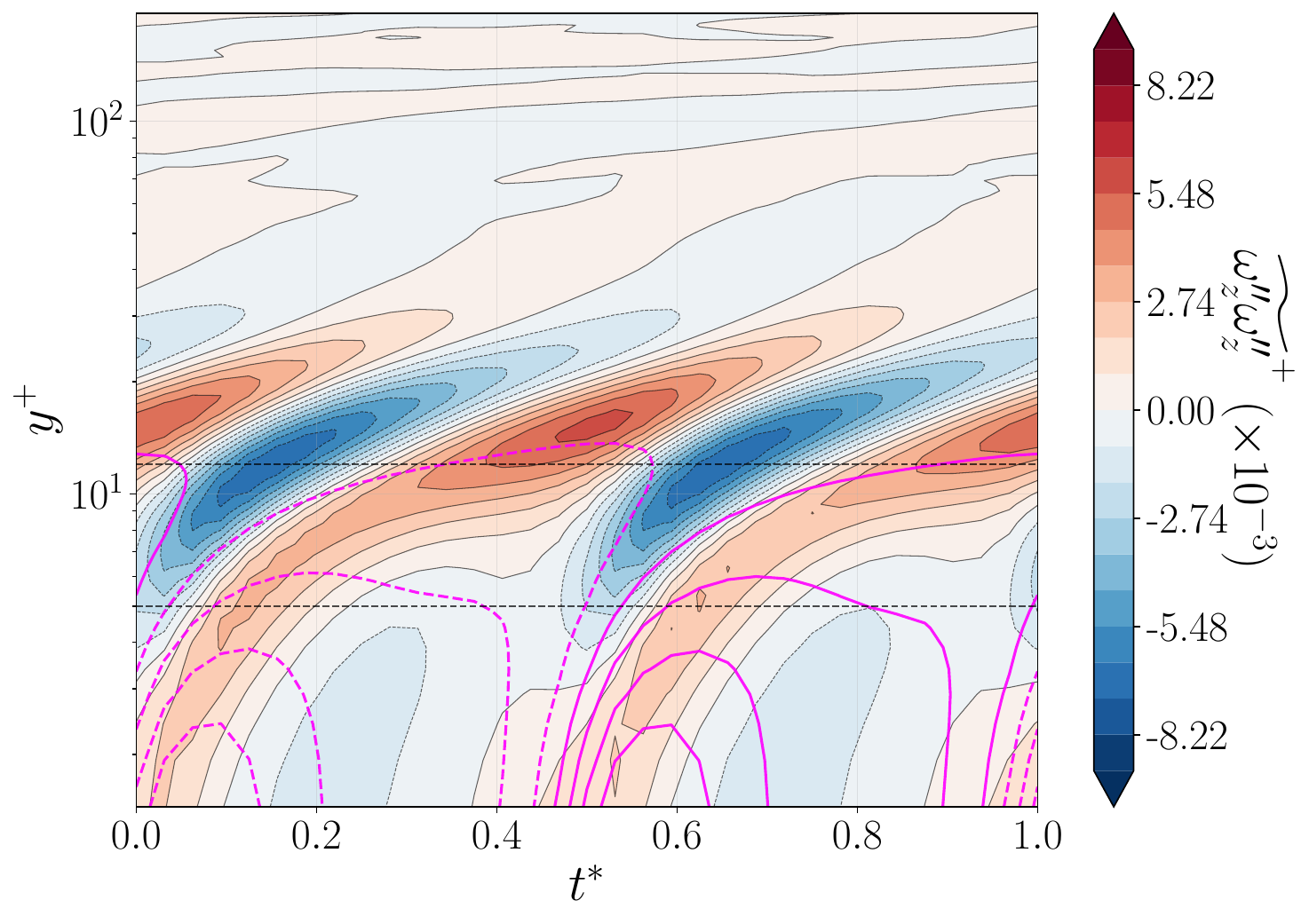}
		\centerline{(f)}
	\end{minipage}
	\caption{Phase-coherent vorticity variances $\widetilde{\omega_i^{\prime\prime}\omega_i^{\prime\prime}}^+$ \prfrev{(colour maps) with superimposed isolines of the Stokes strain $\widetilde{\omega_x}^+$ (magenta; dashed: negative values). Reference lines at $y^+=5$ and 12 delimit the near-wall comparison band.}
		Top row: streamwise component (a, b). Middle row: wall-normal component (c, d). Bottom row: spanwise component (e, f).
		Left column: sinusoidal actuation; right column: shape-optimised actuation.}
	\label{fig:vorticity_stokes_layer}
\end{figure}

\prfrev{To examine how the waveform shape modulates the turbulent fluctuations,
the phase-coherent vorticity variances
$\widetilde{\omega_i^{\prime\prime}\omega_i^{\prime\prime}}^+$ are presented
in Figure~\ref{fig:vorticity_stokes_layer} as functions of actuation phase
$t^*$ and wall-normal distance $y^+$, with superimposed isolines of the Stokes
strain $\widetilde{\omega_x}^+$ (\S\ref{sec:temporal_modulation}). As for the
Reynolds stresses (\S\ref{sec:reynolds_stress}), negative values correspond to phases
during which the variance is below the time average
(\S\ref{sec:phase_analysis}). The colour range of each component is the same for both waveforms, and the colour intensity therefore compares the magnitudes of
their modulations.}

\prfrev{Comparison between the two waveforms reveals that their principal
difference resides in the distribution of the modulation over the cycle. It is
instructive to examine this difference first through the streamwise vorticity
variance (Figure~\ref{fig:vorticity_stokes_layer}(a,b)), which measures the
intensity of the streamwise vorticity fluctuations, associated with the
wall-normal transport of momentum. For both waveforms, this variance exceeds its time
average over less than half of the cycle, an asymmetry consistent with the
hysteresis in turbulence properties and skin friction identified by
\citet{agostini2014spanwise} under sinusoidal actuation. For the
shape-optimised waveform, the phases of above-average variance are shorter and
the modulation is stronger: averaged over $5\le y^+\le12$, the variance exceeds
its time average during 42\% of the cycle, against 47\% for the sinusoidal
waveform, and the r.m.s.\ amplitude of its modulation reaches 48\% of the time
average, against 40\%. The above-average variance is thus concentrated in shorter and more intense
intervals, separated by longer intervals of below-average variance.
The wall-normal variance (Figure~\ref{fig:vorticity_stokes_layer}(c,d))
exhibits the same contrast, with nearly the same fractions of the cycle,
whereas the reverse is observed for the spanwise variance
(Figure~\ref{fig:vorticity_stokes_layer}(e,f)), which exceeds its time average
during approximately 60\% of the cycle for the shape-optimised waveform, against 52\% for
the sinusoidal waveform.}

\prfrev{The wall-normal and spanwise modulations are, moreover, anti-correlated
in phase for both waveforms. At each wall-normal location, their correlation
coefficient over the cycle,
\[
\rho_{ab}=\frac{\overline{a\,b}}{\big(\overline{a^2}\;\overline{b^2}\big)^{1/2}},\qquad
a=\widetilde{\omega_y^{\prime\prime}\omega_y^{\prime\prime}},\quad
b=\widetilde{\omega_z^{\prime\prime}\omega_z^{\prime\prime}},
\]
is at most $-0.69$ throughout the band $5\le y^+\le12$, with band means of $-0.90$ and
$-0.85$ for the sinusoidal and shape-optimised waveforms, respectively. Whether one component grows at the expense of the other, or
the two respond to separate changes in their production, cannot be decided
from the variances alone; the answer requires the component transport equations,
which are examined in \S\ref{sec:enstrophy}, after the phases of the
modulation have been related to the actuation intervals in
\S\ref{sec:duty_cycle}.}

\subsection{\prfrev{Reversal and Displacement intervals}}
\label{sec:duty_cycle}

\begin{figure}
	\centering
	\begin{minipage}{0.48\textwidth}
		\centering
		\includegraphics[width=\textwidth]{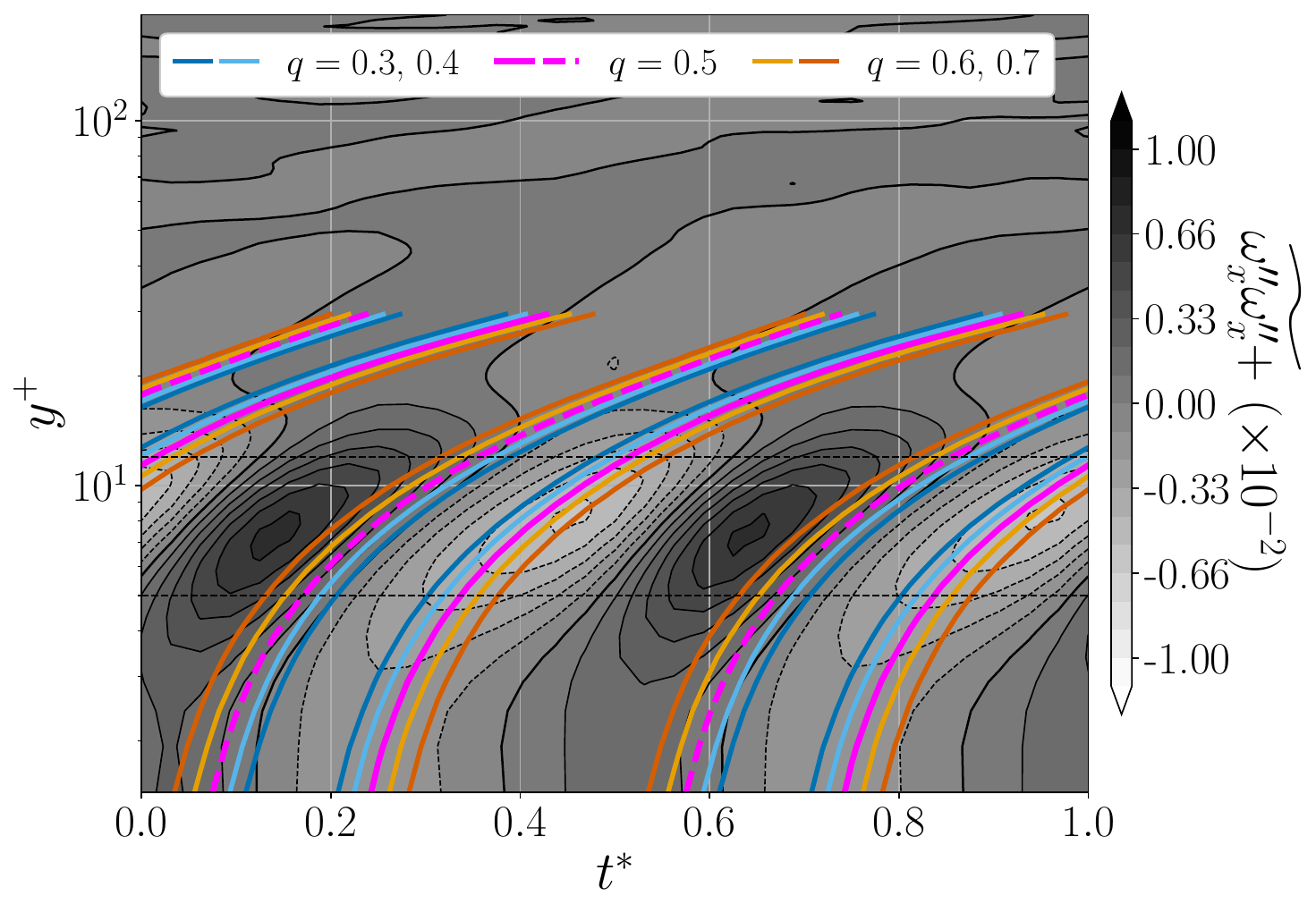}
		\centerline{(a)}
	\end{minipage}
	\hfill
	\begin{minipage}{0.48\textwidth}
		\centering
		\includegraphics[width=\textwidth]{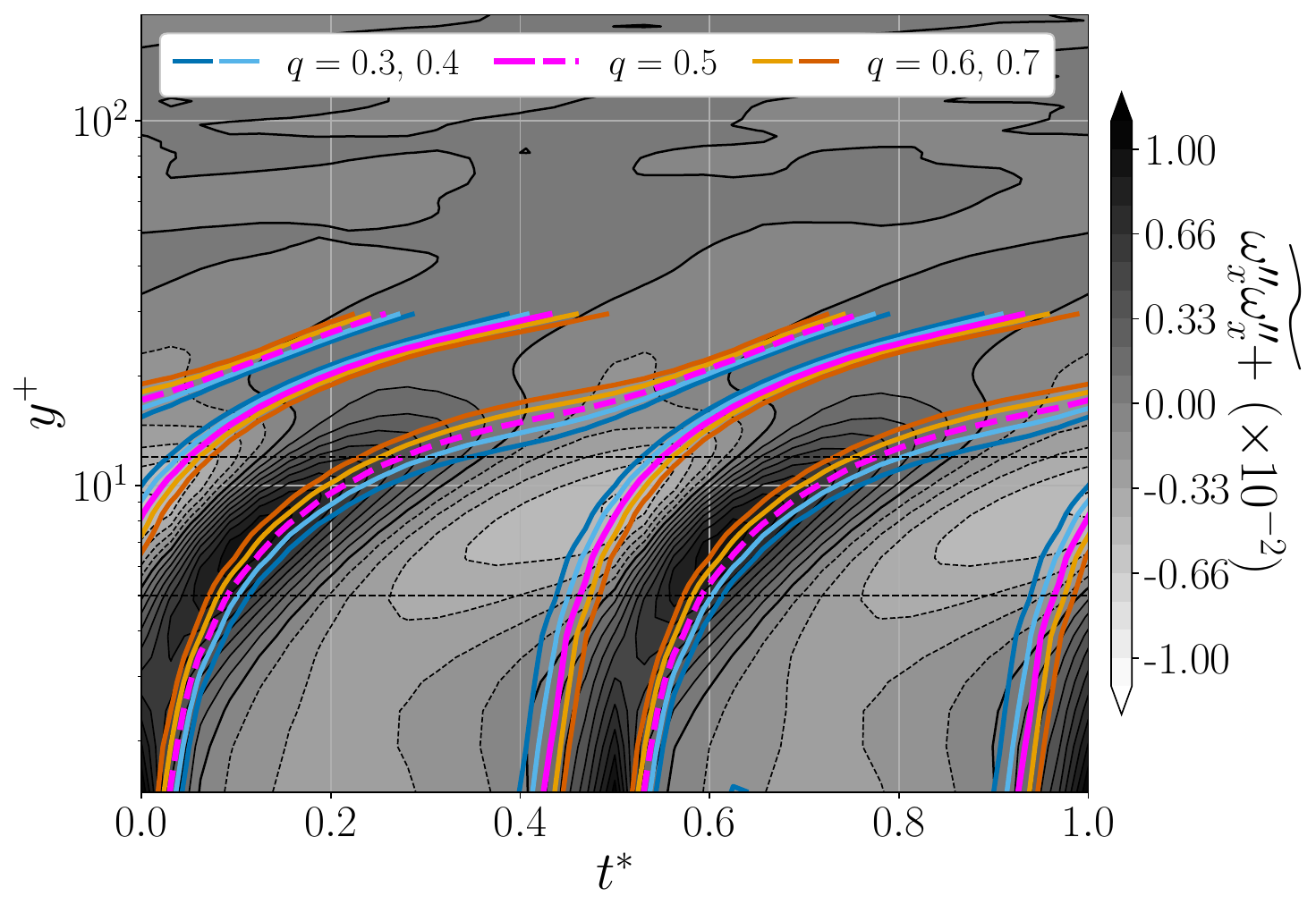}
		\centerline{(b)}
	\end{minipage}

	\begin{minipage}{0.48\textwidth}
		\centering
		\includegraphics[width=\textwidth]{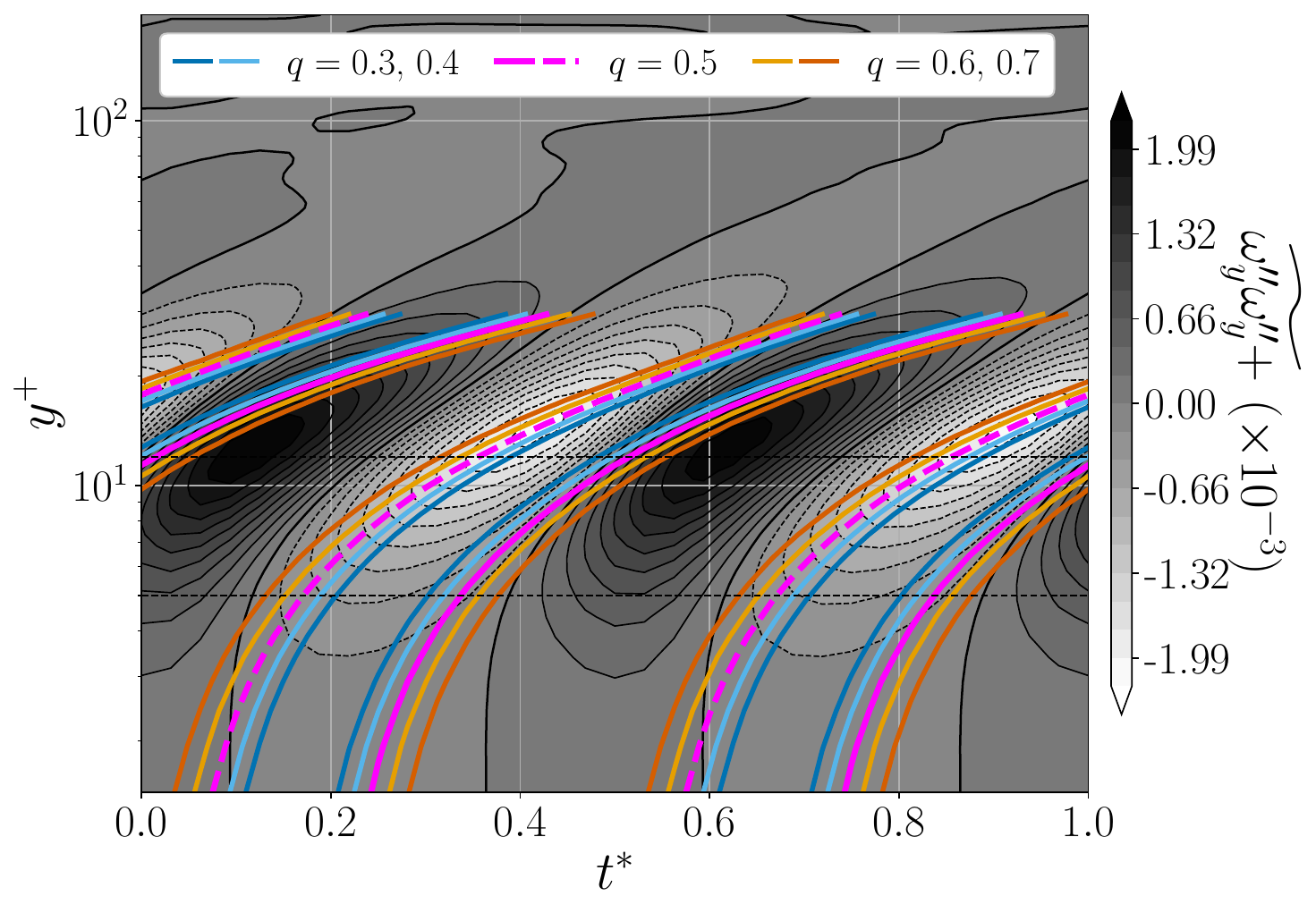}
		\centerline{(c)}
	\end{minipage}
	\hfill
	\begin{minipage}{0.48\textwidth}
		\centering
		\includegraphics[width=\textwidth]{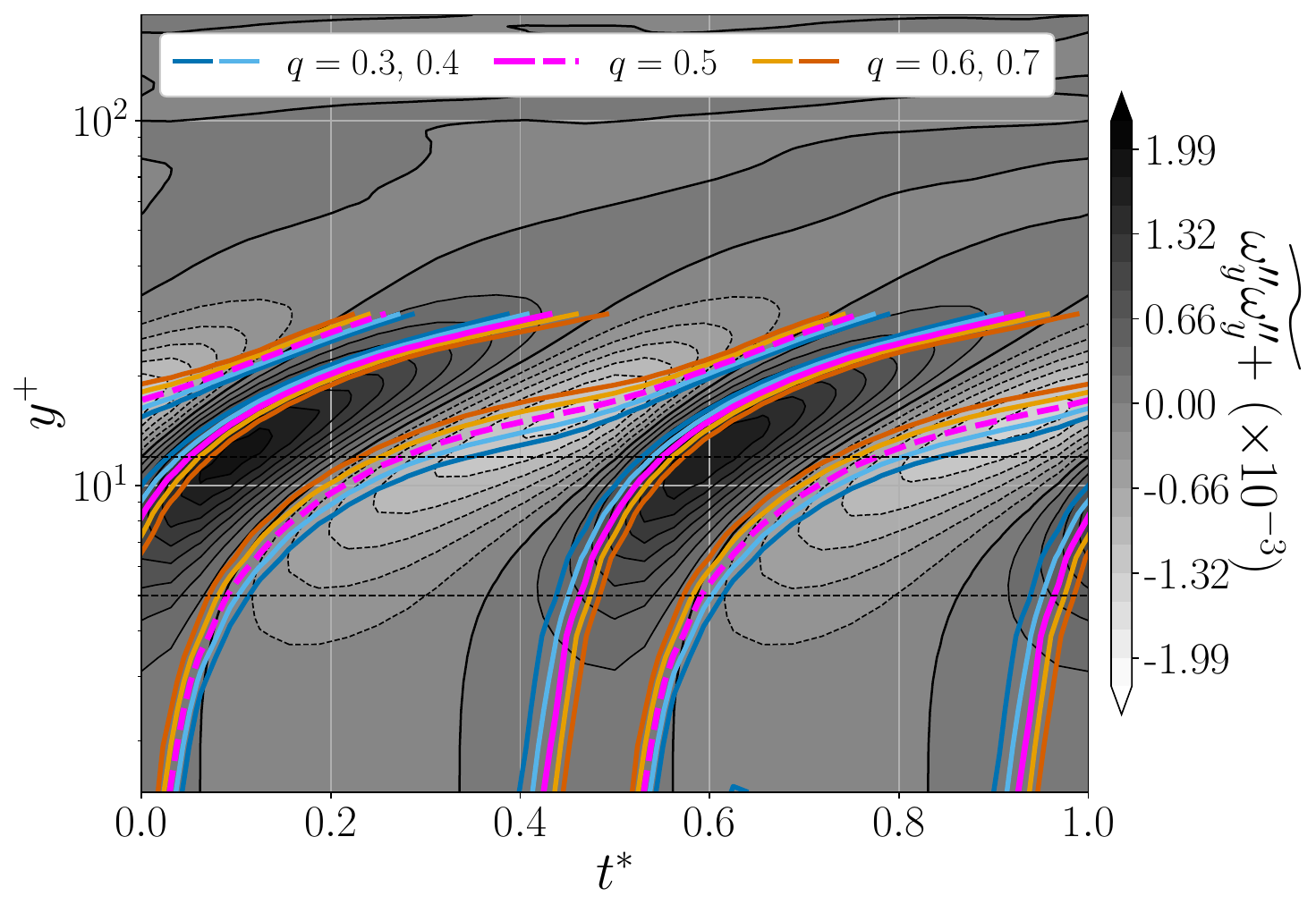}
		\centerline{(d)}
	\end{minipage}

	\begin{minipage}{0.48\textwidth}
		\centering
		\includegraphics[width=\textwidth]{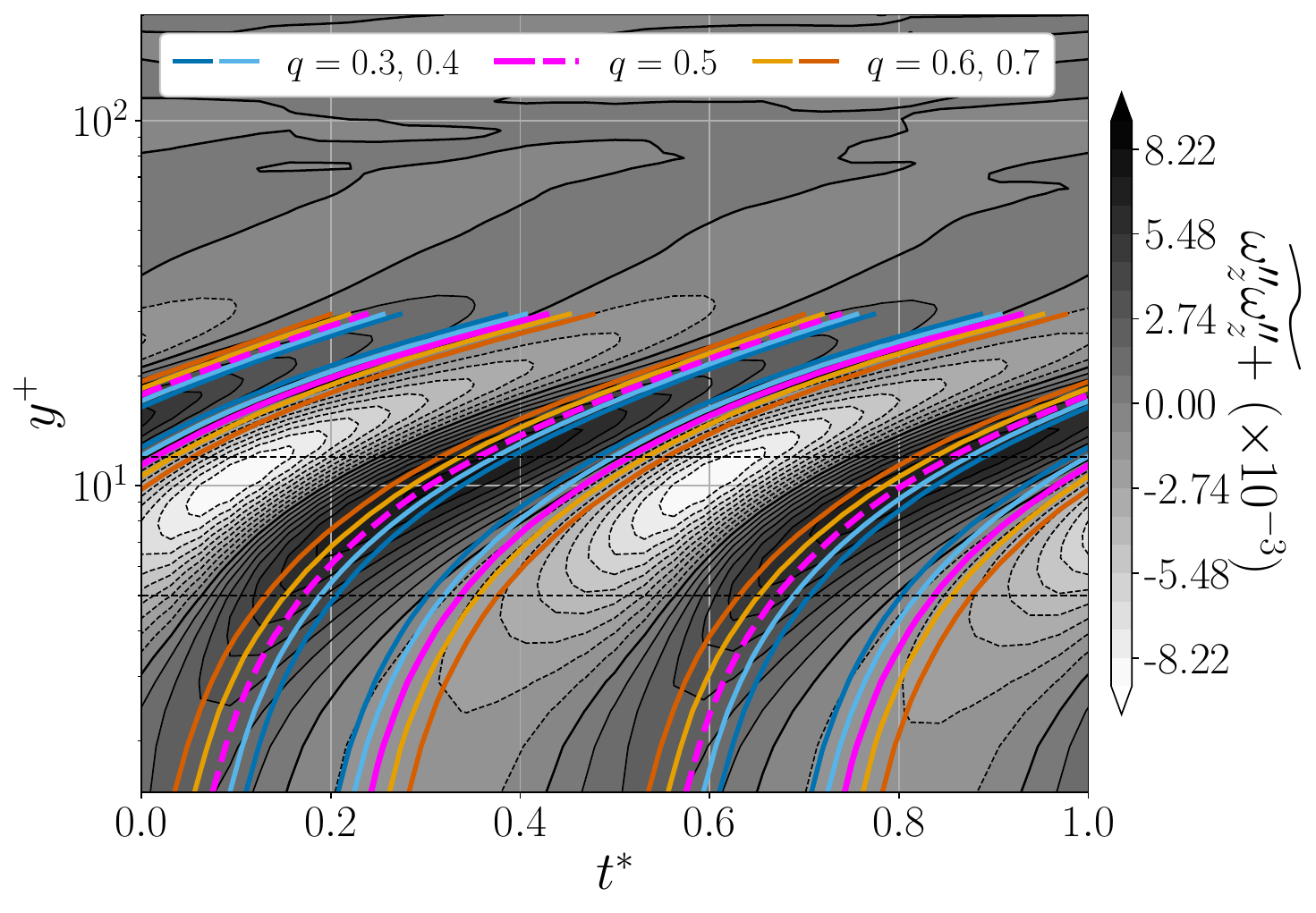}
		\centerline{(e)}
	\end{minipage}
	\hfill
	\begin{minipage}{0.48\textwidth}
		\centering
		\includegraphics[width=\textwidth]{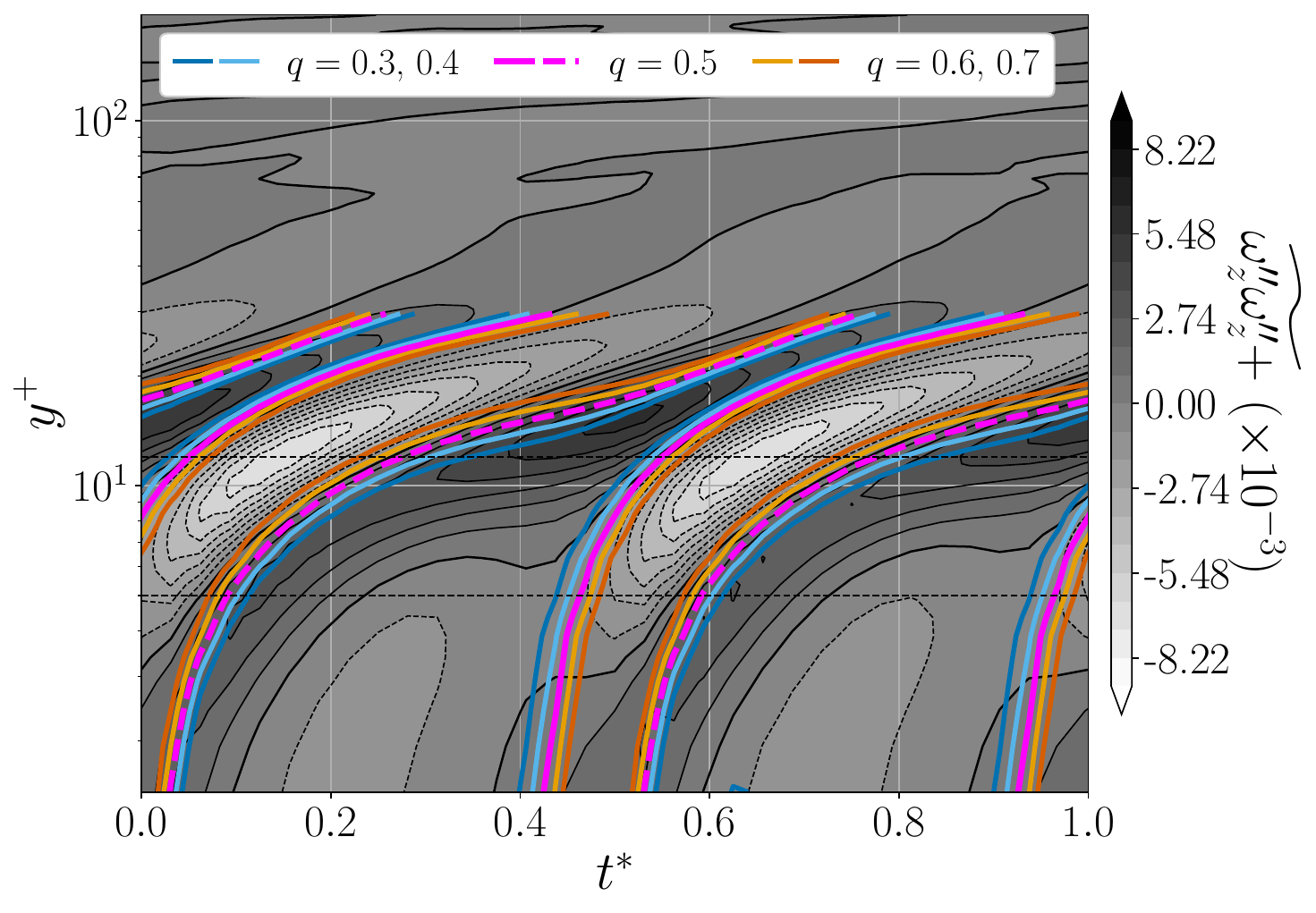}
		\centerline{(f)}
	\end{minipage}
	\caption{\prfrev{Phase-coherent vorticity variances $\widetilde{\omega_i^{\prime\prime}\omega_i^{\prime\prime}}^+$ (greyscale). Solid/dashed lines: start/end of Reversal, where $|\partial_tG|$, normalised by its maximum over the cycle at each $y^+$, rises above/falls below the threshold $q$; thick magenta: $q=0.5$; thin blue: $q=0.3$ and 0.4; thin orange: $q=0.6$ and 0.7. The coloured boundaries are shown for $y^+\le30$. Reference lines at $y^+=5$ and 12 delimit the near-wall comparison band.
		Top row: streamwise component (a, b); middle row: wall-normal (c, d); bottom row: spanwise (e, f).
		Left column: sinusoidal actuation; right column: shape-optimised.}}
	\label{fig:vorticity_grayscale_correlation}
\end{figure}

\prfrev{The Reversal and Displacement intervals introduced in
\S\ref{sec:actuation} are defined at each wall-normal location from the rate
of change of the Stokes strain, $|\partial_tG|$, normalised by its maximum over
the cycle at the same wall-normal location; the definition is therefore
independent of the amplitude of $G$, which decreases with distance from
the wall. Reversal comprises the phases during which this normalised rate
exceeds a threshold $q$, and Displacement the complementary phases, their
fractions of the cycle being designated $f_{\mathrm{Rev}}$ and $f_{\mathrm{Disp}}$. Each
Reversal interval is further divided into its first and second halves, R1 and
R2, during which $G$ weakens towards its change of sign and grows again with
the opposite sign, respectively. \prfrev{The outer boundaries of a Reversal interval are fixed by the rate threshold alone: they bracket, without coinciding with, the zero crossing of $G$ that separates R1 from R2, and they are distinct from the wall-velocity reversals, from which the local strain is separated by the viscous phase shift (\S\ref{sec:actuation}).} The partition classifies the phases of a continuous
response and provides no independent measure of SSP activity. The boundaries of the Reversal intervals are superimposed on the
phase-coherent vorticity variances in
Figure~\ref{fig:vorticity_grayscale_correlation} (greyscale) for the
thresholds $q=0.3$ to 0.7, each Reversal interval extending from a solid line
to the following dashed one: the thin blue lines correspond to $q=0.3$ and
0.4, the thin orange lines to $q=0.6$ and 0.7, and the thick magenta lines to
$q=0.5$, the threshold adopted in all subsequent phase maps and conditional
statistics. The influence of $q$ on the two durations is examined below.
All boundaries are shown up to $y^+=30$, where the envelope of the spanwise
velocity has decayed to 1.2\,\% and 1.4\,\% of $W^+$ for the sinusoidal and
shape-optimised waveforms, respectively (\S\ref{sec:temporal_modulation} and
Figure~\ref{fig:vorticity_phase_variation}(a,b)); this height limits the
representation and is not a physical boundary of the influence of the Stokes
layer.}

\prfrev{Two averages are used in what follows, in order to separate the level
of a quantity within an interval from the contribution of that interval to the
time average. For a phase-averaged quantity $A(y,t^*)$ and an interval $I$ occupying the
fraction $f_I$ of the cycle, i.e.\ of duration $f_I$ for a period of unity, the
conditional mean and the decomposition of the time average are
\[
\overline{A}^{\,I}=\frac{1}{f_I}\int_IA\,\mathrm{d}t^*,\qquad
\overline{A}=\sum_If_I\,\overline{A}^{\,I},
\]
in which the sum runs over the intervals of a disjoint partition of the cycle,
$I\in\{\mathrm{Disp},\mathrm{Rev}\}$ or
$I\in\{\mathrm{Disp},\mathrm{R1},\mathrm{R2}\}$. The conditional mean
$\overline{A}^{\,I}$ thus measures the level of $A$ over the interval,
independently of its duration, whereas $f_I\overline{A}^{\,I}$ is the
contribution of the interval to the time average. The band mean $[\,\cdot\,]_b$ is the mean over the wall-normal grid points in
$5\le y^+\le12$
(Appendix~\ref{app:verification}).}

\begin{figure}
	\centering
	\includegraphics[width=0.55\textwidth]{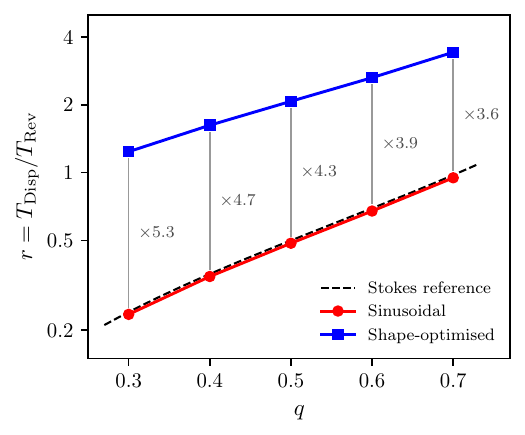}
	\caption{\prfrev{Ratio $r$ of the Displacement to the Reversal duration,
	computed from the Displacement fraction averaged over $5\le y^+\le12$, as a
	function of the threshold $q$, for the
	sinusoidal and shape-optimised waveforms; dashed line: laminar Stokes
	reference, Eq.~(\ref{eq:threshold_harmonic}). Labels give
	$r_{\mathrm{shaped}}/r_{\sin}$ at each $q$. Logarithmic ordinate. The
	estimator is described in Appendix~\ref{app:verification}.}}
	\label{fig:duration_ratio}
\end{figure}

\prfrev{The relative durations of the two intervals are compared next, through
the ratio $r=f_{\mathrm{Disp}}/f_{\mathrm{Rev}}$ at each wall-normal location,
or equivalently through the fraction of the cycle occupied by Displacement,
$f_{\mathrm{Disp}}=r/(1+r)$. The band mean $[\,f_{\mathrm{Disp}}]_b$ is evaluated first
and converted afterwards into a ratio,
$r=[\,f_{\mathrm{Disp}}]_b/(1-[\,f_{\mathrm{Disp}}]_b)$, designated $r_{\sin}$ and
$r_{\mathrm{shaped}}$ for the sinusoidal and shape-optimised waveforms
(Appendix~\ref{app:verification}). These ratios
are presented in Figure~\ref{fig:duration_ratio} for the thresholds
$q=0.3$--0.7, together with the reference given by the laminar Stokes solution
for sinusoidal wall motion (Appendix~\ref{app:verification}). At $q=0.5$,
$r_{\sin}\approx0.5$, i.e.\ Displacement occupies $1/3$ of the cycle, whereas
$r_{\mathrm{shaped}}\approx2$, i.e.\ $2/3$ of the cycle.
The Stokes strain, however, has nearly the same magnitude during Displacement
for the two waveforms, the band mean of its Displacement conditional mean
being $[\,\overline{|G^+|}^{\,\mathrm{Disp}}]_b\approx0.87$ and 0.88; the two
waveforms therefore differ in the time over which this strain level is
maintained and in the sharpness of the reversals, the peak curvature and change rate of the strain being larger by factors of about 1.7 and 2.4
(\S\ref{sec:temporal_modulation}).}

\prfrev{The boundaries for the five thresholds lie close to one another in
Figure~\ref{fig:vorticity_grayscale_correlation}, a first visual indication that
the partition depends only moderately on $q$.
Figure~\ref{fig:duration_ratio} quantifies this dependence and brings to light a strong and robust
contrast between the two waveforms: the shape-optimised ratio is 5.3 times the
sinusoidal ratio at $q=0.3$ and 3.6 times at $q=0.7$, the contrast persisting
at every threshold whilst decreasing moderately as $q$ increases. As the
sinusoidal ratios lie on the curve of the laminar reference, which itself
varies with $q$, the threshold is a descriptive convention rather than a
physical switching constant, and the two waveforms are compared at the same
threshold.}

\prfrev{The modulation of the variances is now located within these intervals.
The variance lobes in Figure~\ref{fig:vorticity_grayscale_correlation} are
inclined approximately as the interval boundaries. Within the band
$5\le y^+\le12$, the wall-normal variance
(Figure~\ref{fig:vorticity_grayscale_correlation}(c,d)) is enhanced mainly
during R1 and attenuated during Displacement. The spanwise variance
(Figure~\ref{fig:vorticity_grayscale_correlation}(e,f)) exhibits the opposite
organisation, consistently with the anti-correlation of the two components
noted in \S\ref{sec:phase_turbulent}. The streamwise variance
(Figure~\ref{fig:vorticity_grayscale_correlation}(a,b)) follows the same
organisation as the wall-normal variance with a phase lag, its highest
levels occurring around the end of R2.}

\prfrev{This organisation is quantified by the conditional means
$[\,\overline{A}^{\,I}]_b$ at $q=0.5$. They confirm the signs of the departures from the time average, which are unchanged at the
other thresholds examined (Appendix~\ref{app:verification}), and they place the Reynolds shear stress
and the production of Figure~\ref{fig:phase_comparison}(a--d) in the same
organisation as the wall-normal and streamwise variances:
\[
\begin{aligned}
[\,\overline{A}^{\,\mathrm{Rev}}]_b>[\,\overline{A}\,]_b>
[\,\overline{A}^{\,\mathrm{Disp}}]_b&\qquad\text{for}\qquad
A=\langle\omega_y''^2\rangle,\ \langle\omega_x''^2\rangle,\
-\langle u''v''\rangle\ \text{and}\ P_{uu},\\
[\,\overline{A}^{\,\mathrm{Rev}}]_b<[\,\overline{A}\,]_b<
[\,\overline{A}^{\,\mathrm{Disp}}]_b&\qquad\text{for}\qquad
A=\langle\omega_z''^2\rangle.
\end{aligned}
\]
As the time average is the sum of the two contributions,
$\overline{A}=f_{\mathrm{Disp}}\,\overline{A}^{\,\mathrm{Disp}}
+f_{\mathrm{Rev}}\,\overline{A}^{\,\mathrm{Rev}}$ with
$f_{\mathrm{Disp}}+f_{\mathrm{Rev}}=1$, the departures from the time average
satisfy
\[
f_{\mathrm{Disp}}\big(\overline{A}^{\,\mathrm{Disp}}-\overline{A}\big)
=-f_{\mathrm{Rev}}\big(\overline{A}^{\,\mathrm{Rev}}-\overline{A}\big),
\]
i.e. they have opposite signs and the shorter interval carries the larger one: for the
shape-optimised waveform, the wall-normal and streamwise variances exceed
their time averages most strongly within the short Reversal, visible as the
narrow bands between the solid and dashed magenta lines of
Figure~\ref{fig:vorticity_grayscale_correlation}(b,d), whereas for the
sinusoidal waveform the wall-normal variance is most strongly attenuated
within the short Displacement
(Figure~\ref{fig:vorticity_grayscale_correlation}(c)). For the wall-normal
variance, the departures, expressed as percentages of the time average,
$100\,\big([\,\overline{A}^{\,I}]_b/[\,\overline{A}\,]_b-1\big)$, are:}
\begin{itemize}\setlength{\itemsep}{0pt}\setlength{\parskip}{0pt}
\item \prfrev{sinusoidal waveform: $+16\,\%$ during Reversal and $-33\,\%$ during
Displacement;}
\item \prfrev{shape-optimised waveform: $+36\,\%$ during Reversal and $-20\,\%$ during
Displacement.}
\end{itemize}
\prfrev{A conditional mean measures the level of a quantity over an interval,
not its rate of change. The departures listed above therefore describe the phase
organisation of the variances; they indicate neither growth nor decay, which
follow from the budgets of \S\ref{sec:enstrophy}, nor the terms that produce
this organisation.}

	
	\section{Stochastic enstrophy budget analysis}
	\label{sec:enstrophy}
	The phase-resolved analysis of \S\ref{sec:results} \prfrev{shows} that the stochastic
	vorticity variances $\langle\omega_i^{\prime\prime}\omega_i^{\prime\prime}\rangle^+$
	respond to the Stokes-strain state \prfrev{consistently} with the duty-cycle
	modulation of the self-sustaining process.  \prfrev{The variances, however, measure
	the level of the vorticity fluctuations at each phase; they do not identify the
	production and dissipation processes of the fluctuations. In order to identify these processes, the
	terms of the budget equations of the variances, i.e.\ production, stretching,
	transport, diffusion and dissipation, are examined throughout the actuation cycle.}
	\prfrev{These} equations, derived from the fluctuating
	vorticity transport equations under the triple decomposition of
	\S\ref{sec:methodology}, are summarised in Appendix~\ref{app:enstrophy_equations}.

\prfrev{The budget terms are compared between the Reversal and Displacement
intervals defined in \S\ref{sec:duty_cycle}. In the phase-resolved maps of
\S\S\ref{sec:enstr_wy}--\ref{sec:enstr_wx}, the Reversal interval at each
$y^+$ extends from the solid to the dashed magenta isoline, at $q=0.5$, as in
Figure~\ref{fig:vorticity_grayscale_correlation}. These isolines, and
the red line described below, extend over the same range, $y^+\le30$. \prfrev{The red line in the maps is drawn through the midpoint of each
Reversal interval, i.e. at $q\approx1$, and separates R1 from R2, as defined in
\S\ref{sec:duty_cycle}.}}

\prfrev{In order to establish how each interval contributes to the time
average of a budget term $A$, the decomposition of \S\ref{sec:duty_cycle},
$\overline{A}=\sum_If_I\overline{A}^{\,I}$, is applied at each wall-normal
location, with the partition $I\in\{\mathrm{Disp},\mathrm{R1},\mathrm{R2}\}$.
As the budget terms are rates of change of enstrophy, the conditional mean
$\overline{A}^{\,I}$ is referred to as the conditional rate of the term, and $f_I\overline{A}^{\,I}$ as its contribution; both are quoted as band means (\S\ref{sec:duty_cycle}). Each term is taken with
the sign that it carries in its budget equation, positive when it increases
the enstrophy and negative when it reduces it, and the averages retain this
sign; a negative conditional rate thus indicates that the term reduces the
enstrophy, on average, over the interval. }

\begin{figure}
	\centering
	\includegraphics[width=\textwidth]{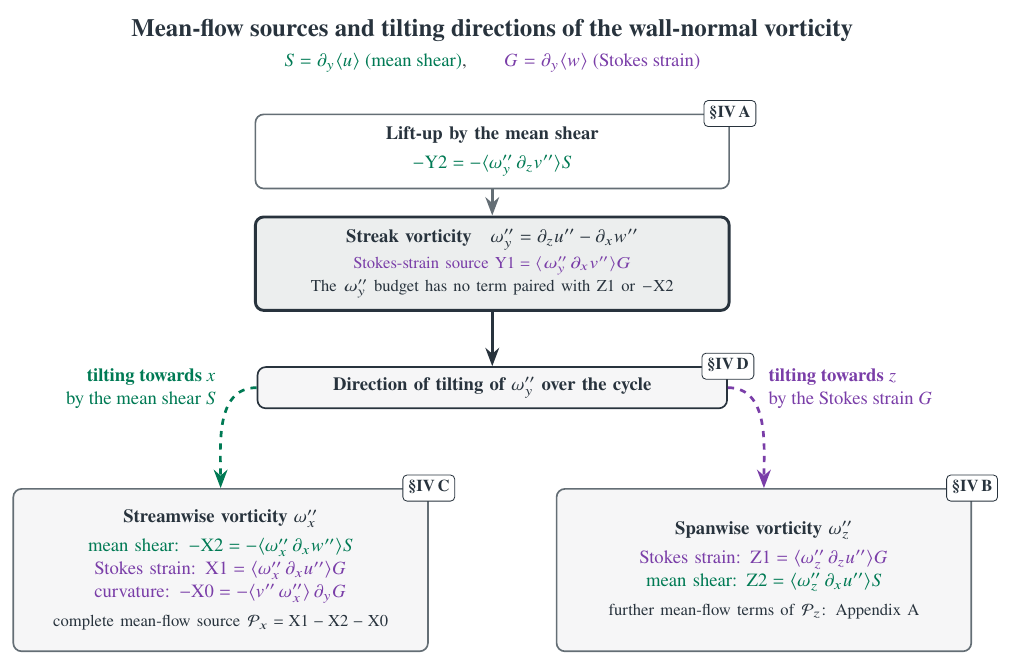}
	\caption{\prfrev{Mean-flow sources of the three enstrophy budgets and tilting
	directions of the wall-normal vorticity, with the subsections in which they
	are examined; green: terms involving the mean shear $S$, purple: terms
	involving the Stokes strain $G$. The sources are derived in
	Appendix~\ref{app:enstrophy_equations}; Figure~\ref{fig:enstrophy_sketch}
	completes this figure with the results of
	\S\S\ref{sec:enstr_wy}--\ref{sec:enstr_tilting}.}}
	\label{fig:budget_frame}
\end{figure}

\prfrev{The three budgets are then examined component by component. In terms of the mean shear $S$ and the Stokes strain $G$
(\S\ref{sec:methodology}), their dominant sources are:}
\begin{itemize}\setlength{\itemsep}{0pt}\setlength{\parskip}{0pt}
\item \prfrev{the mean-shear production of wall-normal enstrophy by lift-up,
$-\mathrm{Y2}=-\langle\omega_y''\partial_zv''\rangle S$;}
\item \prfrev{the Stokes-strain tilting production of spanwise enstrophy,
$\mathrm{Z1}=\langle\omega_z''\partial_zu''\rangle G$;}
\item \prfrev{the mean-shear production of streamwise enstrophy,
$-\mathrm{X2}=-\langle\omega_x''\partial_xw''\rangle S$;}
\item \prfrev{the Stokes-curvature production of streamwise enstrophy,
$-\mathrm{X0}=-\langle v''\omega_x''\rangle\partial_yG$.}
\end{itemize}
\prfrev{Z1 and $-\mathrm{X2}$ contain the tilting of the wall-normal vorticity
$\omega_y''$ towards $z$ by the Stokes strain and towards $x$ by the mean
shear, respectively (\S\S\ref{sec:enstr_wz} and~\ref{sec:enstr_wx}). The tilted vorticity could
be regarded as subtracted from $\omega_y''$, the spanwise and
streamwise enstrophy then growing at the expense of the wall-normal enstrophy.
The wall-normal budget, however, contains no term equal to Z1 or
$-\mathrm{X2}$ with the opposite sign (Eq.~(\ref{eq:Py})); the tilting therefore produces spanwise and streamwise enstrophy without an equal loss of
wall-normal enstrophy. The analysis therefore examines the direction in which
the mean-flow gradients tilt $\omega_y''$.
Figure~\ref{fig:budget_frame} represents these sources and the two tilting
directions, together with the subsections in which they are examined, and
\S\ref{sec:enstr_synthesis} completes it with the results of
\S\S\ref{sec:enstr_wy}--\ref{sec:enstr_tilting}. The derivation of the sources, and
the remaining terms of the three budgets, are given in
Appendix~\ref{app:enstrophy_equations}.}

	\subsection{\prfrev{Wall-normal enstrophy ($\omega_y^{\prime\prime}$): lift-up and Stokes-strain sources}}
	\label{sec:enstr_wy}

\leavevmode\prfrev{The wall-normal component is examined first. Its fluctuation,
$\omega_y''=\partial_zu''-\partial_xw''$, is associated with the streaks through the
spanwise gradient of the streamwise velocity fluctuations, and its variance,
$\langle\omega_y''\omega_y''\rangle$, can be regarded as a surrogate for their amplitude. The questions
addressed are which mean-flow gradient produces $\omega_y''$, how this
production is organised over the cycle, and what modulates it.}

\begin{figure}
		\centering
		\begin{minipage}{0.48\textwidth}
			\centering
			\includegraphics[width=\textwidth]{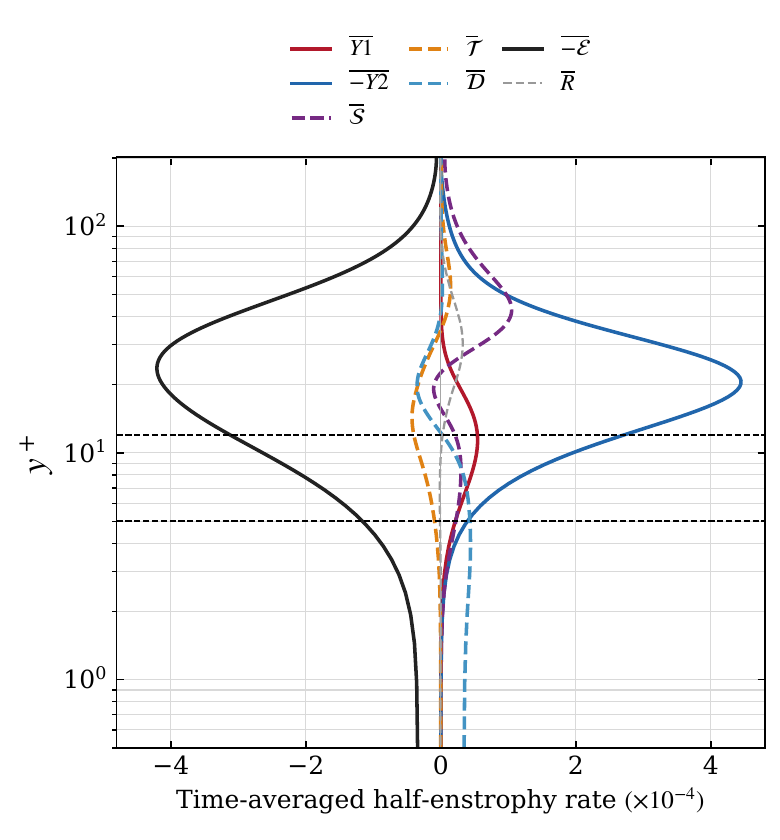}
			\centerline{(a)}
		\end{minipage}\hfill
		\begin{minipage}{0.48\textwidth}
			\centering
			\includegraphics[width=\textwidth]{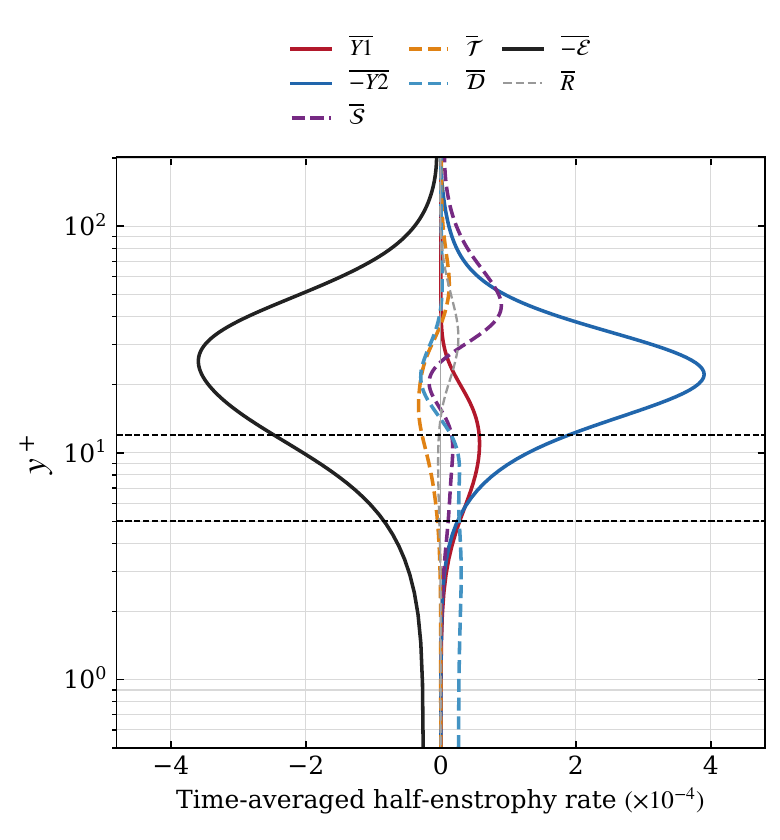}
			\centerline{(b)}
		\end{minipage}
		\caption{\prfrev{Time-averaged wall-normal half-enstrophy budget: signed rates for sinusoidal (a) and shape-optimised (b) actuation. Overbars in the legend denote the time averages of the terms of Eq.~(\ref{eq:enstr_generic}), the component subscript $y$ being omitted on $\mathcal S$, $\mathcal T$, $\mathcal D$ and $\mathcal E$; $R$ is the residual. Each product was formed at every phase of the 32-phase ensemble before averaging over the cycle (Eq.~(\ref{eq:phase_covariance})).}}
\label{fig:budget_wy_tavg}
	\end{figure}

\prfrev{In the equation of $\omega_y''$, the mean-flow source arises from the
wall-normal transport of the mean momentum. A wall-normal velocity $v''$ changes
the streamwise and spanwise velocity fluctuations at the rates $-Sv''$ and
$-Gv''$, and the curl of these rates, $\partial_z(-Sv'')-\partial_x(-Gv'')$, is
the source $G\partial_xv''-S\partial_zv''$ of
$\omega_y''=\partial_zu''-\partial_xw''$. Its multiplication by $\omega_y''$ and
phase averaging give the mean-flow production of the wall-normal
half-enstrophy, $\tfrac12\langle\omega_y''\omega_y''\rangle$ (Eq.~(\ref{eq:Py}) of Appendix~\ref{app:enstrophy_equations}),}
\begin{equation}
  \label{eq:Py_text}
  \mathcal P_y=
  \underbrace{\langle\omega_y''\,\partial_xv''\rangle\,G}_{\mathrm{Y1}}
  \;-\;
  \underbrace{\langle\omega_y''\,\partial_zv''\rangle\,S}_{\mathrm{Y2}} .
\end{equation}
\prfrev{The source $-\mathrm{Y2}$ is the lift-up mechanism: where $v''$ varies
along $z$, as on the flanks of the streaks, the streamwise velocity is changed
unevenly in $z$, which builds the spanwise shear $\partial_zu''$. The source Y1
is the same mechanism acting on the spanwise velocity, and is the only direct
contribution of the Stokes strain to this budget. The
time-averaged budget (Figure~\ref{fig:budget_wy_tavg}(a,b)) is dominated by
viscous dissipation and by $\overline{-\mathrm{Y2}}$; over $5\le y^+\le12$, $\overline{\mathrm{Y1}}$ is about
3 and 2 times weaker than $\overline{-\mathrm{Y2}}$ for the sinusoidal and
shape-optimised waveforms, respectively; $\overline{\mathrm{Y1}}$ is about 12\,\% higher for the shape-optimised
waveform, whereas $\overline{-\mathrm{Y2}}$ is about 35\,\% lower.}

\begin{figure}
		\centering
		\begin{minipage}{0.48\textwidth}
			\centering
			\includegraphics[width=\textwidth]{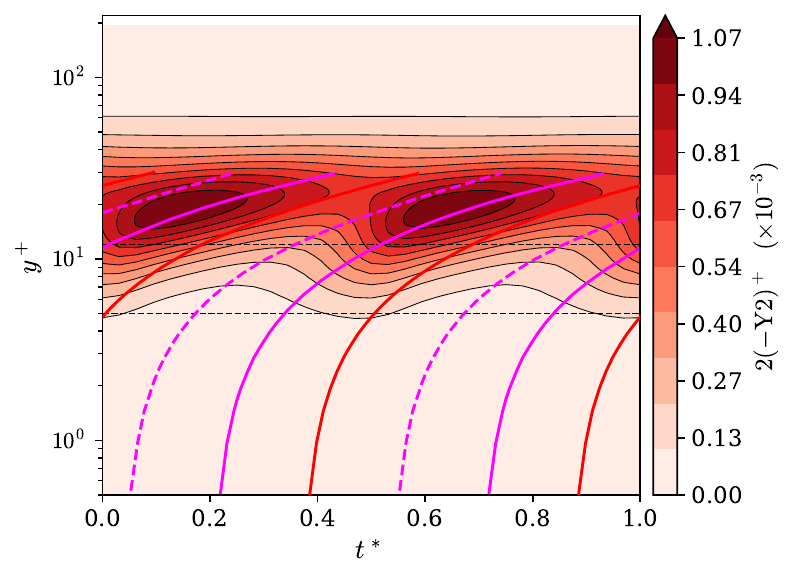}
			\centerline{(a)}
		\end{minipage}\hfill
		\begin{minipage}{0.48\textwidth}
			\centering
			\includegraphics[width=\textwidth]{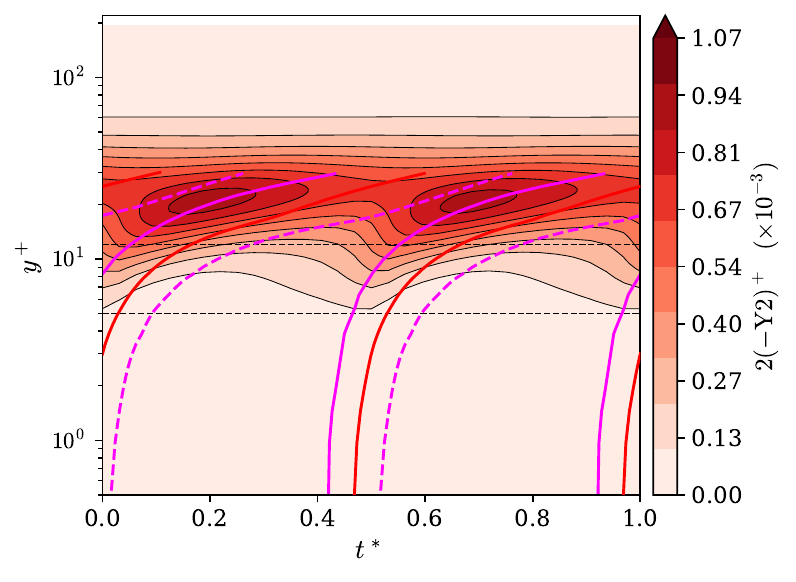}
			\centerline{(b)}
		\end{minipage}
		\begin{minipage}{0.48\textwidth}
			\centering
			\includegraphics[width=\textwidth]{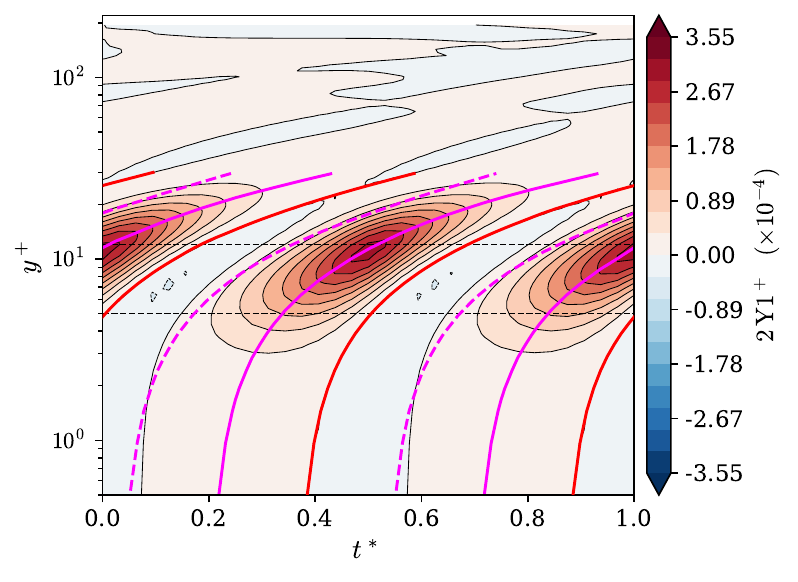}
			\centerline{(c)}
		\end{minipage}\hfill
		\begin{minipage}{0.48\textwidth}
			\centering
			\includegraphics[width=\textwidth]{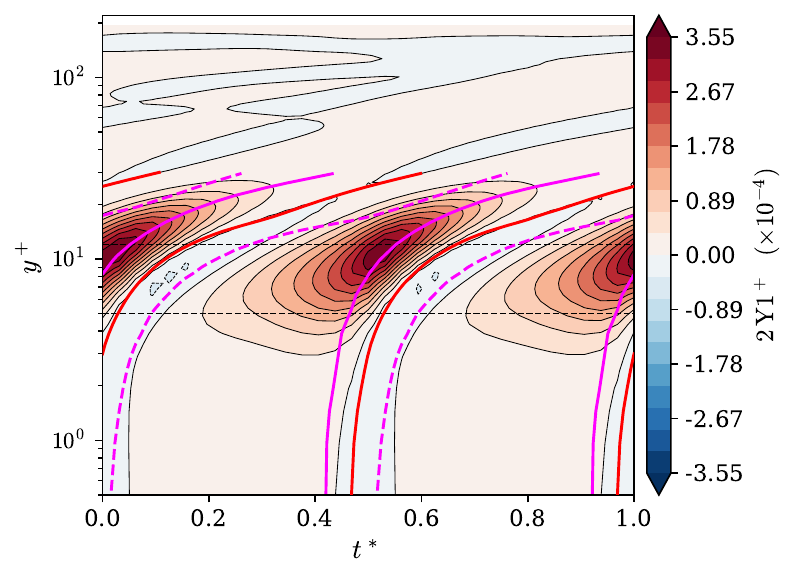}
			\centerline{(d)}
		\end{minipage}
		\caption{Wall-normal enstrophy $\langle\omega_y^{\prime\prime 2}\rangle^+$ budget:
			\prfrev{phase-resolved production terms as functions of $t^*$ and $y^+$, for
			sinusoidal (left column) and shape-optimised (right column) actuation.
			(a,\,b)~Mean-shear production $-\mathrm{Y2}=-\langle\omega_y''\partial_zv''\rangle S$, plotted as $2(-\mathrm{Y2})$.
			(c,\,d)~Stokes-strain source $\mathrm{Y1}=\langle\omega_y''\partial_xv''\rangle G$, plotted as $2\mathrm{Y1}$.
			Isoline convention as described in the text (solid/dashed magenta:
			start/end of Reversal, where the normalised rate of change of the Stokes
			strain rises above/falls below $q=0.5$; red line: division between R1
			and R2, at $q\approx1$). All superimposed lines are shown for $y^+\leq30$.}
			Reference lines at $y^+ = 5$ and $y^+ = 12$ \prfrev{delimit the near-wall comparison band}.
			\prfrev{Colour: full-variance rate, twice the half-variance term of Appendix~\ref{app:enstrophy_equations}.}}
		\label{fig:budget_wy_phase}
	\end{figure}

\prfrev{Figure~\ref{fig:budget_wy_phase}(a,b) conveys the phase maps of
$-\mathrm{Y2}$ in relation to the Reversal boundaries. Within
$5\le y^+\le12$, the maximum of $-\mathrm{Y2}$ is reached in R1 for the sinusoidal
waveform. For the shape-optimised waveform, $-\mathrm{Y2}$ rises above its time
average during the last part of Displacement and peaks at the start of
Reversal, i.e.\ at the beginning of R1. The conditional rate of $-\mathrm{Y2}$
is accordingly highest in R1, and at $q=0.5$ it is approximately 30\,\% and
20\,\% lower in Displacement than in Reversal for the sinusoidal and
shape-optimised waveforms, respectively; the contrast arises from R1, as the
rate in R2 differs from that in Displacement by less than 10\,\%. Near the
wall, lift-up thus follows the variation of
$\langle\omega_y''\omega_y''\rangle$ over the cycle in
Figure~\ref{fig:vorticity_grayscale_correlation}(c,d).}

\prfrev{Above $y^+=12$, the largest
values of $-\mathrm{Y2}$ are reached near $y^+=20$, close to the maximum of its
time average (Figure~\ref{fig:budget_wy_tavg}). At this wall-normal location,
the lobes precede the local start of Reversal, i.e.\ they occur in
Displacement; the higher conditional level of $-\mathrm{Y2}$ in
Reversal is thus confined to $5\le y^+\le12$. The phase delay of $-\mathrm{Y2}$
increases with distance from the wall, and the phase delay of $G$ increases 1.5
to 2 times as rapidly; the production lobes are therefore less inclined than the
Reversal boundaries. The inclination of the boundaries arises from the viscous
propagation of $G$ away from the wall (\S\ref{sec:temporal_modulation}),
whereas $-\mathrm{Y2}$ is carried by the streaks and by the wall-normal motions
that extend over the buffer layer, which respond more rapidly. As the amplitude
of $G$ at $y^+=20$ is less than 1/10 of its value at $y^+=5$, it is arguable
that lift-up above $y^+=12$ is organised by the stronger strain closer to the
wall, and that the position of the lobes in Displacement, relative to the
boundaries, reflects the different propagation rates, not a different response
to the actuation.}

\prfrev{Figure~\ref{fig:budget_wy_phase}(c,d) conveys the Stokes-strain source
Y1. The correlation $\langle\omega_y''\partial_xv''\rangle$ follows the sign of
$G$ with a phase lag $\Delta t^*\approx0.14$
(Appendix~\ref{app:verification}). Y1 therefore grows during Displacement, in
which $|G|$ remains close to its maximum, peaks at the start of Reversal,
decreases during R1 and is negative in R2, where $G$ has changed sign before the
correlation. Over $5\le y^+\le12$, the conditional rate of Y1 is about twice as
large in Displacement as in Reversal, and Displacement contributes about
50\,\% and 80\,\% of $\overline{\mathrm{Y1}}$ for the sinusoidal and
shape-optimised waveforms, respectively. Lift-up nevertheless exceeds Y1 in
every interval and at every threshold, the ratio of Y1 to $-\mathrm{Y2}$ being
largest in Displacement, about 60\,\% and 70\,\% at $q=0.5$ for the
sinusoidal and shape-optimised waveforms, respectively.}

\prfrev{A high conditional rate of production does not, by itself, imply that the
variance increases, as its tendency results from the balance of all the terms of
the budget (Table~\ref{tab:cyy_balance}). Over $5\le y^+\le12$,
$\langle\omega_y''\omega_y''\rangle$ follows the sum of the two mean-flow
sources, $-\mathrm{Y2}+\mathrm{Y1}$, and the dissipation $\mathcal E_y$ follows
$\langle\omega_y''\omega_y''\rangle$, with delays $\Delta t^*\le0.1$ in both
cases (correlation coefficients above 0.95 at the shortest delay). The band mean
of the conditional mean of $\partial_t\langle\omega_y''\omega_y''\rangle$ is
positive in Displacement and R1 and negative only in R2. In R2, $\mathcal E_y$
exceeds the sum of the mean-flow sources and of the nonlinear stretching
$\mathcal S_y=\langle\omega_y''\omega_j''\partial_jv''\rangle$
(Appendix~\ref{app:enstrophy_equations}), diffusion reducing the excess only
partially; the decrease is consistent with the delayed response, the dissipation
still reflecting the variance reached in R1 after lift-up has weakened. In
Displacement, $-\mathrm{Y2}+\mathrm{Y1}$ remains slightly below
$\mathcal E_y$, and the positive tendency arises mainly from viscous diffusion,
$\mathcal D_y$. The growth of the variance in this interval therefore requires
contributions beyond the mean-flow sources and does not, by itself,
establish a regeneration of the streaks within the interval.}

\prfrev{The two waveforms are compared next. The question is whether the
lower time-averaged lift-up of the shape-optimised waveform arises from the
durations of the intervals or from the conditional rates. Over
$5\le y^+\le12$, the time average $\overline{-\mathrm{Y2}}$ is about 35\,\% lower than
for the sinusoidal waveform (Table~\ref{tab:source_verification}), and
$-\mathrm{Y2}$ is lower at every sampled phase and wall-normal location up to
$y^+\approx33$, i.e.\ well beyond the maximum of its time average
(Appendix~\ref{app:verification}). The conditional rates are lower by
approximately 24 to 33\,\% in each of the three intervals
(Table~\ref{tab:interval_rates}). In the duration--rate accounting at $q=0.5$
(Table~\ref{tab:accounting}), these lower rates reduce the time average by
about 27\,\% of the sinusoidal value, whereas the different durations reduce it
by about 8\,\%. The lower lift-up is therefore attributable mainly to a lower
production within each interval rather than to the shorter Reversal.}

\prfrev{The question is then what modulates lift-up. At each wall-normal
location within $5\le y^+\le12$, $S$ varies by about 2\,\% over the cycle,
whereas the maximum of $-\mathrm{Y2}$ over the cycle is, on average, about 3
times its minimum for both waveforms; the modulation is therefore carried by
the correlation $\langle\omega_y''\partial_zv''\rangle$ rather than by $S$. In
terms of Figure~\ref{fig:budget_frame}, the wall-normal budget thus completes
the upper part of the figure: the Stokes strain enters it directly only through
Y1. How $G$, absent from $-\mathrm{Y2}$, modifies the correlation is not
revealed by this budget. The three following subsections examine instead the
production of spanwise and streamwise enstrophy and the direction in which $S$
and $G$ tilt $\omega_y''$ over the cycle, i.e.\ the two lower boxes and the
central part of the figure.}

	\subsection{\prfrev{Spanwise enstrophy ($\omega_z^{\prime\prime}$): Stokes-strain production}}
	\label{sec:enstr_wz}

\leavevmode\prfrev{The Stokes-strain source of the spanwise enstrophy is
$\mathrm{Z1}=\langle\omega_z''\partial_zu''\rangle G$, which contains the
tilting of $\omega_y''$ towards $z$ by $G$. The questions
addressed are whether Z1 is the dominant mean-flow source of the spanwise
budget, and how it is organised over the cycle and altered by the waveform.}

\begin{figure}
		\centering
		\begin{minipage}{0.48\textwidth}
			\centering
			\includegraphics[width=\textwidth]{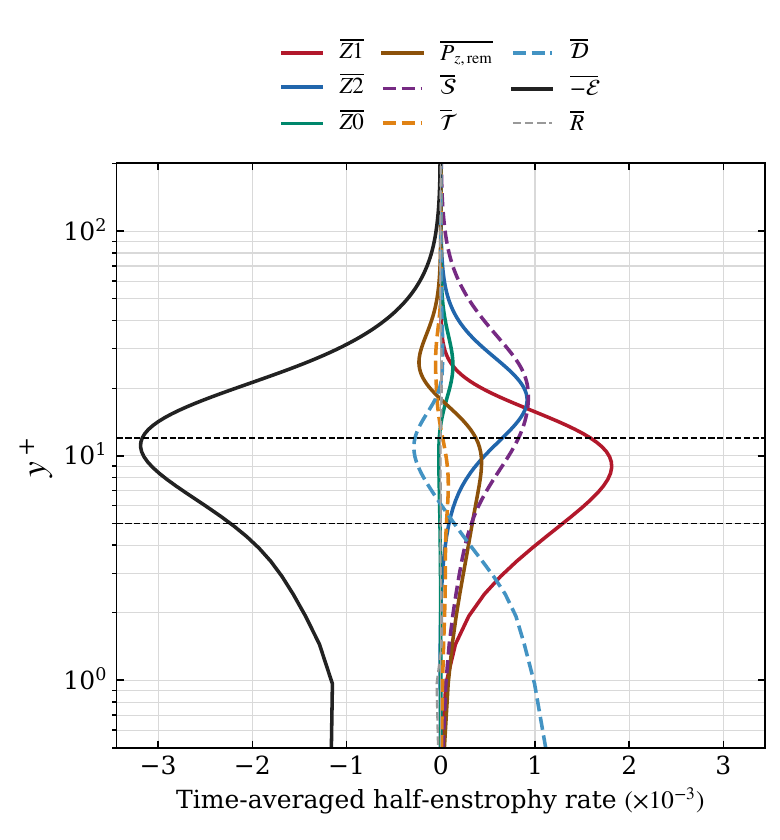}
			\centerline{(a)}
		\end{minipage}\hfill
		\begin{minipage}{0.48\textwidth}
			\centering
			\includegraphics[width=\textwidth]{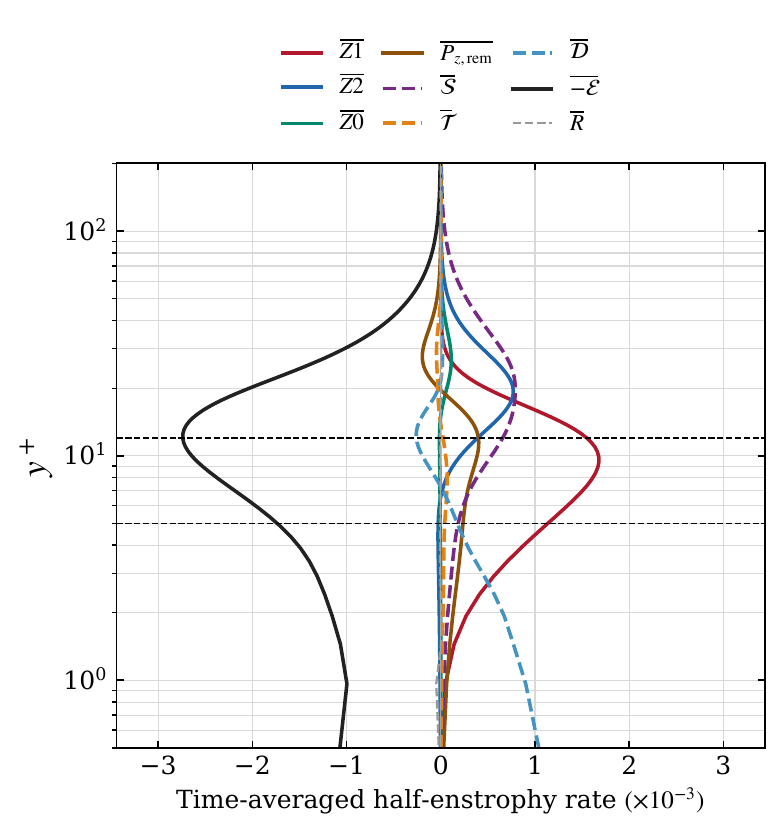}
			\centerline{(b)}
		\end{minipage}
		\caption{\prfrev{Time-averaged spanwise half-enstrophy budget: signed rates for sinusoidal (a) and shape-optimised (b) actuation. Overbars in the legend denote the time averages of the terms of Eq.~(\ref{eq:enstr_generic}), the component subscript $z$ being omitted on $\mathcal S$, $\mathcal T$, $\mathcal D$ and $\mathcal E$; $P_{z,\mathrm{rem}}$ denotes the wall-normal stretching part of Eq.~(\ref{eq:Pz}); $R$ is the residual. Each product was formed at every phase of the 32-phase ensemble before averaging over the cycle (Eq.~(\ref{eq:phase_covariance})).}}
\label{fig:budget_wz_tavg}
	\end{figure}

\begin{figure}
		\centering
		\begin{minipage}{0.48\textwidth}
			\centering
			\includegraphics[width=\textwidth]{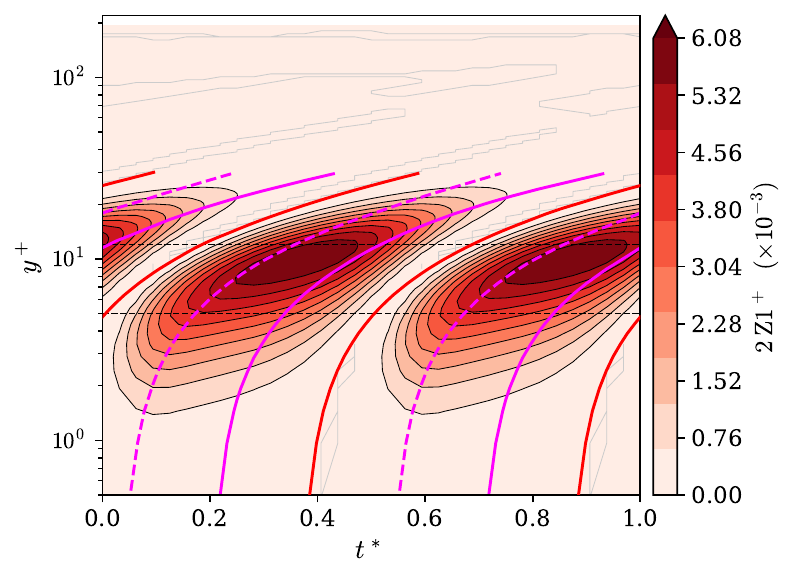}
			\centerline{(a)}
		\end{minipage}\hfill
		\begin{minipage}{0.48\textwidth}
			\centering
			\includegraphics[width=\textwidth]{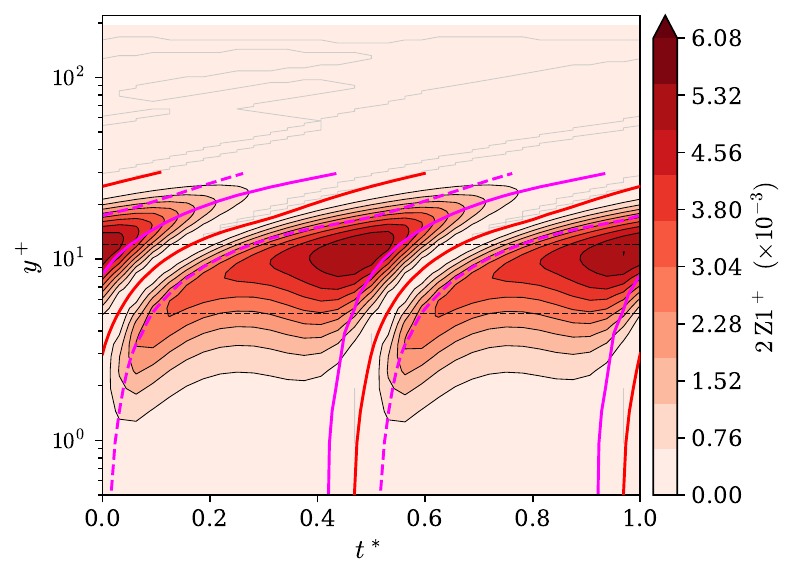}
			\centerline{(b)}
		\end{minipage}
		\begin{minipage}{0.48\textwidth}
			\centering
			\includegraphics[width=\textwidth]{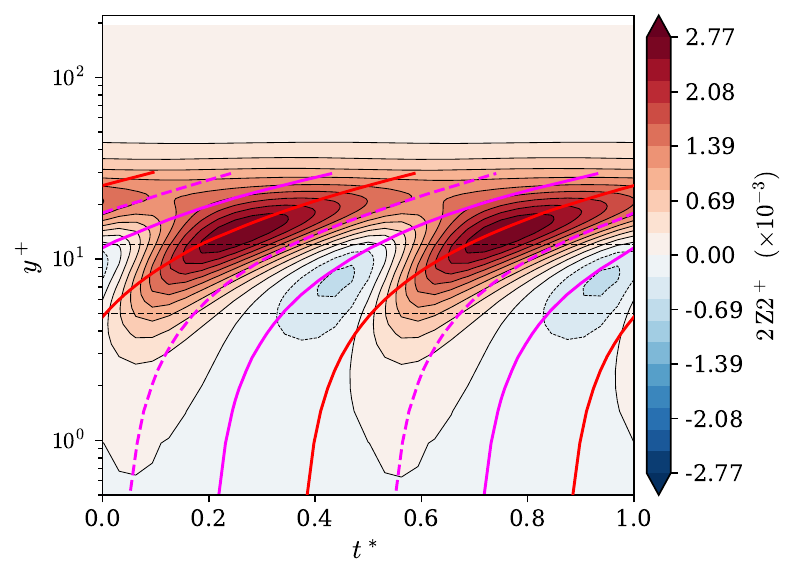}
			\centerline{(c)}
		\end{minipage}\hfill
		\begin{minipage}{0.48\textwidth}
			\centering
			\includegraphics[width=\textwidth]{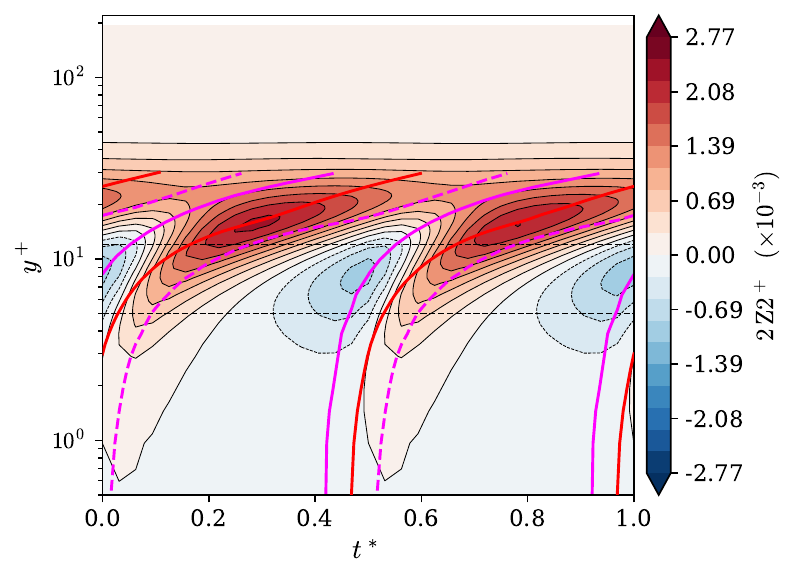}
			\centerline{(d)}
		\end{minipage}
		\begin{minipage}{0.48\textwidth}
			\centering
			\includegraphics[width=\textwidth]{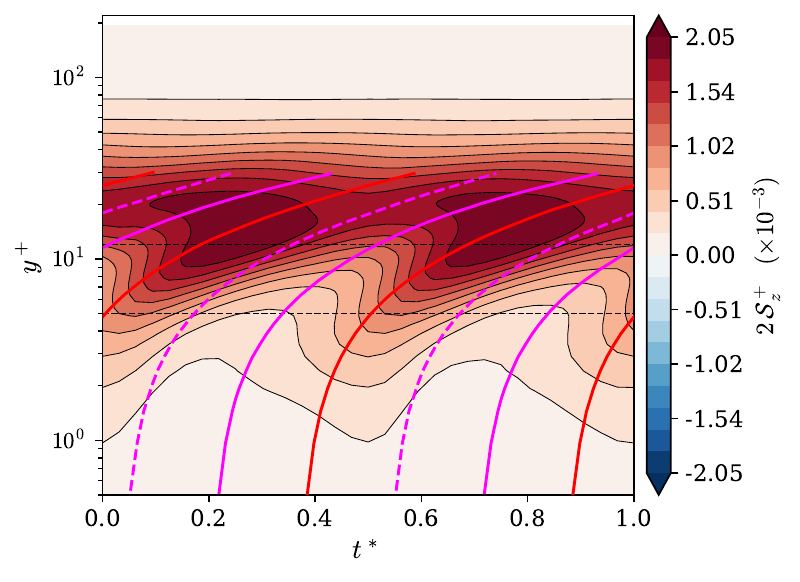}
			\centerline{(e)}
		\end{minipage}\hfill
		\begin{minipage}{0.48\textwidth}
			\centering
			\includegraphics[width=\textwidth]{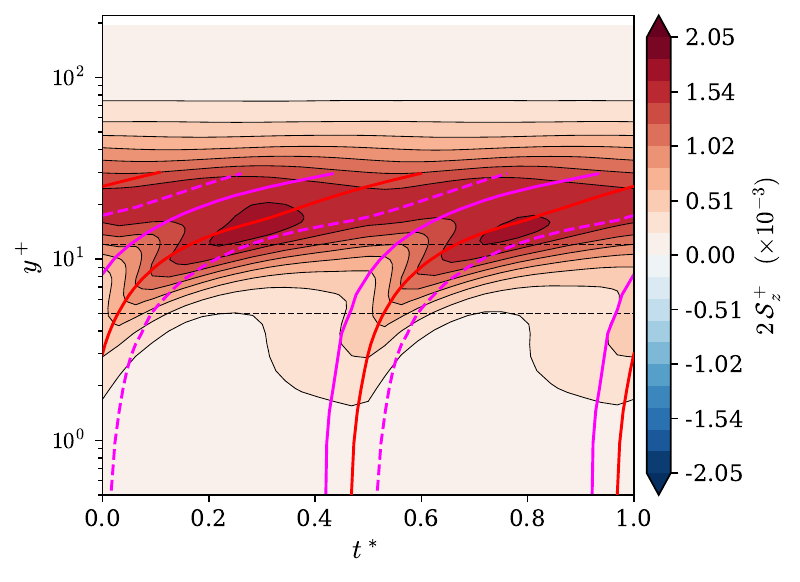}
			\centerline{(f)}
		\end{minipage}
		\caption{Spanwise enstrophy $\langle\omega_z^{\prime\prime 2}\rangle^+$ budget:
			phase-resolved mean-flow production terms (a--d) and nonlinear stretching (e,\,f) for sinusoidal (left column) and shape-optimised
			(right column) actuation.
			(a,\,b)~Stokes-induced tilting term \prfrev{$\mathrm{Z1}=\langle\omega_z''\partial_zu''\rangle G$, plotted as $2\mathrm{Z1}$.}
			(c,\,d)~Streamwise-gradient part of the stretching of the mean spanwise vorticity \prfrev{$\mathrm{Z2}=\langle\omega_z''\partial_xu''\rangle S$, plotted as $2\mathrm{Z2}$.}
			\prfrev{(e,\,f)~Turbulent stretching term $\mathcal S_z$
			(Appendix~\ref{app:enstrophy_equations}).}
			\prfrev{Isoline convention} and reference lines as in Figure~\ref{fig:budget_wy_phase}.
			\prfrev{Colour: full-variance rate, twice the half-variance term of Appendix~\ref{app:enstrophy_equations}.}}
		\label{fig:budget_wz_phase}
	\end{figure}

\prfrev{The mean-flow production of the spanwise enstrophy is}
\begin{equation}
\prfrev{\mathcal P_z=\underbrace{\langle\omega_z''\,\partial_zu''\rangle\,G}_{\mathrm{Z1}}
+\underbrace{\langle\omega_z''\,\partial_xu''\rangle\,S}_{\mathrm{Z2}}
+\underbrace{\langle\omega_z''\,\partial_yv''\rangle\,S}_{P_{z,\mathrm{rem}}}
+\underbrace{\langle v''\omega_z''\rangle\,\partial_y^2\langle u\rangle}_{\mathrm{Z0}},}
\label{eq:Pz_text}
\end{equation}
\prfrev{in which Z0 is the curvature term (Eq.~(\ref{eq:Pz}) of
Appendix~\ref{app:enstrophy_equations}). Z2 and $P_{z,\mathrm{rem}}$ arise from
the stretching of the mean spanwise vorticity, $\langle\omega_z\rangle=-S$, by
the fluctuating spanwise strain, ${-\langle\omega_z''\partial_zw''\rangle S}$,
which continuity, $\partial_zw''=-\partial_xu''-\partial_yv''$, divides between
the two. The time-averaged budget (Figure~\ref{fig:budget_wz_tavg}(a,b)) conveys
that Z1 is the dominant mean-flow source of the spanwise enstrophy, balanced
mainly by dissipation. Over $5\le y^+\le12$, $\overline{\mathrm{Z2}}$ and $\overline{P_{z,\mathrm{rem}}}$ are
about 5 and 4 times weaker than $\overline{\mathrm{Z1}}$ for the sinusoidal waveform, and about 14 and
5 times weaker for the shape-optimised one, their sum reaching about 43\,\% and
29\,\% of $\overline{\mathrm{Z1}}$. Z2 is thus part of a stretching and not a tilting term. It is
concentrated mainly in Reversal (Figure~\ref{fig:budget_wz_phase}(c,d)), and the
maximum of its time average, reached further from the wall near
$y^+\approx18$--20, is approximately half the maximum of $\overline{\mathrm{Z1}}$.}

\prfrev{As $\partial_zu''=\omega_y''+\partial_xw''$, Z1 splits exactly into
${\langle\omega_y''\omega_z''\rangle G+\langle\omega_z''\partial_xw''\rangle G}$. In the time average over
$5\le y^+\le12$, the first part is about 1.2 and 1.3 times $\overline{\mathrm{Z1}}$ for the sinusoidal and
shape-optimised waveforms, respectively, the second part being negative, and Z1
is thus carried mainly by the part in which $G$ tilts $\omega_y''$ towards $z$.}

\prfrev{Figure~\ref{fig:budget_wz_phase}(a,b) conveys the phase maps of Z1. Within
$5\le y^+\le12$, Z1 is enhanced during Displacement and reduced around Reversal;
its conditional mean is about 2.2 and 3.0 times larger in Displacement than in
Reversal at $q=0.5$, for the sinusoidal and shape-optimised waveforms,
respectively, and Displacement contributes about 50\,\% and 85\,\% of the
time-averaged Z1. The conditional levels of Z1 and $-\mathrm{Y2}$
(Figure~\ref{fig:budget_wy_phase}(a,b)) are therefore ordered oppositely, the
latter being lower in Displacement. At fixed wall-normal location, however,
their phase variations are anti-correlated for the sinusoidal waveform only,
with a correlation coefficient averaging $-0.63$ over the band, and
uncorrelated for the shape-optimised waveform.
As both sources are formed from the same velocity fluctuations and respond to
the same wall motion, their opposite organisation neither establishes that one
process ceases when the other strengthens nor provides an independent measure of
SSP activity.}

\prfrev{For the shape-optimised waveform, Displacement, in which Z1 is enhanced,
is twice as long, whereas the peak of $\overline{\mathrm{Z1}}$ is lower
(Figure~\ref{fig:budget_wz_tavg}(a,b)) and the phase-resolved lobes are weaker
(Figure~\ref{fig:budget_wz_phase}(a,b)). During Displacement, $G$ has nearly the
same magnitude for the two waveforms (\S\ref{sec:duty_cycle}); the lower
conditional rate of Z1 therefore arises from a lower magnitude of the
correlation $\langle\omega_z''\partial_zu''\rangle$ and not from a weaker
strain. In the duration--rate accounting at $q=0.5$
(Table~\ref{tab:accounting}), the longer Displacement raises $\overline{\mathrm{Z1}}$
by about 27\,\% of the sinusoidal value, whereas the lower conditional rates
reduce it by about 36\,\%, a net reduction of about 8\,\%. The larger drag
reduction of the shape-optimised waveform is therefore not accompanied by a
larger time-averaged production of spanwise enstrophy.}

\prfrev{The nonlinear stretching
$\mathcal S_z=\langle\omega_z''\omega_j''\partial_jw''\rangle$
(Figure~\ref{fig:budget_wz_phase}(e,f)) involves the stochastic fluctuations
alone and is therefore a turbulent rather than a mean-flow source. Its time
average $\overline{\mathcal S_z}$ over $5\le y^+\le12$ is approximately $1/3$ and $1/4$ of $\overline{\mathrm{Z1}}$ for
the sinusoidal and shape-optimised waveforms, respectively. Its conditional mean
is 0.8 and 0.9 times its time average in Displacement, rising to about 1.3 and
1.5 times in R2, in which the maximum is reached shortly after the red line at
nearly every wall-normal location of the band. At fixed wall-normal location,
$\mathcal S_z$ follows $\langle\omega_y''\omega_y''\rangle$
(\S\ref{sec:enstr_wy}) with a delay close to $1/10$ of the cycle, the
correlation coefficient being 0.99 and 0.95 at this delay, against 0.24 and 0.18
at zero delay, and the maximum of $\langle\omega_y''\omega_y''\rangle$, i.e.\
the largest streak amplitude, is reached near the red line
(Appendix~\ref{app:verification}). It is arguable, therefore, that the
fluctuating velocity gradients stretch $\omega_z''$ more strongly as the streaks
intensify.}

\prfrev{In terms of Figure~\ref{fig:budget_frame}, the spanwise budget thus
completes the right-hand box of the figure, Z1 being the dominant mean-flow
source and enhanced during Displacement. The left-hand box, i.e.\ the production
of streamwise enstrophy, is examined in \S\ref{sec:enstr_wx}, and the central
part, i.e.\ the balance of the two tilting directions, in
\S\ref{sec:enstr_tilting}.}

	\subsection{Streamwise enstrophy ($\omega_x^{\prime\prime}$): \prfrev{mean-shear production and Reversal duration}}
	\label{sec:enstr_wx}

\begin{figure}
		\centering
		\begin{minipage}{0.48\textwidth}
			\centering
			\includegraphics[width=\textwidth]{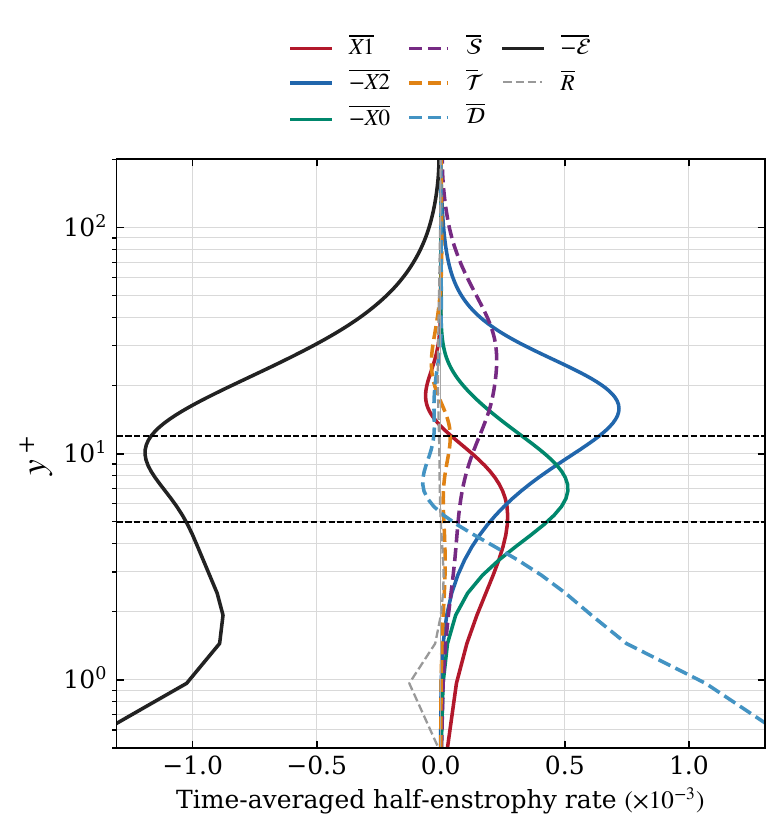}
			\centerline{(a)}
		\end{minipage}\hfill
		\begin{minipage}{0.48\textwidth}
			\centering
			\includegraphics[width=\textwidth]{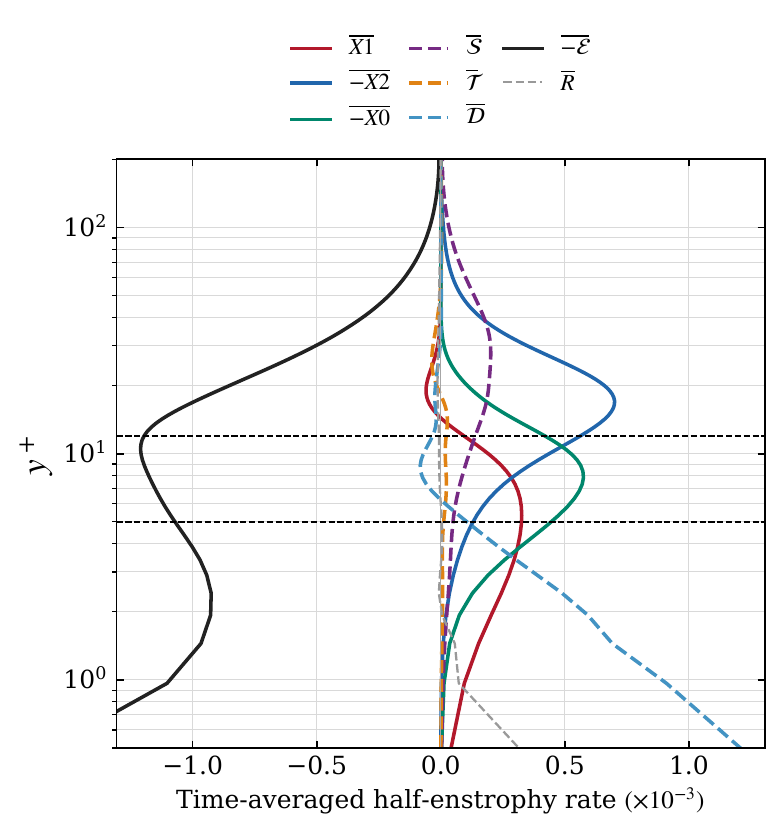}
			\centerline{(b)}
		\end{minipage}
		\caption{\prfrev{Time-averaged streamwise half-enstrophy budget: signed rates for sinusoidal (a) and shape-optimised (b) actuation. Overbars in the legend denote the time averages of the terms of Eq.~(\ref{eq:enstr_generic}), the component subscript $x$ being omitted on $\mathcal S$, $\mathcal T$, $\mathcal D$ and $\mathcal E$; $R$ is the residual. Each product was formed at every phase of the 32-phase ensemble before averaging over the cycle (Eq.~(\ref{eq:phase_covariance})).}}
\label{fig:budget_wx_tavg}
	\end{figure}

\leavevmode\prfrev{The mean-flow production of the streamwise enstrophy comprises
three sources,}
\begin{equation}
\prfrev{\mathcal P_x=\underbrace{\langle\omega_x''\,\partial_xu''\rangle\,G}_{\mathrm{X1}}
-\underbrace{\langle\omega_x''\,\partial_xw''\rangle\,S}_{\mathrm{X2}}
-\underbrace{\langle v''\omega_x''\rangle\,\partial_yG}_{\mathrm{X0}},}
\label{eq:Px_text}
\end{equation}
\prfrev{in which X1 is the Stokes-strain source and $-\mathrm{X0}$ the
Stokes-curvature source, $\partial_yG=\partial_y^2\langle w\rangle$ being the
curvature of the Stokes profile. Both vanish in the absence of spanwise actuation.
The mean-shear source $-\mathrm{X2}$ contains the tilting of $\omega_y''$ towards
$x$ by $S$, i.e.\ the second tilting direction of Figure~\ref{fig:budget_frame}.
In the time-averaged budget (Figure~\ref{fig:budget_wx_tavg}(a,b)), dissipation is
balanced by $\overline{-\mathrm{X2}}$, which peaks near $y^+\approx16$--17, and by
$\overline{-\mathrm{X0}}$, which peaks near $y^+\approx7$--8 and is of the same order as
$-\mathrm{X2}$ within $5\le y^+\le12$; X1 contributes to a lesser extent. The
questions addressed are which part of $-\mathrm{X2}$ the tilting of $\omega_y''$
towards $x$ carries, in which interval each source contributes, and how the
shorter Reversal of the shape-optimised waveform affects the budget.}

\prfrev{Substitution of $\partial_xw''=\partial_zu''-\omega_y''$ into $-\mathrm{X2}$
gives the exact split}
\begin{equation*}
\prfrev{-\mathrm{X2}=\langle\omega_x''\omega_y''\rangle S-\langle\omega_x''\partial_zu''\rangle S.}
\end{equation*}
\prfrev{The first part arises from the term $S\omega_y''$, through which $S$ tilts
$\omega_y''$ towards $x$. The remainder arises from the term $-S\partial_zu''$,
through which $\partial_zu''$ tilts the mean spanwise vorticity,
$\langle\omega_z\rangle=-S$, towards $x$. In the time average over
$5\le y^+\le12$, the first part is about 1.2 and 0.8 times $\overline{-\mathrm{X2}}$ for the
sinusoidal and shape-optimised waveforms, respectively, and approximates it to
within 20\,\%. The tilting of $\omega_y''$ towards $x$ is therefore the principal
part of $\overline{-\mathrm{X2}}$. The approximation fails in Displacement,
where the first part is only 2\,\% and 6\,\% of $-\mathrm{X2}$ and the remainder
carries almost all of it. It also fails in the comparison of the waveforms, as the
remainder changes sign between them: the first part decreases by 47\,\% for the
shape-optimised waveform, whereas $-\mathrm{X2}$ decreases by 23\,\%. The complete term $-\mathrm{X2}$ is therefore retained throughout. As
$\partial_xw''=0$ for streamwise-uniform modes, $-\mathrm{X2}$ is carried by
streamwise-varying fluctuations alone, in line with the association of the source
with the sinuous displacement of the streaks (\S\ref{sec:intro};
\citealp{brooke1993origin,Schoppa2002}).}

\begin{figure}
		\centering
		\begin{minipage}{0.48\textwidth}
			\centering
			\includegraphics[width=\textwidth]{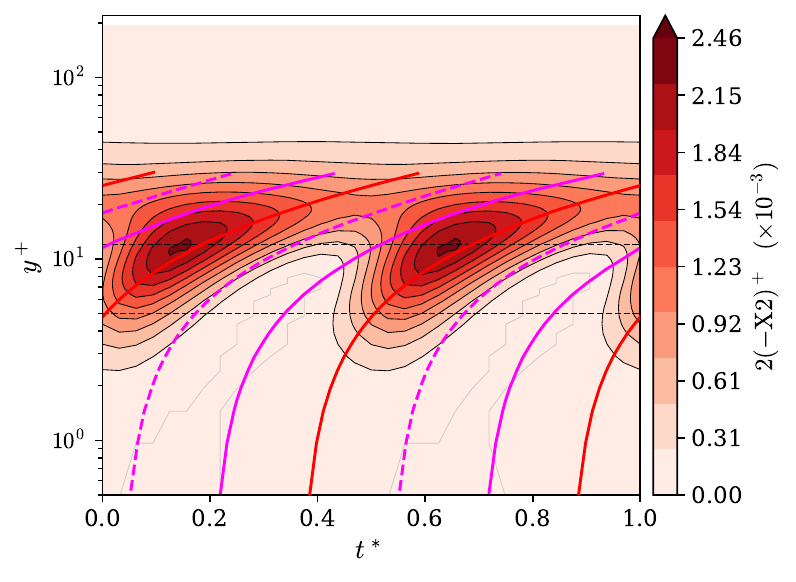}
			\centerline{(a)}
		\end{minipage}\hfill
		\begin{minipage}{0.48\textwidth}
			\centering
			\includegraphics[width=\textwidth]{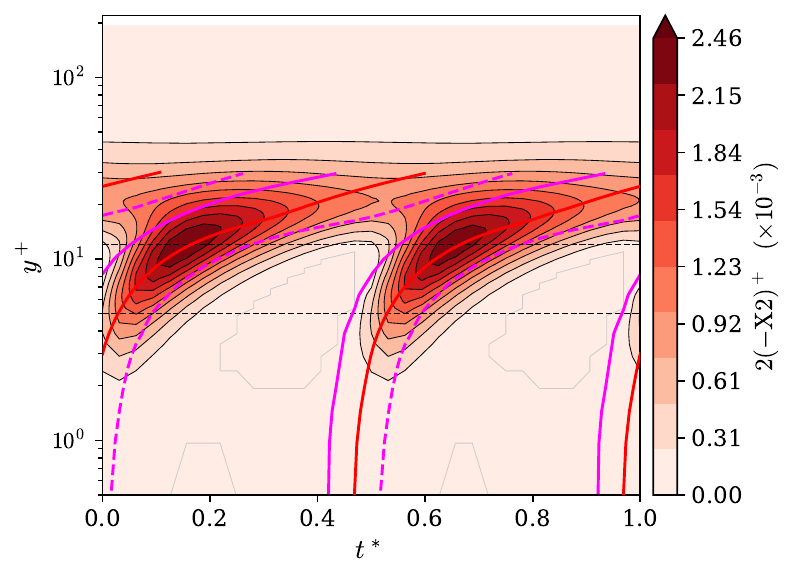}
			\centerline{(b)}
		\end{minipage}
		\begin{minipage}{0.48\textwidth}
			\centering
			\includegraphics[width=\textwidth]{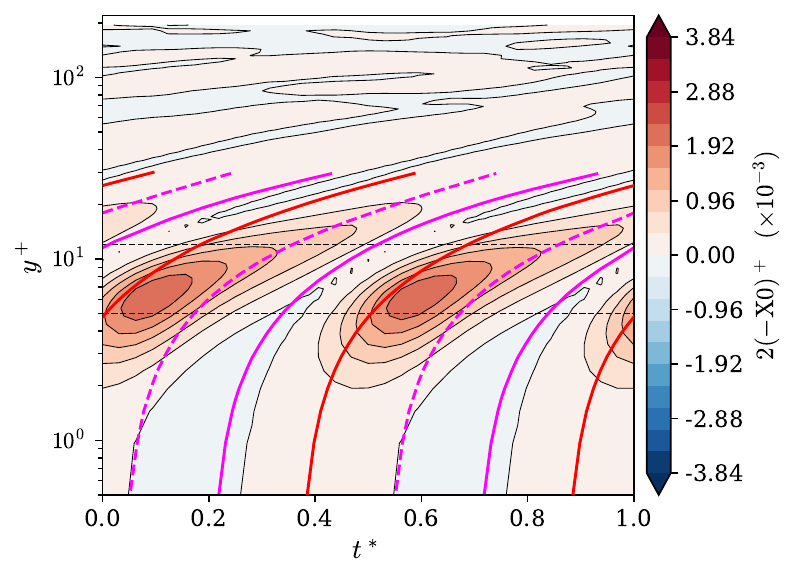}
			\centerline{(c)}
		\end{minipage}\hfill
		\begin{minipage}{0.48\textwidth}
			\centering
			\includegraphics[width=\textwidth]{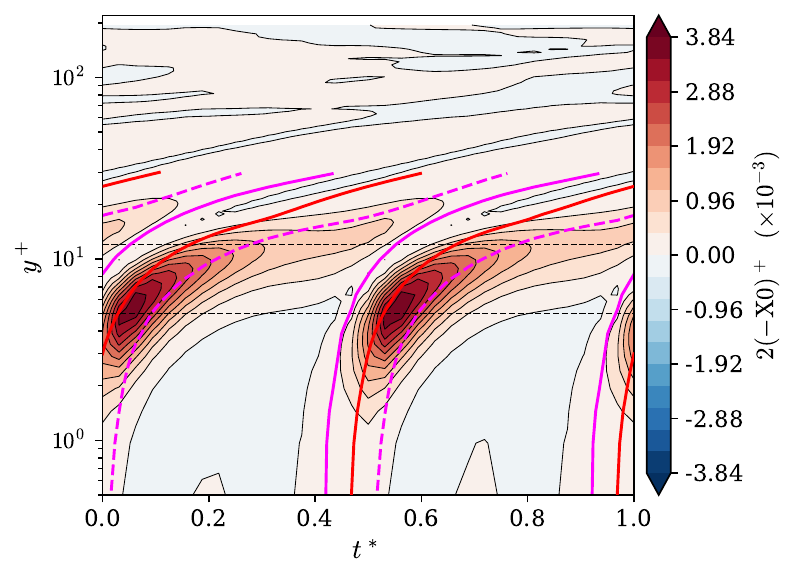}
			\centerline{(d)}
		\end{minipage}
		\begin{minipage}{0.48\textwidth}
			\centering
			\includegraphics[width=\textwidth]{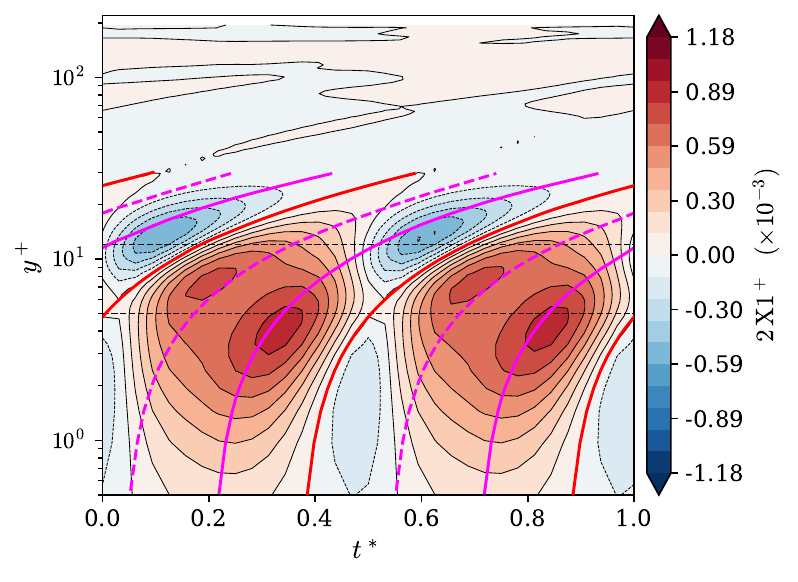}
			\centerline{(e)}
		\end{minipage}\hfill
		\begin{minipage}{0.48\textwidth}
			\centering
			\includegraphics[width=\textwidth]{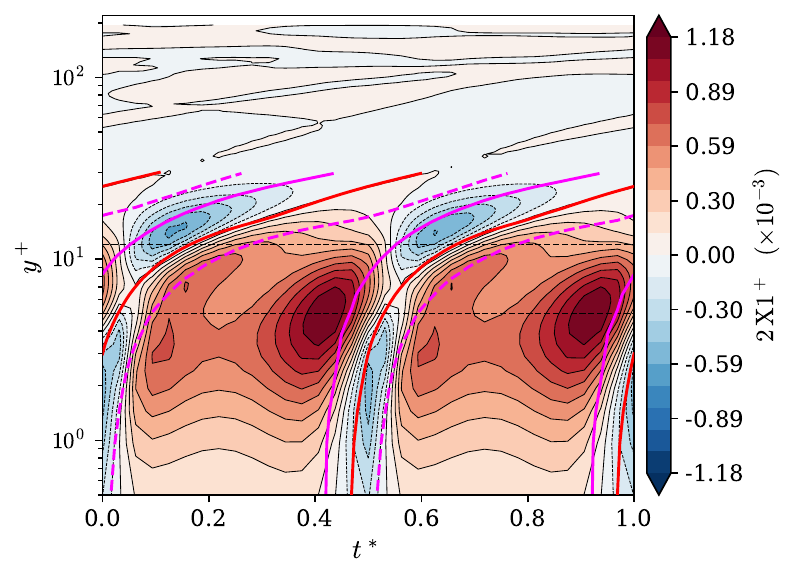}
			\centerline{(f)}
		\end{minipage}
		\caption{Streamwise enstrophy $\langle\omega_x^{\prime\prime 2}\rangle^+$ budget:
			phase-resolved production terms for sinusoidal (left column) and shape-optimised
			(right column) actuation.
			(a,\,b)~Production term \prfrev{$-\mathrm{X2}=-\langle\omega_x''\partial_xw''\rangle S$
			(regrouped mean-shear production), plotted as $2(-\mathrm{X2})$.}
			(c,\,d)~Production term \prfrev{$-\mathrm{X0}=-\langle v''\omega_x''\rangle\partial_yG$
			(Stokes-profile curvature contribution), plotted as $2(-\mathrm{X0})$.}
			\prfrev{(e,\,f)~Stokes-strain source $\mathrm{X1}=\langle\omega_x''\partial_xu''\rangle G$, plotted as $2\mathrm{X1}$.}
			\prfrev{Isoline convention} and reference lines as in Figure~\ref{fig:budget_wy_phase}.
			\prfrev{Colour: full-variance rate, twice the half-variance term of Appendix~\ref{app:enstrophy_equations}.}}
		\label{fig:budget_wx_phase}
	\end{figure}

\prfrev{The phase organisation of the sources is considered at $q=0.5$. Within
$5\le y^+\le12$, $-\mathrm{X2}$ is concentrated in Reversal
(Figure~\ref{fig:budget_wx_phase}(a,b)), which contributes about 93\,\% and
80\,\% of its time average for the sinusoidal and shape-optimised waveforms,
respectively, whilst occupying 68\,\% and 34\,\% of the cycle
(Table~\ref{tab:interval_rates}). X1 (Figure~\ref{fig:budget_wx_phase}(e,f))
exhibits the opposite organisation, Displacement, which occupies 33\,\% and
66\,\% of the cycle, contributing approximately 50\,\% and 90\,\% of its time
average, as for Y1 and Z1 (\S\S\ref{sec:enstr_wy} and~\ref{sec:enstr_wz}). The
shape-optimised waveform halves the duration of Reversal, whereas the peak of
$-\mathrm{X2}$ over the cycle is 4\,\% higher and its conditional rate is higher
in every interval (Table~\ref{tab:interval_rates}). The time average
$\overline{-\mathrm{X2}}$ is nevertheless 23\,\% lower, as $-\mathrm{X2}$ is 6 to 7 times
weaker in Displacement than in Reversal. In the duration--rate accounting
(Table~\ref{tab:accounting}), the shorter Reversal reduces
$\overline{-\mathrm{X2}}$ by about 44\,\% of the sinusoidal value, and the higher conditional
rates raise it by about 21\,\%. The lower time average $\overline{-\mathrm{X2}}$ is therefore
accounted for by the shorter Reversal, not by weaker conditional rates. Over
$5\le y^+\le12$, the higher time averages $\overline{\mathrm{X1}}$ and $\overline{-\mathrm{X0}}$, by about
33\,\% and 16\,\%, more than balance this decrease: the complete source
$\mathcal P_x=\mathrm{X1}-\mathrm{X2}-\mathrm{X0}$ (Eq.~(\ref{eq:Px_text})) is
approximately 4\,\% higher, whereas the time average
$\overline{\langle\omega_x''^2\rangle}$ is 2\,\% lower
(Figure~\ref{fig:vorticity_comparison}(b)).}

\prfrev{In terms of Figure~\ref{fig:budget_frame}, the streamwise budget thus
completes the left-hand box: $-\mathrm{X2}$, principally the tilting of
$\omega_y''$ towards $x$ by $S$, is concentrated in Reversal, an interval shorter
for the shape-optimised waveform, whose complete mean-flow source is nevertheless
not lower, as X1 and $-\mathrm{X0}$ compensate the decrease of $-\mathrm{X2}$.}

	\subsection{\prfrev{Effective tilting rates of the wall-normal vorticity}}
	\label{sec:enstr_tilting}

\leavevmode\prfrev{The mean-flow sources of the equations of $\omega_z''$ and
$\omega_x''$ are $G\partial_zu''$ and $-S\partial_xw''$, respectively.
Premultiplied by $\omega_z''$ and $\omega_x''$ and averaged, they give Z1 and
$-\mathrm{X2}$ (Eqs.~(\ref{eq:Pz_text}) and~(\ref{eq:Px_text})), the production
terms of the spanwise and streamwise enstrophy, which are not rates of tilting of
$\omega_y''$. The premultiplication selects the part of each source correlated
with $\omega_z''$ or $\omega_x''$, which carries vorticity of other origins. After
premultiplication, Z1 contains $\langle\omega_y''\omega_z''\rangle G$, from the
tilting of $\omega_y''$, and $\langle\omega_z''\partial_xw''\rangle G$, from the
tilting of the mean streamwise vorticity (\S\ref{sec:enstr_wz}). Likewise,
$-\mathrm{X2}$ contains $\langle\omega_x''\omega_y''\rangle S$ and
$-\langle\omega_x''\partial_zu''\rangle S$ (\S\ref{sec:enstr_wx}). The rates of the
two tilting processes are therefore derived next. The questions addressed are how
the tilting towards $z$ compares with that towards $x$ over the cycle, and how
their balance relates to $|G|$ and to the interpretation of
\citet{agostini_turbulence_2015}, in which a strong tilting towards $z$
accompanies the reduction of the mean-shear production $-\mathrm{Y2}$ of
$\omega_y''$.}

\begin{figure}
		\centering
		\begin{minipage}{0.48\textwidth}
			\centering
			\includegraphics[width=\textwidth]{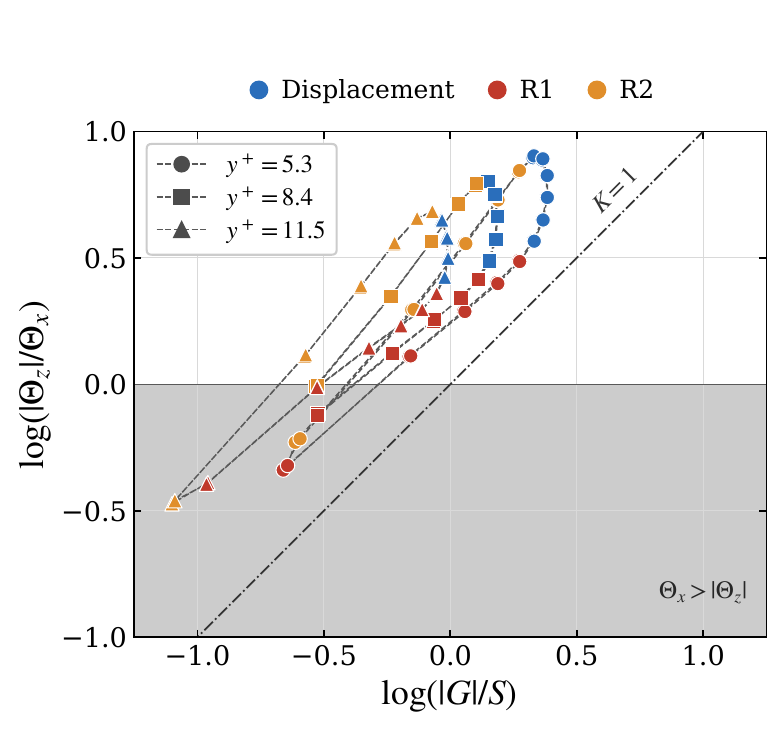}
			\centerline{(a)}
		\end{minipage}\hfill
		\begin{minipage}{0.48\textwidth}
			\centering
			\includegraphics[width=\textwidth]{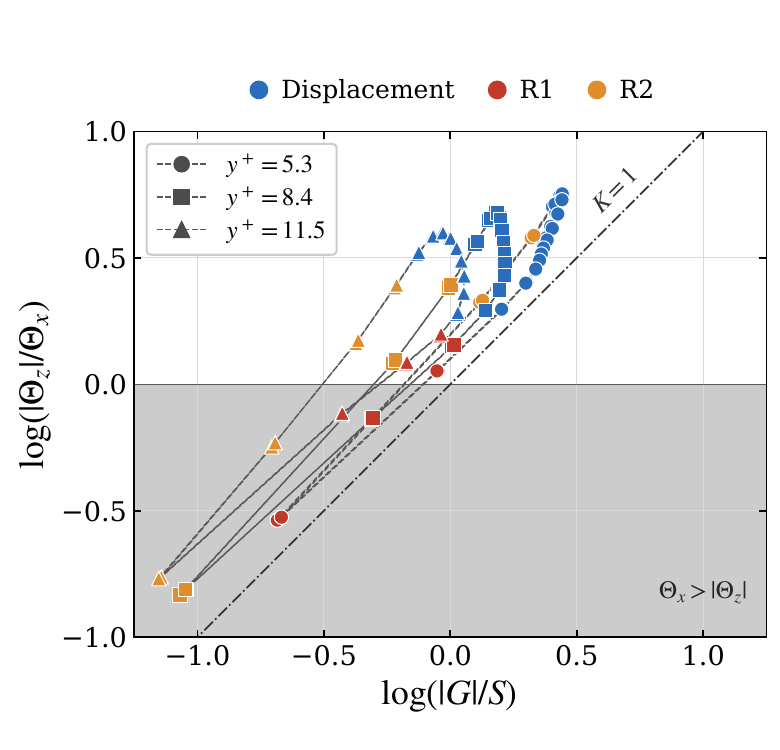}
			\centerline{(b)}
		\end{minipage}
		\caption{\prfrev{Ratio of the effective tilting rates, $|\Theta_z|/\Theta_x$, against $|G|/S$ (Eq.~(\ref{eq:tilting_ratio})) over the cycle, for sinusoidal (a) and shape-optimised (b) actuation. Symbols: $y^+=5.3$, 8.4 and 11.5; colour: Displacement, R1 and R2 at $q=0.5$; dashed lines join consecutive phases; dash-dotted line: $K=1$; grey region: $\Theta_x>|\Theta_z|$. log denotes the decimal logarithm. Points at which $|G|$ is below 5\,\% of its maximum over the cycle at that wall-normal location are omitted.}}
		\label{fig:tilting_ratio}
	\end{figure}

\prfrev{Each source separates into a term that tilts $\omega_y''$ and a term that
tilts a mean vorticity component. As
$\partial_zu''=\omega_y''+\partial_xw''$, the source $G\partial_zu''$ of
$\omega_z''$ is $G\omega_y''$, the tilting of $\omega_y''$ by $G$, plus
$G\partial_xw''$, the tilting of the mean streamwise vorticity,
$\langle\omega_x\rangle=G$, by $\partial_xw''$. The source $-S\partial_xw''$ of
$\omega_x''$ is $S\omega_y''$, the tilting of $\omega_y''$ by $S$, plus
$-S\partial_zu''$, the tilting of the mean spanwise vorticity,
$\langle\omega_z\rangle=-S$, by $\partial_zu''$. Premultiplication of the two
sources by $\omega_y''$, instead of $\omega_z''$ and $\omega_x''$, followed by
averaging, provides their contributions to the budgets of
$\langle\omega_x''\omega_y''\rangle$ and $\langle\omega_y''\omega_z''\rangle$
(Appendix~\ref{app:verification}). No other mean-flow term of these budgets
contains the tilting of $\omega_y''$. These contributions are designated the
effective tilting rates of $\omega_y''$ towards $x$ by $S$ and towards $z$ by $G$,}
\begin{equation}
\prfrev{\Theta_x\equiv-S\langle\omega_y''\partial_xw''\rangle,\qquad
\Theta_z\equiv G\langle\omega_y''\partial_zu''\rangle.}
\label{eq:effective_tilting}
\end{equation}
\prfrev{Over the phase distribution, $\Theta_x$ is positive, and $\Theta_z$ has the
sign of $G$, as $\langle\omega_y''\partial_zu''\rangle$ is positive. The ratio of
the rates therefore uses $|\Theta_z|$, and
Eq.~(\ref{eq:effective_tilting}) gives}
\begin{equation}
\prfrev{\frac{|\Theta_z|}{\Theta_x}=\frac{|G|}{S}\,K,\qquad
K=\frac{\langle\omega_y''\partial_zu''\rangle}{-\langle\omega_y''\partial_xw''\rangle},}
\label{eq:tilting_ratio}
\end{equation}
\prfrev{in which $K$ is the ratio of the two velocity-gradient correlations.
The logarithm of Eq.~(\ref{eq:tilting_ratio}) is}
\begin{equation*}
\prfrev{\log\frac{|\Theta_z|}{\Theta_x}=\log\frac{|G|}{S}+\log K.}
\end{equation*}
\prfrev{The balance of the two tilting directions therefore follows $|G|/S$ if the
variation of $\log K$ over the cycle is small compared with that of
$\log(|G|/S)$. Each point of
Figure~\ref{fig:tilting_ratio} corresponds to one phase of the cycle at one
wall-normal location, $y^+=5.3$, 8.4 or 11.5, for the sinusoidal (a) and
shape-optimised (b) waveforms. Points of constant $K$ are aligned on a line
parallel to the line $K=1$. In R1, the points follow closely such a line, i.e.\ $K$
is nearly constant, whereas the departures from constant $K$ occur mainly in
Displacement, where the points move almost vertically. These departures are small
compared with the variation of $|G|/S$ over the cycle, which therefore appears to set most of
the variation of the balance.}

\prfrev{The grey region of Figure~\ref{fig:tilting_ratio}, below the line
$\log(|\Theta_z|/\Theta_x)=0$, in which $\Theta_x$ exceeds $|\Theta_z|$, gives the
phases at which the tilting towards $x$ prevails. It contains only points of
Reversal, around the change of sign of $G$. Above the grey region, the tilting
towards $z$ prevails; this region contains all the points of Displacement and 85 to
87\,\% of the phases of the cycle for both waveforms
(Appendix~\ref{app:verification}).}

\prfrev{The interval means over $5\le y^+\le12$ at $q=0.5$
(Table~\ref{tab:interval_rates}) average over the phases at which the tilting
towards $z$ prevails and those at which the tilting towards $x$ prevails. $|\Theta_z|$ is
about 4 and 3 times $\Theta_x$ in Displacement for the sinusoidal and
shape-optimised waveforms, respectively, and only 1.6 and 1.2 times $\Theta_x$ in
R1, as $\Theta_x$ increases by a factor of about 2.4 and 2.5 whereas $|\Theta_z|$
remains about 0.95 of its Displacement value; in R2, after the change of sign of
$G$, the ratio is 2.2 and 1.1. The tilting towards $z$ nevertheless
prevails in every interval mean; its margin depends on $q$
(Appendix~\ref{app:verification}). Hence $\omega_y''$ is tilted mainly towards $z$
in Displacement, where $|G|$ remains near its maximum, and increasingly towards $x$
in R1, as $G$ approaches its change of sign. This shift of the balance, with both
directions active in every interval, is referred to as the redirection of the
tilting of $\omega_y''$.}

\prfrev{The interpretation of \citet{agostini_turbulence_2015} associated the
reduction of the mean-shear production of $\omega_y''$ with strong Stokes-strain
tilting of $\omega_y''$ towards $z$, and the phases of drag increase with weak
spanwise tilting (\S\ref{sec:intro}). That association rested on the phase
relationships of the component-enstrophy terms; the tilting towards $x$ was not
examined. The tilting is more directed towards $z$ in Displacement than in R1, and
$-\mathrm{Y2}$ is lower in Displacement, whereas $\Theta_x$ is strongest in R1, where
$-\mathrm{Y2}$ has its highest conditional level (\S\ref{sec:enstr_wy}). At fixed
wall-normal location, the balance $|\Theta_z|/\Theta_x$ is anti-correlated with
$-\mathrm{Y2}$ (correlation coefficients $-0.80$ and $-0.62$), in line with the
earlier association. This does not establish a causal relation: the
anti-correlation is consistent with the strong association of $\Theta_x$ with
$-\mathrm{Y2}$ (0.98 and 0.89) and with $\langle\omega_y''\omega_y''\rangle$ (0.93
and 0.99), and the ordering is equally compatible with $\Theta_x$ responding to the
streak amplitude built by lift-up (Appendix~\ref{app:verification}). A connection
with the drag reduction would involve the SSP loop, which is not measured here
(\S\ref{sec:enstr_wx}).}

\prfrev{In terms of Figure~\ref{fig:budget_frame}, the effective tilting rates
thus complete the central box: the balance of the two tilting directions follows
$|G|/S$, with the tilting towards $x$ prevailing only around the change of sign of
$G$.}

\subsection{\prfrev{Synthesis: two tilting paths for the wall-normal vorticity associated with the streaks}}
\label{sec:enstr_synthesis}
\begin{figure}[p]
\centering
\includegraphics[width=\textwidth,height=0.90\textheight,keepaspectratio]{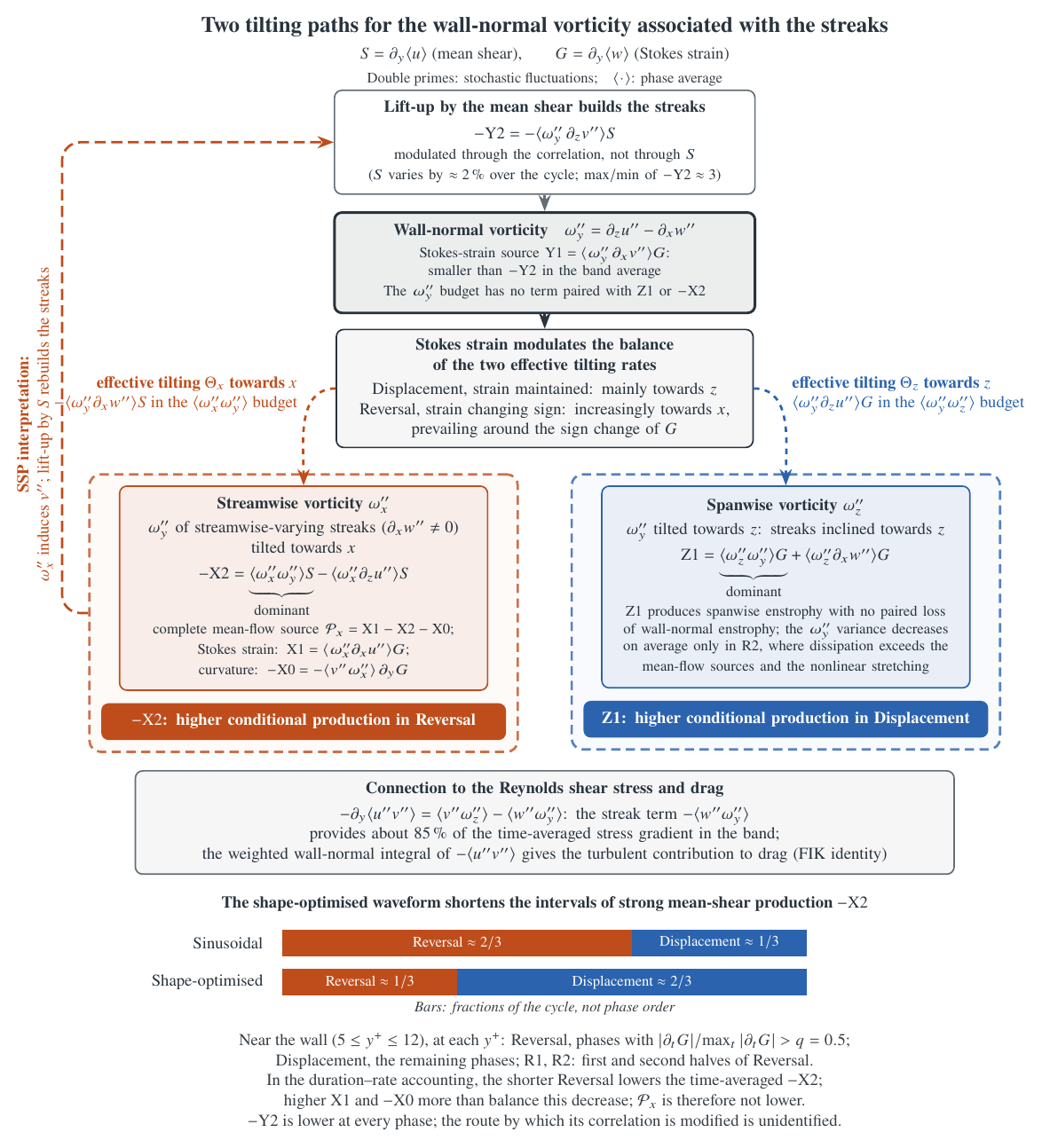}
\caption{\prfrev{Two tilting paths for the wall-normal vorticity
associated with the near-wall streaks, $5\le y^+\le12$: the frame of Figure~\ref{fig:budget_frame}
completed with the results of \S\S\ref{sec:enstr_wy}--\ref{sec:enstr_tilting}. $S=\partial_y\langle u\rangle$ and $G=\partial_y\langle w\rangle$ are the mean shear and the Stokes strain. Warm and cool colours refer to Reversal and Displacement (\S\ref{sec:duty_cycle}, $q=0.5$), respectively; the underbraces mark the
dominant parts in the time average over the band, not in every interval; the red dashed loop is an
interpretation, not measured here.}}
\label{fig:enstrophy_sketch}
\end{figure}

\prfrev{The budgets of the three component enstrophies are combined in
Figure~\ref{fig:enstrophy_sketch}, which completes the frame of
Figure~\ref{fig:budget_frame} with the results of
\S\S\ref{sec:enstr_wy}--\ref{sec:enstr_tilting} for the near-wall band
$5\le y^+\le12$. The questions addressed are through which paths $\omega_y''$
contributes to the streamwise and spanwise budgets, how the Stokes strain
shifts the balance of these paths over the cycle, and how the waveform alters
their contributions. The mean-flow sources of the three budgets are labelled X,
Y and Z, respectively (Appendix~\ref{app:enstrophy_equations}), and the
intervals are those of \S\ref{sec:duty_cycle} at $q=0.5$.}

\prfrev{The wall-normal vorticity, $\omega_y''=\partial_zu''-\partial_xw''$, is
associated with the streaks, and its variance can be regarded as a surrogate for
their amplitude. It is built by lift-up, $-\mathrm{Y2}=-\langle\omega_y''\partial_zv''\rangle S$,
the Stokes strain entering its budget directly only through Y1
(\S\ref{sec:enstr_wy}). Both mean-flow gradients also tilt $\omega_y''$, $S$
towards $x$ and $G$ towards $z$, at the effective rates
$\Theta_x=-\langle\omega_y''\partial_xw''\rangle S$ and
$\Theta_z=\langle\omega_y''\partial_zu''\rangle G$
(Eq.~(\ref{eq:effective_tilting})); the tilting towards $x$ requires
$\partial_xw''\ne0$ and thus concerns the wall-normal vorticity of
streamwise-varying streaks. The balance of the two directions follows the ratio
$|G|/S$ over the cycle, the correlation between the logarithms being 0.90 and
0.89 (Figure~\ref{fig:tilting_ratio}). The tilting towards $z$ prevails
throughout Displacement, the ratio $|\Theta_z|/\Theta_x$ being about 4 and 3 for
the sinusoidal and shape-optimised waveforms, respectively, and the tilting
towards $x$ gains weight in R1, where the ratio is 1.6 and 1.2 and $\Theta_x$ is
about 2.4 and 2.5 times its Displacement value
(Table~\ref{tab:interval_rates}). The tilting towards $x$ exceeds the tilting
towards $z$ only around the change of sign of $G$, during approximately 15\,\% of
the actuation cycle for both waveforms, all in Reversal
(Appendix~\ref{app:verification}). Both directions therefore contribute within each interval ($\Theta_z$ vanishes only where $G=0$), and as $|G|$ decreases from Displacement towards its change of sign in
R1, the tilting of $\omega_y''$ is redirected from $z$ towards $x$. The two rates are not linked by an enstrophy conservation law: neither occurs in the budget of $\langle\omega_y''\omega_y''\rangle$. They are linked kinematically, $\Theta_z/G+\Theta_x/S=\langle\omega_y''^2\rangle$ (Eq.~(\ref{eq:theta_identity})), so that the two normalised rates sum to the wall-normal enstrophy, both being positive over the phases and heights examined, whilst the variation of their balance over the cycle is carried by $|G|/S$ and by the partition of $\langle\omega_y''^2\rangle$ between the two correlations (Eq.~(\ref{eq:tilting_ratio})).}

\prfrev{Each path appears in a budget as the principal part of a mean-flow
source. In the time average over the band, the principal part of the mean-shear
production $-\mathrm{X2}=-\langle\omega_x''\partial_xw''\rangle S$ is
$\langle\omega_x''\omega_y''\rangle S$, in which $S$ tilts $\omega_y''$ towards
$x$, although this part carries almost none of $-\mathrm{X2}$ in Displacement
(\S\ref{sec:enstr_wx}), and the principal part of the Stokes-strain production
$\mathrm{Z1}=\langle\omega_z''\partial_zu''\rangle G$ is
$\langle\omega_y''\omega_z''\rangle G$, in which $G$ tilts $\omega_y''$ towards
$z$ (\S\ref{sec:enstr_wz}). Reversal, which occupies about $2/3$ and $1/3$ of
the cycle for the sinusoidal and shape-optimised waveforms, respectively,
contributes about 93 and 80\,\% of the time-averaged $-\mathrm{X2}$, whereas
Displacement, which occupies about $1/3$ and $2/3$, contributes about 50 and
85\,\% of the time-averaged Z1 (\S\S\ref{sec:enstr_wz} and~\ref{sec:enstr_wx}).
These contributions are consistent with $\Theta_x$ strengthening in Reversal and
$|\Theta_z|$ remaining near its Displacement level until $G$ changes sign. The
waveform acts on the two paths through different factors
(Table~\ref{tab:accounting}): on the mean-shear path through the duration of
Reversal, and on lift-up and the Stokes-strain production through the
conditional rates (\S\S\ref{sec:enstr_wy}--\ref{sec:enstr_wx}). The drag
reduction of the shape-optimised waveform, larger by about 2.3 percentage
points, is thus accompanied by a time-averaged lift-up $\overline{-\mathrm{Y2}}$ about 35\,\% lower and a time-averaged $\overline{\mathrm{Z1}}$
about 8\,\% lower, whilst the complete streamwise mean-flow production
$\overline{\mathcal{P}_x}$ is about 4\,\% higher. The near-wall
variances are not stationary over the analysed cycles, and these time averages
are therefore conditional on that window (Appendix~\ref{app:verification}).}

\prfrev{The two paths are ordered oppositely to lift-up over the cycle. The
tilting is more directed towards $z$ in Displacement, where $-\mathrm{Y2}$ is
lower, whereas $\Theta_x$ is strongest in R1, where $-\mathrm{Y2}$ has its
highest conditional level, and at fixed wall-normal location
$|\Theta_z|/\Theta_x$ is anti-correlated with $-\mathrm{Y2}$, the correlation
coefficients being $-0.80$ and $-0.62$. This is the ordering proposed by
\citet{agostini_turbulence_2015}, in which a reduction of lift-up accompanies a
strong Stokes-strain tilting towards $z$. The anti-correlation does not,
however, establish a causal relation, as the same ordering follows from
$\Theta_x$ responding to the streak amplitude built by lift-up
(\S\ref{sec:enstr_tilting}).}

\prfrev{Two elements of the figure are not part of the enstrophy budgets. The
lower box relates the vorticity to the wall-normal gradient of the Reynolds
shear stress, $-\langle u''v''\rangle$, through the kinematic identity
$-\partial_y\langle u''v''\rangle=\langle v''\omega_z''\rangle-\langle w''\omega_y''\rangle$
of Eq.~(\ref{eq:shear_stress_vorticity}), in which the streak term
$-\langle w''\omega_y''\rangle$ provides about 85\,\% of the time-averaged
stress gradient in the band for both waveforms
(Appendix~\ref{app:verification}); the weighted wall-normal integral of the time average of
$-\langle u''v''\rangle$ gives the turbulent contribution to the time-averaged drag
\citep{fukagata2002}. The red dashed loop is the self-sustaining process (SSP),
in which $\omega_x''$ induces $v''$ and lift-up by $S$ rebuilds the streaks; it
is not measured here. Within each waveform, Reversal, where the tilting towards
$x$ and $-\mathrm{X2}$ are strongest, accompanies conditional means of the
streamwise and wall-normal vorticity variances and of the Reynolds shear stress
above their time averages, whereas Displacement, where Z1 is enhanced,
accompanies conditional means below them. These means are relative to each case's own
time average; they are read as a reduced activity in Displacement and a recovery in
Reversal, and establish neither attenuation relative to the unactuated flow nor cessation
of the SSP. The variances and the lift-up production remain non-zero in both intervals, and the
lift-up production is about 30\,\% and 20\,\% lower in Displacement than in Reversal
(\S\ref{sec:enstr_wy}).
The interpretation of this temporal
organisation in terms of duration is examined in \S\ref{sec:diagnostic_tool}.}

\clearpage
\section{Discussion}
\label{sec:discussion}

\subsection{The shape-optimised waveform as a diagnostic tool}
\label{sec:diagnostic_tool}
\prfrev{The shape-optimised waveform is not proposed here as a route to a net
energy saving, as its actuation cost is higher (\S\ref{sec:intro}); it serves to
separate two factors that a single sinusoidal case cannot vary independently.
For the sinusoidal signal, the fractions of the cycle occupied by Reversal and
Displacement are set by the signal itself. The quasi-square waveform reverses
these fractions whilst the magnitude of the Stokes strain during Displacement
remains nearly unchanged (\S\ref{sec:duty_cycle}); the duration of an
interval can then be distinguished from the conditional rate of a source within it,
as in the duration--rate accounting of Table~\ref{tab:accounting}
(\S\ref{sec:enstr_synthesis}). This distinction is not an isolated test of
duration, as the sharpness of the strain reversals also changes
(\S\S\ref{sec:temporal_modulation} and \ref{sec:duty_cycle}), and a phase
relation under a common actuation is not a selective intervention in the
governing dynamics of the kind examined by \citet{lozano2021cause}. The
accounting therefore guides the interpretation of the budgets without
establishing an effect of duration on the drag. It shows that the waveform acts
on each path through a different factor: on the mean-shear production
$-\mathrm{X2}$ through the duration of Reversal, which the shape-optimised
waveform approximately halves at $q=0.5$, and on the lift-up and the Stokes-strain production through the
conditional rates within the intervals (Table~\ref{tab:accounting}). The
contrast may be described as follows: the sinusoidal signal varies the strain continuously and spends about
two thirds of the cycle in Reversal, whereas the shape-optimised signal holds a
nearly constant strain during a Displacement that occupies about two thirds of
the cycle and passes rapidly through the reversals.}

\subsection{\prfrev{From vorticity budgets to momentum transport}}
\label{sec:causal_chain}
\prfrev{The enstrophy budgets were examined in preference to the stress budgets
for two reasons: enstrophy is associated more directly with vortical structures,
and thus with mixing, and the pressure gradient has zero curl and provides no
source in the enstrophy budgets, whereas the pressure terms of the stress budgets
depend on the flow over a region, of the order of the integral scale, and not on
the local state alone, which makes them harder to interpret. The connection to
the wall friction is consequently indirect. The vorticity and the Reynolds shear
stress are related by the kinematic identity of
Eq.~(\ref{eq:shear_stress_vorticity}),
$-\partial_y\langle u''v''\rangle=\langle v''\omega_z''\rangle-\langle w''\omega_y''\rangle$,
which constrains the gradient of the stress and not the rate at which vorticity
is tilted. Within $5\le y^+\le12$, the correlations $\langle v''\omega_z''\rangle$
and $\langle w''\omega_y''\rangle$, obtained from the phase ensemble
(\S\ref{sec:phase_analysis}), reproduce the directly differentiated stress
gradient (Appendix~\ref{app:verification}), and $-\langle w''\omega_y''\rangle$,
associated with the streaks, provides about 85\,\% of its time average; only this
term is considered below. Each correlation is the product of a correlation
coefficient and of the r.m.s. values of the two fluctuations,}
\begin{equation}
\prfrev{\langle w''\omega_y''\rangle=\rho_{w\omega_y}\,
 \langle w''^2\rangle^{1/2}\langle\omega_y''^2\rangle^{1/2}.}
\end{equation}
\prfrev{For the shape-optimised waveform, the time average of
$\langle w''\omega_y''\rangle$ is 0.73 times the sinusoidal value. The ratios of the time-averaged factors, whose product is approximately
this ratio, are 0.84 for the r.m.s. of $\omega_y''$, 0.90 for the magnitude of
the correlation coefficient and 0.96 for the r.m.s. of $w''$. The reduction of
the dominant term of the stress gradient is therefore accounted for chiefly by the weaker
wall-normal vorticity of the streaks, consistently with the lower lift-up
(\S\ref{sec:enstr_wy}), and to a lesser extent the weaker correlation.}

\prfrev{The wall friction follows from the phase-averaged streamwise-momentum
equation. With the pressure and the wall shear stress $\tau_w$ each divided by
the density and $\Pi(t)=-\partial_x\langle p\rangle$,}
\begin{equation}
\prfrev{\partial_t\langle u\rangle=\Pi(t)+\nu\partial_y^2\langle u\rangle
 -\partial_y\langle u''v''\rangle,\qquad
 \tau_w(t)=\nu\left.\partial_y\langle u\rangle\right|_{y=0},}
\end{equation}
\prfrev{where $\Pi(t)$ is evaluated at constant flow rate from the wall shear
stress averaged over the two walls. The balance closes to 1.0 and 1.2\,\% of the
r.m.s. stress gradient in the band, whilst the tendency $\partial_t\langle u\rangle$
amounts to 17 and 15\,\% of that gradient (Appendix~\ref{app:verification}). This
tendency is phase-locked to the actuation and not numerical: as the flow rate is
fixed, its wall-normal integral vanishes, and it measures a reshaping of the
near-wall profile by the phase variation of the stress gradient. The wall
friction is defined by the wall-normal gradient of $\langle u\rangle$ at the wall,
which responds to the stress gradient integrated over height and over the
preceding part of the cycle. A change of $-\langle w''\omega_y''\rangle$ at a
given phase within $5\le y^+\le12$ need not be accompanied by a proportional
change of $\tau_w$ at that phase.}

\prfrev{The budgets exclude a term-for-term subtraction for either tilting path: the budget
of $\langle\omega_y''^2\rangle$ contains only Y1 and $-\mathrm{Y2}$ as mean-flow sources
(Eq.~(\ref{eq:Py})) and no term equal and opposite to Z1 or to $-\mathrm{X2}$
(\S\ref{sec:enstrophy}), and the streamwise budget (Eq.~(\ref{eq:Px})) contains no
mean-flow term carrying spanwise enstrophy into it; the production of spanwise
enstrophy therefore does not act on the streaks by subtracting directly from
their enstrophy. The budgets also exclude a direct action of the Stokes strain on the
streaks within their own budget, since the only such term, Y1, is a source in the cycle average (negative in R2), remains smaller
than $-\mathrm{Y2}$ in every interval and at every threshold, and is not lower for the
shape-optimised waveform (\S\ref{sec:enstr_wy}); a modulation of lift-up by the mean
shear, which varies by about 2\,\% over the cycle whereas $-\mathrm{Y2}$ varies by a factor
of about three; and a net destruction of wall-normal enstrophy during Displacement, in
which the measured tendency is positive. They remain compatible with a weaker
replenishment accompanied by a dissipation that falls proportionally less than the
production (Table~\ref{tab:cyy_balance}, for the phases of decreasing
$\langle\omega_y''^2\rangle$), with changes of both the r.m.s. and the correlation
coefficient of $\omega_y''$, with nonlinear stretching, turbulent transport, viscous
diffusion and pressure-mediated changes of the velocity field, none of which is isolated
here, and with an ordering in which $\Theta_x$ responds to the streak amplitude built by
lift-up (\S\ref{sec:enstr_tilting}). The pressure
gradient has zero curl and provides no direct vorticity source within the flow,
but it acts on the velocity-gradient correlations that the budgets contain. Three
elements of a route from the tilting to the drag are nevertheless supported individually: $|G|/S$
orients the tilting of $\omega_y''$ (\S\ref{sec:enstr_synthesis}); the r.m.s. of
$\omega_y''$ and the correlation coefficient account for most of the reduction of
$-\langle w''\omega_y''\rangle$, which provides the dominant part of the stress
gradient through the identity above; and the drag follows from the weighted
wall-normal integral of the time-averaged stress (\S\ref{sec:enstr_synthesis}). The link between the
redirection of the tilting and the weaker $\omega_y''$, namely a weakening of the
streaks through a reduction of their regeneration, rests on the
interpretation of Figure~\ref{fig:enstrophy_sketch}.}

\subsection{\prfrev{Stokes-layer response and relation to earlier studies}}
\label{sec:literature_connections}
\prfrev{Viscous diffusion makes the phase-averaged near-wall spanwise flow within the Stokes layer depend on the preceding
wall motion, and not on the instantaneous wall velocity alone. In the laminar Stokes
approximation, a Fourier harmonic of order $n$ has penetration depth
$\delta_n=\sqrt{2\nu/(n\omega)}$, $\omega=2\pi/T$ being the angular frequency of the
actuation, and a height-dependent phase shift. The tilting balance of \S\ref{sec:enstr_tilting} uses the strain at each height, and therefore this phase-shifted strain and not the instantaneous wall velocity; the measured offset of $\langle\omega_y''\partial_xv''\rangle$ behind $G$, about 0.14 cycle (Appendix~\ref{app:verification}), is shorter than a complete Reversal of the shape-optimised waveform, which occupies about one third of the cycle. A change in the waveform harmonics
therefore changes the shear and the curvature experienced at each wall-normal location,
even at fixed $W^+$ and $T^+$. The wall-velocity reversal, the local zero-crossing of the
shear and the threshold boundaries of the Reversal interval are thus distinct instants, which are
distinguished in the definition of the Reversal and Displacement intervals
(\S\ref{sec:duty_cycle}), and the turbulence adds its own finite adjustment time. The
acceleration scaling recalled in \S\ref{sec:intro} \citep{Ding2024} rests on $W^+/T^+$,
which by definition is the same for both waveforms. For the shape-optimised waveform,
however, the temporal distribution of the acceleration is not uniform, and thus differs
from that of the sinusoidal actuation (\S\ref{sec:actuation}). The cycle-mean acceleration magnitudes differ by approximately 5\,\% (0.54 and 0.56), whereas the r.m.s. values differ by 53\,\% and the peaks by a factor of about 3.1; the larger drag reduction of the shape-optimised waveform is therefore compatible with a role of the concentration of the acceleration around the reversals, and not of its mean value. If the parameter of \citet{Ding2024} is generalised to the cycle-mean acceleration magnitude, equal to $4W^+/T^+$ for a sinusoid, the shape-optimised waveform raises it by only approximately 5\,\%, so that the increment measured here is unlikely to be attributable to a higher cycle-mean acceleration alone; neither the mean acceleration nor its concentration around the reversals is isolated here as the cause.}

\prfrev{A comparable organisation of the strain has been observed under steady spatial square-wave actuation, away from the optimal wavelength. At a post-optimal wavelength, \citet{Knoop2025} observed attenuation following
abrupt strain changes and recovery during the subsequent slowly varying interval, whereas at the suboptimal and near-optimal wavelengths the response was quasi-homogeneous in the streamwise direction. Under the convective transformation used by these authors, their near-optimal wavelength corresponds to a period close to $T^+\approx100$, near the present one, at which the spatial forcing gives a quasi-homogeneous response whereas the temporal forcing shows a marked phase dependence; the correspondence between the two forcings is therefore partial. These observations concern spatial variations of the Reynolds stresses and of the skin friction; they are compatible with the present temporal organisation, without identifying the conditional enstrophy-production rates. Under streamwise-travelling waves at $Re_\tau=180$,
\citet{umair2023vorticity} identified the attenuation of the spanwise enstrophy at the wall,
connected to the fluctuating wall shear stress, as the primary effect of the forcing, the
attenuation of the wall-normal enstrophy being subordinate. For large amplitudes, this
attenuation arises from the direct straining by $\partial W/\partial x$, which is absent in
the homogeneous oscillation considered here; for the latter, they report the spanwise
enstrophy damped in the viscous sublayer, with its maximum near $y^+\approx10$. Their
conclusion, drawn for a different forcing and amplitude, therefore does not directly concern the role of $\omega_y''$ in the present streaks. In their numerical experiments, suppressing the near-wall spanwise vorticity also attenuates $\omega_y''$, the two components being connected in that regime; the direction of the influence is not tested by the present budgets. The tilting of $\omega_y''$ by the Stokes
strain, which gives Z1 here, belongs to the forcing-related production terms of their
spanwise-enstrophy budget.}

\prfrev{In plane Couette flow at $Re_\tau\approx33$, \citet{bengana2022exact} found that
sinusoidal wall oscillation stabilises the upper-branch exact coherent states and reduces
the turbulence production. The reduction appears mainly in the period-averaged lift-up
production of the streamwise-uniform field. The inhibition of the lift-up is proposed, with
the reservation `presumably', as the central mechanism of the stabilisation. For the drag
reduction, a mixing between the low- and high-speed streaks is proposed, the near-wall part
of each streak being displaced beneath its neighbour by the Stokes layer. The pair thereby
becomes more symmetric and the wall shear is lowered, and the two proposals are presented
as consistent. The comparison of the authors is between a controlled and an uncontrolled
state, whereas the present budgets compare two drag-reducing waveforms, for which
$-\mathrm{Y2}$ is lower for the waveform with the larger drag reduction
(\S\ref{sec:enstr_wy}). The present results are therefore compatible with theirs, and extend a lower production associated with lift-up, measured here in the wall-normal enstrophy budget and not in the energy budget, to a pair of controlled states, resolved in phase and in wall-normal distance. For the streamwise-uniform component of their decomposition,
$\omega_y''=\partial_zu''$, and the contribution of these modes to Z1 equals their
contribution to $\langle\omega_y''\omega_z''\rangle G$, in which $G$ tilts $\omega_y''$
towards $z$. The direct contribution of the Stokes strain to the wall-normal vorticity equation, Y1, vanishes for these modes (\S\ref{sec:enstr_wy}; Appendix~\ref{app:enstrophy_equations}), and so do X1, $-\mathrm{X2}$ and $\Theta_x$, which require streamwise variation (\S\S\ref{sec:enstr_wx} and~\ref{sec:enstr_tilting}), whereas Z1, $\Theta_z$ and $-\mathrm{X0}$ remain. The Stokes strain still forces the spanwise velocity of these modes through the term $-Gv''$ of the spanwise momentum equation; this term does not enter the equation of $\omega_y''$, which it reaches through the velocity field, linearly, and not only through the nonlinear terms. A streamwise-uniform description therefore retains the spanwise tilting path and cannot contain the streamwise one.}

\prfrev{The two tilting paths offer a possible explanation of a lower lift-up. This
explanation rests on the durations of the paths and not on their relative magnitude, as the
tilting towards $z$ prevails over most of the cycle for both waveforms
(\S\ref{sec:enstr_tilting}). The tilting towards $z$, a direction outside the
mean-flow route of the sketched self-sustaining process (Figure~\ref{fig:enstrophy_sketch}), prevails in Displacement,
whereas the tilting towards $x$ strengthens during Reversal. No mean-flow term returns spanwise enstrophy to the wall-normal or streamwise budget, and Z1 is balanced mainly by dissipation (\S\ref{sec:enstr_wz}); the spanwise component is not thereby removed from the coupled dynamics, in which it takes part through the nonlinear stretching $\mathcal S_z$ and through $\langle v''\omega_z''\rangle$, about 15\,\% of the time-averaged stress gradient in the band (Appendix~\ref{app:verification}). The shape-optimised waveform
approximately halves Reversal at $q=0.5$, and $-\mathrm{X2}$, which carries the tilting towards $x$, is 23\,\% lower for that reason, although the complete mean-flow source of streamwise enstrophy is approximately 4\,\% higher (\S\ref{sec:enstr_wx}). A less active regeneration of the streamwise vorticity, and a lower
lift-up with it, is therefore an interpretation, resting on
Figure~\ref{fig:enstrophy_sketch} (\S\ref{sec:causal_chain}), and not a measured result. The streak mixing proposed by
\citet{bengana2022exact} is compatible with this reading, as the spanwise displacement of
the near-wall part of each streak by the Stokes layer is the counterpart, in the velocity
field, of the tilting of $\omega_y''$ towards $z$ by $G$, which prevails in Displacement.
Linear analyses about the oscillating base flow show a comparable organisation in time. The
optimal perturbation keeps a nearly constant orientation during a half-period before
changing it abruptly \citep{blesbois2013pattern}, and the growth of the streamwise-elongated
streak mode is reduced whereas that of a tilted mode is generally increased \citep{Yakeno2021}. The
orientation held over a half-period corresponds to the strain maintained in Displacement,
although the phases of the two have not been compared. As noted by Yakeno, however, the linear growth of these modes does not correlate with the drag reduction, which should also involve the nonlinear interaction with the streamwise vortices.}

\prfrev{The present results are also consistent with the earlier phase-averaged analyses of
oscillating-wall flow at $T^+=200$ and $Re_\tau=1025$
\citep{agostini2014spanwise,agostini_turbulence_2015}. These studies related the streaks and
the wall-normal enstrophy to a Stokes strain that lingers at $y^+\approx10$--15 in the
phases of drag increase, and also described the tilting towards $z$, which acts mainly within $5\lesssim y^+\lesssim10$, where the near-wall skewness is high. Both features are found here: the
largest time-averaged values of $-\mathrm{Y2}$ are reached near $y^+=20$, above the band,
whereas the tilting towards $z$ prevails at the wall-normal locations sampled within $5\le y^+\le12$
(\S\S\ref{sec:enstr_wy} and \ref{sec:enstr_tilting}). Within the band, $-\mathrm{Y2}$ peaks at the start of Reversal, whereas the lobes near $y^+=20$ occur later, in Displacement, with a phase delay that increases with the distance from the wall (\S\ref{sec:enstr_wy}). The
wall-normal vorticity may thus be generated close to the wall, strengthened over Reversal
and found farther from the wall during Displacement. This reading is an interpretation, as
the phase maps do not follow individual structures, and the two studies differ in Reynolds
number and period. The recovery of the streaks during the slowly varying interval at $T^+=200$
(\S\ref{sec:intro}) and the lower lift-up in Displacement at $T^+\approx111$ are therefore not
descriptions of the same state. The following comparison of the two periods is an interpretation. The response to a slowly varying strain depends on its magnitude
and on the time available for the streaks to regenerate. At $T^+=200$, the Stokes layer is thicker
and the strain near the wall weaker, and the longer interval leaves more time for recovery
\citep{agostini2014spanwise,agostini_turbulence_2015,blesbois2013pattern,quadrio_critical_2004}; at
$T^+\approx111$, the optimum of the present search and within the reported sinusoidal optimum range, the strain held in Displacement is stronger, and,
within $5\le y^+\le12$, the growth of the wall-normal variance arises mainly from viscous diffusion
and does not establish a regeneration of the streaks (\S\ref{sec:enstr_wy}). A strain held for
longer than the present $2/3$ of the period is not tested.}

\subsection{Reynolds-number scope and large-scale footprints}
\label{sec:reynolds_scope}
\prfrev{Whether the present description persists as the Reynolds number increases is not tested here, and the following distinction delimits what may be expected to be retained. At fixed waveform, $T^+$ and $W^+$, the imposed wall motion, and with it the Stokes strain and its distribution over the cycle, are prescribed in wall units, and the ordering of the two tilting paths follows the strain. The measured duration ratios, the conditional production rates and the drag-reduction increment are, in contrast, responses of the turbulence, the drag-reduction increment being the response documented to change with $Re_\tau$. For purely temporal oscillation, the optimal period decreases in wall units, from $T^+=100$ at $Re_\tau=200$ to $T^+=79$ at $Re_\tau=2000$, whereas the maximum drag reduction falls to $23.2\,\%$ at the latter Reynolds number \citep{ricco_review_2021}. The decrease of the drag reduction is slower than the power law in $Re_\tau^{-0.2}$ suggested empirically by early studies \citep{choi2002drag}. It depends partly on the metric, as the shift of the additive constant of the logarithmic law, expressed in actual wall units, is independent of $Re_\tau$ over the Reynolds numbers considered \citep{gatti_reynolds-number_2016,gatti2025skin}.}

\prfrev{The decrease is attributed in part to the footprints of the outer motions on the near-wall small scales \citep{Mathis2009,marusic_high_2010}. In the unactuated flow, the quasi-steady quasi-homogeneous (QSQH) framework accounts for the footprints, as the small-scale statistics become universal once normalised by the local large-scale friction velocity $u_{\tau,LS}$ \citep{zhang_quasisteady_2016,agostini_validity_2016,chernyshenko_extension_2021}. The following use of the effective period for the actuated flow is a hypothesis, not a tested extension. Under oscillation, the normalisation by $u_{\tau,LS}$ no longer suffices, as $u_{\tau,LS}$ also changes the effective period $T u_{\tau,LS}^2/\nu$ \citep{agostini_statistical_2021}, and the large-scale ejections and sweeps create patches in which the effective period departs from the nominal one. Under an ejection, the effective period falls below the optimum and modifies little the already weak turbulent structures of the patch. Under a sweep, the effective period exceeds the optimum and the turbulent motion resumes, with a response that increases faster than linearly with the large-scale friction, in contrast with the nearly linear modulation of the unactuated flow \citep{agostini_statistical_2021}. As the effective period varies with the square of $u_{\tau,LS}$, its distribution is skewed towards long periods. With increasing Reynolds number, the large-scale fluctuations would become larger, which would widen the positive tail and degrade the control efficiency increasingly. At a low Reynolds number, where the large-scale footprints are weak, a period $T^+=325$ can produce a drag increase \citep{guerin2026PBO}; strong sweeps could therefore produce a similar increase locally. A lower nominal period would keep the positive tail closer to the optimum, a possible contribution to the decrease of the optimal period with the Reynolds number. The rescaling of a period is a further assumption, as the large-scale friction velocity must then remain nearly constant over a whole period, which the large-scale motions do not strictly satisfy \citep{zhang_quasisteady_2016}.}

\prfrev{The mechanism proposed here concerns the control of the small scales only, and its efficiency is expected to change with the Reynolds number as the outer motions strengthen. Control of the outer motions is therefore expected to become relevant at high Reynolds number, as the experiments at $4500\le Re_\tau\le15000$ suggest. Inner-scaled actuation reaches up to $25\,\%$ drag reduction, although mostly with a net power loss, whereas outer-scaled actuation gives $5$--$15\,\%$ with a positive net power saving by acting on the inner--outer coupling \citep{chandran2023turbulent,deshpande2023relationship}.}

\section{Conclusion}
\label{sec:conclusions}

\prfrev{The present simulations compare sinusoidal and shape-optimised spanwise wall
oscillations at $Re_{\tau,0}\approx200$, with the same peak wall velocity and period. The gross drag
reduction is higher for the shape-optimised waveform by $2.3\pm0.6$ percentage points, i.e.
approximately 6\,\% in relative terms, whilst the power supplied by the wall motion is more than 1.5
times as large. The principal outcome is therefore not a net energetic gain.}

\prfrev{The shape-optimised waveform approximately halves the duration of Reversal. Within
$5\le y^+\le12$, it holds the Stokes strain $G$ at a nearly constant level during Displacement, which
occupies about $2/3$ of the cycle against $1/3$ for the sinusoidal waveform (\S\ref{sec:duty_cycle},
$q=0.5$; the contrast persists at every threshold examined). The partition organises the measured
activity without implying a switching between distinct states.}

\prfrev{The principal new result, conveyed by Figure~\ref{fig:enstrophy_sketch}, is the
quantification, by the component-enstrophy budgets, of the concurrent action of two tilting paths for the wall-normal
vorticity $\omega_y''$ associated with the near-wall streaks, whereas the earlier association of the Stokes strain
with the phases of drag increase rested on phase relationships \citep{agostini_turbulence_2015}. The mean shear tilts it
towards the streamwise direction, the principal part of $-\mathrm{X2}$ in the time average, a production term of the streamwise enstrophy, and the Stokes strain tilts it towards the spanwise direction, the principal part of Z1 in the time average; the SSP loop itself is not measured. The magnitude of $G$ orders the two effective tilting
rates, the spanwise path prevailing over most of the cycle and the streamwise path only around the
change of sign of $G$. Both paths remain active in every interval, and neither is paired with a loss
in the wall-normal budget, which contains no term corresponding to Z1 or $-\mathrm{X2}$.}

\prfrev{The waveform modifies the two paths through different factors. Over the same band, the time
average of $-\mathrm{X2}$ is 23\,\% lower for the shape-optimised waveform, although its conditional
rate is higher in every interval; the decrease is therefore accounted for by the shorter Reversal in the duration--rate accounting over cycles 31--75. The
complete mean-flow source of streamwise enstrophy,
$\mathcal{P}_x=\mathrm{X1}-\mathrm{X2}-\mathrm{X0}$, is nevertheless approximately 4\,\% higher, as the
increases of X1 and $-\mathrm{X0}$ exceed the decrease of $-\mathrm{X2}$, whereas Z1 is some 8\,\%
lower. The lift-up production $-\mathrm{Y2}$ is lower at every phase, by about 35\,\% in the time
average and principally through lower conditional rates; its modulation over the cycle resides in the
correlation $\langle\omega_y''\partial_zv''\rangle$, as the mean shear $S$ varies by only 2\,\%. The streak term
$-\langle w''\omega_y''\rangle$ provides about 85\,\% of the time-averaged gradient of the Reynolds
shear stress, whose weighted wall-normal integral gives the turbulent contribution to the drag; for the shape-optimised
waveform, its time average is 0.73 times the sinusoidal value, chiefly through the weaker
$\omega_y''$.}

\prfrev{The conditional levels support a duty-cycle description of the near-wall activity, in which
the duration of each interval and the production rate within it are distinguished. At fixed waveform, $T^+$ and
$W^+$, the imposed wall motion is prescribed in wall units and the ordering of the two tilting paths
follows the Stokes strain; the framework may, as a hypothesis, be extended to other waveforms and
Reynolds numbers, for which the duration ratios, the conditional rates and the drag-reduction
increment are the responses to be measured (\S\ref{sec:reynolds_scope}). The shape-optimised waveform serves here as a probe of the changes that accompany the temporal organisation of the actuation, not as an isolated test of duration. The reading of these changes
as a less active regeneration of the near-wall streaks, and the placing of the spanwise path outside the mean-flow route of the sketched self-sustaining process, are interpretations resting on
Figure~\ref{fig:enstrophy_sketch} and not measured results. An independent effect of duration on the drag and a route from
Z1 alone to the drag remain to be established; at high Reynolds number, the control of the outer
motions and of their interaction with the near-wall cycle is expected to become relevant.}

\begin{acknowledgments}
This work was supported by the French National Research Agency (ANR) through the
ANR Young Researchers (JCJC) programme under grant ANR-23-CE46-0004.  The
simulations were performed using the open-source Xcompact3d framework, publicly
available at \url{https://github.com/xcompact3d/Incompact3d}.  The PBO
optimisation algorithm implementation and the data supporting the results of this
study are available from the corresponding author upon reasonable request.  The
author reports no conflict of interest.
\end{acknowledgments}

\appendix
\section{Stochastic enstrophy budget equations}
\label{app:enstrophy_equations}

\prfrev{For the triple decomposition of \S\ref{sec:methodology},
$u_i=\overline{u_i}+\widetilde{u_i}(t^*)+u_i''$, with $\langle\cdot\rangle$ denoting the phase average, the stochastic vorticity is
$\omega_i''=\epsilon_{ijk}\partial_ju_k''$. The phase-averaged velocity is
$(\langle u\rangle,0,\langle w\rangle)$, with
$S=\partial_y\langle u\rangle$ and $G=\partial_y\langle w\rangle$, so that the
phase-averaged vorticity is $\boldsymbol\Omega=(G,0,-S)$. Subtraction of the
phase-averaged vorticity equation from the instantaneous equation gives}
\begin{equation*}
  \partial_t\omega_i''+\langle u_j\rangle\partial_j\omega_i''
  =-v''\,\partial_y\Omega_i+\Omega_j\,\partial_ju_i''
  +\omega_y''\,\partial_y\langle u_i\rangle
  +\big(\omega_j''\partial_ju_i''-u_j''\partial_j\omega_i''\big)'
  +\nu\nabla^2\omega_i'' ,
\end{equation*}
\prfrev{in which the prime on the bracket denotes the removal of its phase average and
$\langle u_i\rangle=(\langle u\rangle,0,\langle w\rangle)$.
Multiplication by $\omega_i''$ and averaging at fixed phase gives, for
each component (no summation over $i$; the mean advection of the variance vanishes, as noted below)}
\begin{equation}
  \label{eq:enstr_generic}
  \frac{1}{2}\frac{\partial\langle\omega_i^{\prime\prime 2}\rangle}
  {\partial t}
  = \mathcal{P}_i + \mathcal{S}_i + \mathcal{T}_i + \mathcal{D}_i - \mathcal{E}_i,
\end{equation}
\prfrev{where $\mathcal P_i=-\langle v''\omega_i''\rangle\,\partial_y\Omega_i
+\Omega_j\langle\omega_i''\partial_ju_i''\rangle
+\langle\omega_i''\omega_y''\rangle\,\partial_y\langle u_i\rangle$
is the mean-flow production by stretching, tilting and curvature.
The remaining terms are nonlinear stretching,
$\mathcal S_i=\langle\omega_i''\omega_j''\partial_ju_i''\rangle$;
turbulent transport,
$\mathcal T_i=-\tfrac12\partial_y\langle v''\omega_i''^2\rangle$;
viscous diffusion,
$\mathcal D_i=\nu\partial_y^2(\tfrac12\langle\omega_i''^2\rangle)$;
and dissipation,
$\mathcal E_i=\nu\langle(\partial_j\omega_i'')^2\rangle\ge0$.
Summation applies to the repeated index $j$ only; $\mathcal S_i$ has no
additional factor $1/2$ under the half-variance convention. The mean
advection $\langle u_j\rangle\partial_j$ of the averaged variance vanishes, as the statistics are
homogeneous in $x$ and $z$ and $\langle v\rangle=0$.}

\prfrev{Homogeneity in $x$ and $z$ leaves the mean gradients
$S$, $G$, $\partial_yS=\partial_y^2\langle u\rangle$ and
$\partial_yG=\partial_y^2\langle w\rangle$, which are retained at each phase.
As $\overline w=0$, $G=\partial_y\widetilde w$ and $\overline G=0$, whereas $S$ varies by about
2\,\% over the cycle (\S\ref{sec:enstr_wy}).
The parallel mean field has $\Omega_y=0$ and $\langle v\rangle=0$. The curvature
production $-\langle v''\omega_y''\rangle\partial_y\Omega_y$ therefore vanishes in
the wall-normal budget, whilst, as no mean velocity gradient involves $v$, no
tilting term of the mean flow enters the $\omega_y''$ equation. No regrouping of the
terms of the other two components can thus produce a counterpart of Z1 or $-$X2 in the
wall-normal budget. Developing boundary layers and curved-wall flows
require separate balances, including mean advection, additional mean gradients
and applicable curvilinear terms, and separate assessment of
the Reversal/Displacement interpretation. The pressure gradient contributes no term to the
vorticity equation, as the curl of a gradient vanishes at constant density; it
acts on the budgets through the velocity-gradient correlations.}

\noindent\textbf{Streamwise component ($\omega_x^{\prime\prime}$).}
\prfrev{Substitution of
$\partial_zu''=\omega_y''+\partial_xw''$ combines two of the four
mean-flow contributions into the single term $-\mathrm{X2}$:}
\begin{equation}
  \label{eq:Px}
  \mathcal{P}_x =
  \underbrace{
    \langle\omega_x^{\prime\prime}\,\frac{\partial u^{\prime\prime}}{\partial x}\rangle
    \frac{\partial\langle w \rangle}{\partial y}
  }_{\text{X1 (Stokes production)}}
  \;-\;
  \underbrace{
    \langle\omega_x^{\prime\prime}\,\frac{\partial w^{\prime\prime}}{\partial x}\rangle
    \frac{\partial\langle u \rangle}{\partial y}
  }_{\text{\prfrev{X2 (mean-shear production)}}}
  \;-\;
  \underbrace{
    \langle v^{\prime\prime}\,\omega_x^{\prime\prime}\rangle
    \frac{\partial^2\langle w \rangle}{\partial y^2}
  }_{\text{X0 (curvature)}} .
\end{equation}
\prfrev{The signed sources $-\mathrm{X2}$ and X1 require streamwise variation, whereas $-\mathrm{X0}$ does not.
The exact split of $-\mathrm{X2}$, namely
$\langle\omega_x''\omega_y''\rangle S-\langle\omega_x''\partial_zu''\rangle S$,
retains its second term (\S\ref{sec:enstr_wx}); $-\mathrm{X2}$ is not replaced
by the first term alone. Over $5\le y^+\le12$, X1 contributes
17\% and 22\% of the complete mean-flow source $\mathcal P_x$ for the sinusoidal and
shape-optimised waveforms, respectively.}

\noindent\textbf{Wall-normal component ($\omega_y^{\prime\prime}$).}
\prfrev{The vanishing mean wall-normal vorticity eliminates curvature
production:}
\begin{equation}
  \label{eq:Py}
  \mathcal{P}_y =
  \underbrace{
    \langle\omega_y^{\prime\prime}\,\frac{\partial v^{\prime\prime}}{\partial x}\rangle
    \frac{\partial\langle w \rangle}{\partial y}
  }_{\text{Y1 (Stokes production)}}
  \;-\;
  \underbrace{
    \langle\omega_y^{\prime\prime}\,\frac{\partial v^{\prime\prime}}{\partial z}\rangle
    \frac{\partial\langle u \rangle}{\partial y}
  }_{\text{\prfrev{Y2 (mean-shear production)}}} .
\end{equation}
\prfrev{The mean-shear source $-\mathrm{Y2}$ dominates the mean-flow production in the time average.
The maximum over $y$ of the time-averaged Y1 is approximately $1/8$ and $1/7$ of that of
$-\mathrm{Y2}$, whose maximum lies near $y^+\approx21$, for the sinusoidal and shape-optimised
waveforms, respectively; the ratios over $5\le y^+\le12$ are
larger (\S\ref{sec:enstr_wy}). Y1 vanishes for streamwise-uniform fluctuations,
but need not vanish for the full field.}

\noindent\textbf{Spanwise component ($\omega_z^{\prime\prime}$).}
\prfrev{The identity
$\omega_y''=\partial_zu''-\partial_xw''$ and incompressibility give}
\begin{equation}
  \label{eq:Pz}
  \begin{aligned}
  \mathcal{P}_z ={}&
  \underbrace{
    \langle\omega_z^{\prime\prime}\,\frac{\partial u^{\prime\prime}}{\partial z}\rangle
    \frac{\partial\langle w \rangle}{\partial y}
  }_{\text{Z1 (Stokes production)}}
  \;+\;
  \underbrace{
    \langle\omega_z^{\prime\prime}\,\frac{\partial u^{\prime\prime}}{\partial x}\rangle
    \frac{\partial\langle u \rangle}{\partial y}
  }_{\text{Z2 (streamwise-gradient stretching)}}
  \\ &\;+\;
  \underbrace{
    \langle\omega_z^{\prime\prime}\,\frac{\partial v^{\prime\prime}}{\partial y}\rangle
    \frac{\partial\langle u \rangle}{\partial y}
  }_{P_{z,\mathrm{rem}}\ \text{(wall-normal-gradient part)}}
  \;+\;
  \underbrace{
    \langle v^{\prime\prime}\,\omega_z^{\prime\prime}\rangle
    \frac{\partial^2\langle u \rangle}{\partial y^2}
  }_{\text{Z0 (curvature)}} .
  \end{aligned}
\end{equation}
\prfrev{Z2 and $P_{z,\mathrm{rem}}$ are the two parts of the stretching of the mean
spanwise vorticity by the fluctuating spanwise strain,
$-\langle\omega_z''\partial_zw''\rangle S$. Over $5\le y^+\le12$, Z1 dominates the mean-flow
production, and Z0, of sign opposite to that of Z1, is about 1\,\% and 0.7\,\% of Z1 for the sinusoidal and shape-optimised waveforms, respectively.}

\noindent\textbf{\prfrev{Cross-covariance tilting terms.}}
\prfrev{The mean-flow contributions to the budgets of
$\langle\omega_x''\omega_y''\rangle$ and
$\langle\omega_y''\omega_z''\rangle$ are defined in
Appendix~\ref{app:verification}. Their effective tilting terms,
$\Theta_x=-S\langle\omega_y''\partial_xw''\rangle$ and $\Theta_z=G\langle\omega_y''\partial_zu''\rangle$, are given in Eq.~(\ref{eq:effective_tilting});
only $\Theta_x$ vanishes for streamwise-uniform fluctuations.
Both covariances enter the source splits of
\S\S\ref{sec:enstr_wz} and~\ref{sec:enstr_wx}. As
$\omega_y''=\partial_zu''-\partial_xw''$, multiplication by $\omega_y''$ and averaging give
the exact relation, for $G\ne0$,}
\begin{equation}
  \label{eq:theta_identity}
  \frac{\Theta_z}{G}+\frac{\Theta_x}{S}=\langle\omega_y''^2\rangle ,
\end{equation}
\prfrev{so that the two normalised rates sum to the wall-normal enstrophy. Both are positive over the phases and heights examined
(\S\ref{sec:enstr_tilting}); the identity constrains only their sum. As shown above, no mean-flow term of the other two budgets has a counterpart in the wall-normal budget.}

\prfrev{None of the mean-flow production terms is sign-definite, whereas $\mathcal E_i$ is. The labels X2, X0 and
Y2 denote the quantities defined in Eqs.~(\ref{eq:Px}) and~(\ref{eq:Py}), which enter their budgets with the
signs shown there.}

\section{Quantitative checks}
\label{app:verification}
\begingroup\color{PRFblue}

\prfrev{Unless stated otherwise, the checks and budget phase maps use
the separate 32-phase, cycle-resolved ensemble of cycles 31--75
(\S\ref{sec:phase_analysis}). Paired values below refer to sinusoidal
and shape-optimised actuation, respectively.}

\prfrev{The near-wall vorticity covariances fail the stationarity test:
the fitted cycle trends of $\langle\omega_y''\omega_y''\rangle$ averaged over $5\le y^+\le12$
have $t$-statistics $-2.24$ and $-4.75$. Standard errors are multiplied
by $\sqrt{2\tau_r}$, where $\tau_r\approx0.57$ and 0.50 cycle are the
integrated autocorrelation times of the residuals. The fitted decreases from cycle 31
to 75 are approximately 16\,\% and 25\,\% of the respective window
means. For the per-cycle mean skin friction of the drag records, the same
test over cycles 31--75, with the autocorrelation time bounded below by 0.5 cycle, gives $t$-statistics of $-1.0$ and $-4.9$,
corresponding to fitted decreases of 0.9\,\% and 2.1\,\% of the respective
means over the window, and no significant trend over cycles 7--75
($|t|<1$).}

\prfrev{All statistical products derive from the same simulations.
In code units, the drag records and solver-accumulated phase statistics cover
$400\le t\le1000$; the 32-phase ensemble covers
$394.8\le t\le987.0$. Its stored phases deviate from their nominal
equally spaced values by at most 0.6\,\% of the period.
The vorticity of Figure~\ref{fig:vorticity_phase_variation} uses
separate snapshots sorted into 32 phase bins, without an established
common averaging window. Comparisons assess consistency, not
independent-run replication. In Table~\ref{tab:source_verification},
each half-window ratio differs from its full-window value by less than
2.5\,\%, and all remain below unity; this half-window check was made for
Table~\ref{tab:source_verification} only, and the drift reported above applies to the covariances
of all the tables. These checks do not establish stationarity.}

\subsection{Drag records and actuation measures}

\prfrev{The controlled drag records span $0.002\le t\le1000$, and the analysis window
is $400\le t\le1000$, the recorded statistics-accumulation time of the solver;
the unactuated record, $0.0045\le t\le720$, is analysed over $225.0045\le t\le720$.
The pressure gradient maintains constant flow rate.
Blocks of at least twice the estimated integrated autocorrelation
time give 18, 8 and 26 blocks for sinusoidal, shape-optimised and
unactuated records. Table~\ref{tab:drag_verification} uses first-order
propagation of block standard errors, including the reference
denominator, with the means of the decimated blocks as point estimates (each window is
decimated by ten, and the incomplete last block is discarded); the means of the complete windows give
38.30\,\% and 40.66\,\% and a difference of 2.36 percentage points, within the intervals.
The variances of the two marginal drag reductions are summed for the
difference, which counts the uncertainty of the common unactuated reference twice and is
therefore conservative; a propagation that retains the common reference gives 1.74--2.87.
The approximate 95\,\% intervals ($\pm1.96$ standard errors, the normal approximation) are
conditional on these records, on the window selection, on the drift described above and on the dependence estimates. With only
eight shape-optimised blocks, the nominal coverage is not established;
run-to-run uncertainty is not measured. The subsets reuse the same runs and are not independent
of the full-window estimate; they provide a window-sensitivity check, not a coverage test.
The difference between the two
drag reductions is 2.28 and 2.38 percentage points on the contiguous
subsets of cycles 31--52 and 53--75, and 2.48, 1.77 and 2.75 percentage
points on the subsets 31--45, 46--60 and 61--75; all lie within the
interval of Table~\ref{tab:drag_verification}.}

\begin{table}[!htbp]\color{PRFblue}
\caption{\prfrev{Gross drag reduction and approximate normal intervals.
$\Delta DR$ is the difference between the two drag reductions, evaluated from the unrounded values.}}
\label{tab:drag_verification}
\centering
\begin{tabular}{l@{\kern 1.5em}r@{\kern 1.5em}r}
\toprule
Quantity & Estimate & Approximate interval\\
\midrule
Sinusoidal DR (\%) & 38.37 & 37.88--38.85\\
Shaped DR (\%) & 40.67 & 40.27--41.08\\
Shaped minus sinusoidal, $\Delta DR$ (percentage points) & 2.31 & 1.68--2.94\\
\bottomrule
\end{tabular}
\end{table}

\prfrev{The 400-bin wall-velocity traces give $W^+=15$ in both cases,
with relative sampled-peak shortfall below $2.4\times10^{-5}$, with respect to the Fourier-interpolated peak. With respect to the nominal value, the shortfall of the sinusoidal waveform is $3.4\times10^{-5}$.
Table~\ref{tab:forcing_verification} uses the accumulated 400-bin
fields. Peak absolute values and RMS values are evaluated over phase
and the near-wall bands of both walls, after differentiation, and are
given in nominal wall units for each waveform together with their
ratio. The curvature $\partial_yG=\partial_y^2\langle w\rangle$ of the
mean spanwise velocity differs from $\partial_y^2G=\partial_y^3\langle w\rangle$.}

\begin{table}[!htbp]\color{PRFblue}
\caption{\prfrev{Peak and RMS values of the actuation measures in
$5\le y^+\le12$, in nominal wall units, for the sinusoidal (sin.) and
shape-optimised (shaped) actuation, with the shaped/sinusoidal ratios (computed before rounding).}}
\label{tab:forcing_verification}
\centering
\begin{tabular}{l@{\kern 1.2em}rrr@{\kern 1.5em}rrr}
\toprule
 & \multicolumn{3}{c}{Peak} & \multicolumn{3}{c}{RMS}\\
\cmidrule(r){2-4}\cmidrule(l){5-7}
Quantity & sin. & shaped & ratio & sin. & shaped & ratio\\
\midrule
$G=\partial_y\langle w\rangle$ & 1.44 & 1.59 & 1.104 & 0.674 & 0.812 & 1.205\\
$\partial_yG$ & 0.343 & 0.569 & 1.657 & 0.160 & 0.199 & 1.250\\
$\partial_tG$ & 0.0814 & 0.199 & 2.448 & 0.0380 & 0.0520 & 1.368\\
\bottomrule
\end{tabular}
\end{table}

\prfrev{Equal wall amplitude and period do not equalise these actuation
measures; Z1 also depends on the velocity--vorticity correlation.}

\subsection{Interval contributions and covariance production}

\prfrev{With $C_{ij}=\langle\omega_i''\omega_j''\rangle$,
$S=\partial_y\langle u\rangle$ and $G=\partial_y\langle w\rangle$, the mean-flow contributions to the
cross-covariance budgets are}
\begin{align}
M_{xy}&=\langle\omega_x''L_y+\omega_y''L_x\rangle,\qquad
\prfrev{M_{yz}=\langle\omega_y''L_z+\omega_z''L_y\rangle,}\\
L_x&=G\partial_xu''-S\partial_xw''-(\partial_yG)v'',\qquad
L_y=G\partial_xv''-S\partial_zv'',\\
\prfrev{L_z}&\prfrev{{}=G\partial_zu''-S\partial_zw''+(\partial_yS)v'',}\nonumber
\end{align}
\prfrev{where $L_i$ contains the mean-flow terms of the
$\omega_i''$ equation, excluding mean advection, which vanishes from
the covariance budgets by homogeneity.
The regrouped $L_x$ and $L_z$ contain the effective tilting terms of
Eq.~(\ref{eq:effective_tilting}) after multiplication by $\omega_y''$
and averaging. Nonlinear stretching, transport, diffusion and
dissipation are excluded from $M_{xy}$ and $M_{yz}$.}

\prfrev{At each height, the threshold $q$ applied to $|\partial_tG|$, normalised by its maximum over the cycle at that height,
defines Displacement and Reversal. Reversal is divided into R1 and R2;
R2 receives the extra bin when Reversal contains an odd number of sampled phases.
Over $5\le y^+\le12$, this division lies next to the local
shear-change-rate maximum. For a signed quantity $Q$, the conditional
rate, interval contribution and time average are}
\begin{equation}
\prfrev{r_w(y)=\overline{Q}^{\,w}(y)},\qquad A_w(y)=f_w(y)r_w(y),\qquad
\overline Q(y)=\sum_w A_w(y).
\end{equation}
\prfrev{The source statistics use all 32 phases without half-period
folding. The classifier for the vorticity budgets, variances and phase
maps uses periodic central differences of the ensemble Stokes strain.
Reynolds shear stress and $P_{uu}$ use the same rule on the accumulated
400-bin fields. Local products are formed at each phase and height, folded between the two walls with the parity of each term, and then averaged over the 32 phases and arithmetically over
the grid heights of $5\le y^+\le12$. Signed sources retain their
signs throughout.}

\prfrev{The duration estimator of Figure~\ref{fig:duration_ratio}
instead uses the 400-bin spanwise velocity, periodic spectral
differentiation in time, linearly interpolated threshold crossings
and trapezoidal averaging over $5\le y^+\le12$, with interpolated end
points. The cycle fraction is averaged before conversion to $r$.
These estimators are not numerically interchangeable: at $q=0.5$, the ratio of the shape-optimised to the sinusoidal duration ratio is 4.26 for the estimator of Figure~\ref{fig:duration_ratio}, 4.09 from the band-mean cycle fractions of Table~\ref{tab:interval_rates} (periodic central difference) and 3.94 for the estimator of the original figure.}

\prfrev{For the laminar sinusoidal reference of
Figure~\ref{fig:duration_ratio}, the Stokes strain at each height is
proportional to $\sin(2\pi t/T+\phi(y))$, with local phase shift
$\phi(y)$. Displacement satisfies
$|\cos(2\pi t/T+\phi(y))|\le q$, giving}
\begin{equation}
\prfrev{f_{\mathrm{Disp}}(q)=\frac{2}{\pi}\arcsin q,\qquad
r(q)=\frac{f_{\mathrm{Disp}}}{1-f_{\mathrm{Disp}}}
=\frac{2\arcsin q}{\pi-2\arcsin q},}
\label{eq:threshold_harmonic}
\end{equation}
\prfrev{independently of height, with $r(0.5)=0.5$.}

\prfrev{At every change of interval membership over $0.3\le q\le0.7$,
both waveforms retain the conditional-departure signs of
\S\ref{sec:duty_cycle}: streamwise and wall-normal vorticity variances,
Reynolds shear stress and $P_{uu}$ are below their time averages in
Displacement and above in Reversal; the spanwise variance has the
reverse ordering. Over $5\le y^+\le12$, the smallest band-mean
departure of Displacement and of Reversal as a whole exceeds 3\,\% of the band-mean time average in magnitude.
The vorticity and velocity quantities use the 32-phase and 400-bin
statistics, respectively.}

\prfrev{For cases $a$ (sinusoidal) and $b$ (shape-optimised), the two
exact duration--rate decompositions are}
\begin{align}
\Delta\overline Q
&=\sum_w r_w^a(f_w^b-f_w^a)+\sum_w f_w^b(r_w^b-r_w^a)\quad\text{and}\\
\Delta\overline Q
&=\sum_w r_w^b(f_w^b-f_w^a)+\sum_w f_w^a(r_w^b-r_w^a).
\end{align}
\prfrev{The first and second sums give the duration and conditional-rate
contributions. Local products precede height averaging; normalisation
by the sinusoidal time average gives the percentages in
Table~\ref{tab:accounting}. Section~\ref{sec:enstrophy} quotes the mean
of the two orders; at $q=0.5$ the two orders differ in magnitude by about 12, 11, 3 and 6 percentage points
for $M_{xy}$, $-\mathrm{X2}$, Z1 and $-\mathrm{Y2}$, respectively, their signed difference being the interaction
$\sum_w(f_w^b-f_w^a)(r_w^b-r_w^a)$ between the changes of duration and of rate. These identities describe observed means,
not independently imposed duration changes.}

\begin{table}[!htbp]\color{PRFblue}
\caption{\prfrev{Ratios of the time averages over $5\le y^+\le12$ (shape-optimised/sinusoidal) and their
sensitivity to the splitting of the verification window. All source products are
formed before averaging; row labels denote the time averages.}}
\label{tab:source_verification}
\centering
\begin{tabular}{l@{\kern 1.5em}r@{\kern 1.5em}r@{\kern 1.5em}r@{\kern 1.5em}r}
\toprule
Cycles & $-\mathrm{X2}$ & Z1 & $-\mathrm{Y2}$ & $M_{xy}$\\
\midrule
31--75 & 0.773 & 0.916 & 0.656 & 0.599\\
31--52 & 0.783 & 0.922 & 0.660 & 0.611\\
53--75 & 0.762 & 0.910 & 0.652 & 0.585\\
\bottomrule
\end{tabular}
\end{table}

\prfrev{At $q=0.5$, with the periodic central-difference classifier, Reversal contributes approximately 97\% and 98\%
of signed $M_{xy}$. Table~\ref{tab:accounting} gives both decomposition
orders: at every threshold, duration reduces $M_{xy}$ and
$-\mathrm{X2}$ and increases Z1; conditional rates oppose the duration
contribution for $-\mathrm{X2}$ and Z1.
Near the $-\mathrm{Y2}$ maximum, $20\le y^+\le23$, the
shape-optimised/sinusoidal mean ratio is 0.878. At $q=0.5$, lower
conditional rates reduce the sinusoidal time mean by about 11\%, and
the duration contribution is 1--2\%; the longer near-wall Displacement
does not extend to these heights.}

\prfrev{Table~\ref{tab:interval_rates} gives cycle fractions and rates
at $q=0.5$. Within Reversal, the $-\mathrm{X2}$ rate ratios also
exceed unity at $q=0.3$ and 0.7; the $M_{xy}$ rate ratios exceed unity in
both parts at $q=0.3$ and 0.5 only (these values are not tabulated).
Over $15\le y^+\le30$, the shape-optimised/sinusoidal time-mean ratios
are 1.008, 1.087, 0.874 and 0.902 for $-\mathrm{X2}$, Z1,
$-\mathrm{Y2}$ and $M_{xy}$, respectively.
For the shape-optimised waveform, $-\mathrm{Y2}$ is lower at every
sampled phase and height up to $y^+\approx33$; further from the wall, its
time average remains lower by less than 3\,\% up to $y^+\approx60$. Over
$5\le y^+\le12$, the time-averaged Y1 is approximately 12\,\% higher for
the shape-optimised waveform (25\,\% in the conditional mean of Table~\ref{tab:cyy_balance}).}

\prfrev{For the exact split
$-\mathrm{Y2}=\langle\omega_x''\omega_y''\rangle S-\langle\omega_y''\partial_yw''\rangle S$
(\S\ref{sec:enstr_wy}), the first part is 3.6 and 2.9 times $-\mathrm{Y2}$
in the time average over $5\le y^+\le12$; the second cancels
approximately 70\,\% and 65\,\% of the first.
For $-\mathrm{X2}=\langle\omega_x''\omega_y''\rangle S-\langle\omega_x''\partial_zu''\rangle S$
(\S\ref{sec:enstr_wx}), the first part is 1.19 and 0.82 times $-\mathrm{X2}$;
the remainder has magnitude approximately $1/5$ of $-\mathrm{X2}$
and opposite signs between waveforms.
In Displacement, the ratio of the first part to $-\mathrm{X2}$, formed from the conditional
means, is small, below 0.1, and sensitive to cancellation between heights.
The shape-optimised/sinusoidal time-mean ratios are 0.53 for
the first part of $-\mathrm{X2}$, 0.77 for $-\mathrm{X2}$ and 0.66 for $-\mathrm{Y2}$.
In the Fourier decomposition over homogeneous $x,z$,
$k_x=0$ gives $\partial_xw''=0$ and $\omega_y''=\partial_zu''$;
the remainder therefore cancels the contribution of these modes to
the first part mode by mode, leaving $-\mathrm{X2}$ at $k_x\ne0$ alone.}

\prfrev{For $\mathcal S_z$ (\S\ref{sec:enstr_wz}), the band-mean
conditional means divided by the band-mean time average over
$5\le y^+\le12$ are 0.80, 0.84, 1.32 (sinusoidal) and
0.87, 1.01, 1.46 (shape-optimised) in Displacement, R1 and R2.
Its time average is 0.35 and 0.27 times the time average of Z1.
Its cycle maximum lies in R2 at 13 of 13 and 12 of 13 heights,
respectively; the remaining maximum lies in R1.
Cycle correlations use mean-removed signals at each height,
with delays restricted to $\pm1/4$ cycle for the half-cycle-periodic
signals, then averaged over the band.
$\mathcal S_z$ follows $\langle\omega_y''\omega_y''\rangle$ by 3.3 and 3.2 phase samples
(approximately 0.10 cycle), with peak correlations 0.99 and 0.95,
against 0.24 and 0.18 at zero delay.
The $\langle\omega_y''\omega_y''\rangle$ maximum lies within 1.5 samples of the R1/R2 division at
12 of 13 sinusoidal heights and all shape-optimised heights;
the remaining sinusoidal maximum precedes it by 2.5 samples.
The maximum lies in R1 at 10 and 13 of the 13 heights, respectively.}

\prfrev{Table~\ref{tab:interval_rates} averages $|\Theta_z|$ after
taking the magnitude at each phase and height, to prevent cancellation
between half-periods. Over $5\le y^+\le12$, the fraction of the
32 phases with $\Theta_x>|\Theta_z|$ ranges from 12.5 to 18.8\,\%
and from 6.3 to 18.8\,\%, with band means 15.1\,\% and 13.5\,\%.
All lie in Reversal, roughly evenly divided between R1 and R2, and at these phases $|G|$ does not exceed 30\,\% and 41\,\% of its local
cycle maximum.
Pooling samples with $|G|$ at least 5\,\% of its local maximum
gives correlations 0.90 and 0.89 between the logarithms of
$|\Theta_z|/\Theta_x$ and $|G|/S$ (Eq.~(\ref{eq:tilting_ratio})). At
$y^+=5.3$, 8.4 and 11.5, the fitted slopes of $\log(|\Theta_z|/\Theta_x)$ against
$\log(|G|/S)$ range from 0.95 to 1.11. As $\log(|\Theta_z|/\Theta_x)=\log(|G|/S)+\log K$ (Eq.~(\ref{eq:tilting_ratio})), these slopes and
correlations measure how little $\log K$ varies compared with $\log(|G|/S)$; they are not an independent scaling law.}

\prfrev{The interval ratio $|\Theta_z|/\Theta_x$ exceeds
unity in every interval for both waveforms at $q=0.3$ and 0.5, with a minimum of
1.13 at $q=0.5$. At $q=0.7$ it is 0.66 in R2 of the shape-optimised waveform and
exceeds unity elsewhere, with a minimum of 1.07. At fixed wall-normal location,
the correlation coefficients, at zero delay and averaged over the band, of
$\Theta_x$ with $-\mathrm{Y2}$ are 0.98 and 0.89 and with
$\langle\omega_y''\omega_y''\rangle$ 0.93 and 0.99; those of $|\Theta_z|$ with
$-\mathrm{Y2}$ are $-0.08$ and 0.53, and those of $|\Theta_z|/\Theta_x$ with
$-\mathrm{Y2}$ are $-0.80$ and $-0.62$, for the sinusoidal and shape-optimised
waveforms, respectively.}

\begin{table}[!htbp]\color{PRFblue}
\caption{\prfrev{Fractions of the cycle occupied by the intervals, and shape-optimised/sinusoidal ratios of the conditional rates and time averages over $5\le y^+\le12$ at $q=0.5$. The last
four rows instead give, within each waveform, the ratio of the interval mean of
$|\Theta_z|$, taken at each phase and wall-normal location, to the interval
mean of $\Theta_x$ (Eq.~(\ref{eq:effective_tilting})), and the interval mean of
$\Theta_x$ relative to its Displacement value. In the ratio rows, the interval columns refer to the conditional rates (interval means) and the last column to the cycle averages. The values use the periodic central-difference classifier; other classifier variants change the fractions by about 0.005 and the Displacement rate ratios by up to about 0.03, so that the third digit depends on the estimator. R2 receives the extra bin of an odd Reversal and the division need not bisect Reversal, so that R1 and R2 are compared through the conditional rates.}}
\label{tab:interval_rates}
\centering
\begin{tabular}{l@{\kern 1.5em}r@{\kern 1.5em}r@{\kern 1.5em}r@{\kern 1.5em}r}
\toprule
 & Displacement & R1 & R2 & Time average\\
\midrule
Cycle fraction, sinusoidal & 0.325 & 0.315 & 0.361 & --\\
Cycle fraction, shape-optimised & 0.663 & 0.154 & 0.183 & --\\
$-\mathrm{X2}$ ratio & 1.179 & 1.146 & 1.364 & 0.773\\
$M_{xy}$ ratio & 0.354 & 1.142 & 1.195 & 0.599\\
Z1 ratio & 0.748 & 0.780 & 0.410 & 0.916\\
$-\mathrm{Y2}$ ratio & 0.764 & 0.695 & 0.668 & 0.656\\
\midrule
$|\Theta_z|/\Theta_x$, sinusoidal & 4.007 & 1.594 & 2.204 & --\\
$|\Theta_z|/\Theta_x$, shape-optimised & 3.090 & 1.166 & 1.128 & --\\
$\Theta_x$ relative to Displacement, sinusoidal & 1 & 2.40 & 1.26 & --\\
$\Theta_x$ relative to Displacement, shape-optimised & 1 & 2.51 & 1.46 & --\\
\bottomrule
\end{tabular}
\end{table}

\subsection{Wall-normal balance and streamwise response}

\prfrev{For $E_y=\langle\omega_y''\omega_y''\rangle/2$, the signed verification balance is}
\begin{equation}
\partial_tE_y=\mathrm{Y1}-\mathrm{Y2}+\mathcal S_y+\mathcal T_y+
\mathcal D_y-\mathcal E_y+R_y,
\end{equation}
\prfrev{with terms defined in Appendix~\ref{app:enstrophy_equations}
and residual $R_y$. Table~\ref{tab:cyy_balance} conditions each case
on its own measured $\partial_tE_y<0$ at each height, then averages
over the grid points of $5\le y^+\le12$.
Diffusion only partly offsets the excess of dissipation over the
remaining contributions; no term equal and opposite to Z1 occurs (Appendix~\ref{app:enstrophy_equations}). The conditioning selects on average 16.0 and 14.1 of the 32 phases at the 13 heights, so that the two columns refer to different parts of the cycle. The terms $-\mathrm{Y2}$, $\mathcal S_y$ and $\mathcal T_y$ are reduced in magnitude for the shape-optimised waveform by similar factors (0.64, 0.66 and 0.64); the table therefore does not isolate a single term.}

\begin{table}[!htbp]\color{PRFblue}
\caption{\prfrev{Mean signed terms during decreasing $E_y$, averaged over
$5\le y^+\le12$, in $10^{-4}$ nominal reference wall units (the code-unit values multiplied by $(\nu/u_{\tau,0}^2)^3\approx1.65\times10^{-3}$ and then by $10^4$). The symbols denote conditional means over these phases, not cycle averages.}}
\label{tab:cyy_balance}
\centering
\begin{tabular}{l@{\kern 1.5em}r@{\kern 1.5em}r}
\toprule
Term & Sinusoidal & Shaped\\
\midrule
$\mathrm{Y1}$ & 0.031 & 0.039\\
$-\mathrm{Y2}$ & 1.195 & 0.767\\
$\mathcal S_y$ & 0.392 & 0.260\\
$\mathcal T_y$ & -0.252 & -0.161\\
$\mathcal D_y$ & 0.582 & 0.535\\
$-\mathcal E_y$ & -2.493 & -1.889\\
$R_y$ & 0.047 & -0.001\\
$\partial_t E_y$ & -0.498 & -0.451\\
\bottomrule
\end{tabular}
\end{table}

\begin{table}[!htbp]\color{PRFblue}
\caption{\prfrev{Duration--rate accounting over $5\le y^+\le12$: contributions to the
difference of the time averages $\overline{(\cdot)}$ (shape-optimised minus sinusoidal), in \% of the
sinusoidal time average. Each pair gives the two decomposition orders; the
duration contributions are summed over Displacement, R1 and R2. For $M_{xy}$, $-\mathrm{X2}$,
Z1 and $-\mathrm{Y2}$, the two contributions sum, within rounding, to
$100(R-1)$, $R$ being the corresponding ratio of
Table~\ref{tab:source_verification}. The $M_{xy}$ conditional-rate contribution is small compared with the duration contribution, and its sign depends on the
decomposition order at $q=0.3$ and 0.7. The values use the periodic central-difference classifier;
the duration entries depend on the classifier variant by up to about 1 percentage point.}}
\label{tab:accounting}
\centering
\begin{tabular}{llccc}
\toprule
Source & Contribution & $q=0.3$ & $q=0.5$ & $q=0.7$\\
\midrule
$M_{xy}$ & duration & $(-39.3,-50.8)$ & $(-42.9,-54.9)$ & $(-36.1,-40.3)$\\
 & conditional rate & $(-0.8,+10.7)$ & $(+2.8,+14.8)$ & $(-4.0,+0.2)$\\
$-\mathrm{X2}$ & duration & $(-34.6,-44.7)$ & $(-38.4,-49.8)$ & $(-33.6,-39.2)$\\
 & conditional rate & $(+11.9,+22.0)$ & $(+15.7,+27.1)$ & $(+10.9,+16.5)$\\
Z1 & duration & $(+25.4,+21.5)$ & $(+28.9,+25.8)$ & $(+27.0,+22.9)$\\
  & conditional rate & $(-33.8,-29.8)$ & $(-37.3,-34.2)$ & $(-35.4,-31.2)$\\
$-\mathrm{Y2}$ & duration & $(-9.7,-4.2)$ & $(-10.5,-4.9)$ & $(-8.8,-4.7)$\\
 & conditional rate & $(-24.7,-30.2)$ & $(-23.8,-29.5)$ & $(-25.6,-29.7)$\\
\bottomrule
\end{tabular}
\end{table}

\prfrev{The residual and the terms are first averaged over $5\le y^+\le12$,
and the RMS is then taken over the 32 phases of the cycle: the residual RMS is 2.1/3.3\% of term-vector RMS (the RMS over the 32 phases of the Euclidean norm of the vector of the terms of the balance) and 11/16\% of
unsteady-term RMS. At $y^+=28$, these ratios are 5.9/5.1\%
and 142/156\%: the weak unsteady term cannot be assigned reliably
to individual contributions. The near-wall balance is therefore
used; the measured variance-level difference remains valid.
For $\langle\omega_x''\omega_x''\rangle$ and $\langle\omega_z''\omega_z''\rangle$, band residuals are 0.4/0.5\% and
0.1/0.2\% of term-vector RMS, respectively, and remain below 1.6\%
at $y^+=28$.}

\prfrev{The lag of $\langle\omega_y''\partial_xv''\rangle=\mathrm{Y1}/G$
behind $G$ (\S\ref{sec:enstr_wy}) uses the 32-phase ensemble.
At each of the 13 heights in $5\le y^+\le12$, phases with
$|G|\le15\,\%$ of its local maximum are excluded.
A constant and fundamental harmonic are fitted by least squares to
both signals over the remaining phases. Their phase difference,
averaged over height, is $\Delta t^*\approx0.14$ cycle (0.144 and 0.136), i.e.\ about 16 and 15 wall-time units,
with height-to-height values from 0.12 to 0.16, which are a spread and not an uncertainty.
These offsets establish neither a streak-reorientation time nor
a causal sequence.}

\prfrev{The streamwise response check uses the full streamwise velocity variance
at its time-mean maximum, $y^+=28$ and 30.
The variance follows production by 12.6 and 13.9 nominal reference
wall-time units at the second harmonic. Per-cycle circular means are
11.7/14.9 and circular standard deviations 5.8/8.7, modulo $T^+/2$ (standard errors of the mean over the 45 cycles of about 0.9 and 1.3). The second-harmonic phase is defined modulo $T^+/2$, the value retained being the principal one.
Including production, transport, diffusion, dissipation and pressure
gives residuals of 0.10/0.53\% of term-vector RMS and 3.9/33\%
of unsteady-term RMS.}

\prfrev{For the phase balance of \S\ref{sec:reynolds_stress}, each
selected sum of measured second-harmonic budget terms is divided by
$\mathrm{i}\omega^+$ and compared with production.
All harmonics are measured in the complete flow.
Dissipation follows production by 3.2 and 3.8 wall-time units,
with amplitudes 29\,\% and 30\,\% of the production amplitude;
pressure, turbulent transport and viscous diffusion have relative
amplitudes 32/38\,\%, 20/18\,\% and 8/9\,\%.
Integration of production alone gives $T^+/8\approx13.9$ wall-time
units; production minus dissipation gives 12.7 and 12.4.
Across all combinations added to production minus dissipation,
pressure and viscous diffusion increase the lag by up to 3.7
and 2.1 wall-time units, and turbulent transport reduces it by up to
3.8; as these shifts are not additive, no share of the lag is assigned to
each term.
All terms give 12.4 and 13.4; the difference from the measured offsets
is of the order of the residual, whose second-harmonic amplitude is
1\,\% and 4\,\% of production (not of the net sum of the terms).
Block-jackknife standard errors over nine groups of five cycles are
below 0.2 wall-time units for production-minus-dissipation lags and
below 0.7 for measured offsets.}

\prfrev{Wall-shear modulation (\S\ref{sec:temporal_modulation}) uses
snapshots with the ensemble phase clock; the pressure-gradient record
has no established phase origin.
Mean wall shears are $\tau_w^+=0.6164$ and 0.5931,
38.4\% and 40.7\% below the nominal unactuated $\tau_w^+=1$,
in agreement with the drag records, and differ by 3.8\%.
Peak-to-peak modulations are 0.324\% and 0.270\% of these means;
second-harmonic amplitudes exceed their standard errors by about an order of magnitude
for both waveforms.}

\prfrev{Over $5\le y^+\le12$, the velocity--vorticity identity,
Eq.~(\ref{eq:shear_stress_vorticity}), reproduces the directly differentiated
stress gradient within 0.05\,\% for both waveforms.
The time-mean shares are approximately 85\,\% from
$-\langle w''\omega_y''\rangle$ and 15\,\% from
$\langle v''\omega_z''\rangle$. The time average of $\langle w''\omega_y''\rangle$
of the shape-optimised waveform is 0.73 times the sinusoidal value, the ratios
of the time-averaged wall-normal-vorticity r.m.s., correlation-coefficient magnitude and
spanwise-velocity r.m.s. being 0.84, 0.90 and 0.96, whose product reproduces the ratio of the time averages as an approximate factorisation, the time average of a product not being the product of the time averages. Within each waveform, the magnitude of
$\langle w''\omega_y''\rangle$ in Displacement is 0.63 (sinusoidal) and 0.66
(shape-optimised) of its value in Reversal at $q=0.5$; the ratios of the
spanwise-velocity and wall-normal-vorticity r.m.s. are 0.80 and 0.75
(sinusoidal) and 0.77 and 0.76 (shape-optimised), whereas the magnitude of the correlation
coefficient is 6 and 14\,\% higher in Displacement. In the phase-averaged
streamwise-momentum balance of \S\ref{sec:causal_chain}, with $\Pi(t)$ obtained
from the wall shear stress, the r.m.s. residual is 1.0 and 1.2\,\% of the r.m.s.
stress gradient in the band for the sinusoidal and shape-optimised waveforms,
respectively, and the r.m.s. tendency $\partial_t\langle u\rangle$ is 17 and 15\,\% of it.}

\endgroup

\end{document}